\documentclass[superscriptaddress,onecolumn, preprintnumbers,amsmath,amssymb,11pt]{revtex4-1}

\usepackage{bm}
\usepackage{latexsym}
\usepackage{dsfont}
\usepackage{graphics}
\usepackage{graphicx}
\usepackage{amsthm}
\usepackage{amsmath}
\usepackage{wasysym}
\usepackage{amssymb}
\usepackage{subfigure}
\usepackage{enumerate}
\usepackage{simplewick}
\usepackage{empheq}
\usepackage[dvips]{epsfig}
\usepackage[colorlinks = true]{hyper ref}
\usepackage{float}
\usepackage{multirow}
\usepackage{cancel}
\usepackage{bm}
\usepackage{cancel}
\DeclareMathOperator{\Tr}{\mathrm{Tr}}

\newcommand{\beq}{\begin{equation}}
\newcommand{\eeq}{\end{equation}}
\newcommand{\nn}{\nonumber \\}

\def\bea{\begin{eqnarray}}
\def\eea{\end{eqnarray}}
\hypersetup{
    bookmarks=true,         
    unicode=false,          
    pdftoolbar=true,        
    pdfmenubar=true,        
    pdffitwindow=false,     
    pdfstartview={FitH},    
    pdfnewwindow=true,      
    colorlinks=true,       
    linkcolor=blue, 
    citecolor=blue,        
    filecolor=magenta,      
    urlcolor=blue           
} 
\begin{document}
\title{Quantum field theory for the chiral clock transition in one spatial dimension}
\author{Seth Whitsitt}
 \affiliation{Department of Physics, Harvard University, Cambridge, MA 02138, USA}
 \affiliation{Joint Quantum Institute, National Institute of Standards and Technology and the University of Maryland, College Park, MD, 20742, USA}
\author{Rhine Samajdar}
 \affiliation{Department of Physics, Harvard University, Cambridge, MA 02138, USA}
\author{Subir Sachdev}
\affiliation{Department of Physics, Harvard University, Cambridge, MA 02138, USA}
\affiliation{Perimeter Institute for Theoretical Physics, Waterloo, Ontario, Canada N2L 2Y5}
\date{\today}

\begin{abstract}
We describe the quantum phase transition in the $N$-state chiral clock model in spatial dimension $d=1$. With couplings chosen to preserve
time-reversal and spatial inversion symmetries, such a model is in the universality class of recent experimental studies of the ordering of pumped Rydberg states in a one-dimensional chain of trapped ultracold alkali atoms. For such couplings and $N=3$, the clock model is expected to have a direct phase transition from a gapped phase with a broken global $\mathbb{Z}_N$ symmetry, to a gapped phase with the $\mathbb{Z}_N$ symmetry restored. The transition has dynamical critical exponent $z \neq 1$, and so cannot be described by a relativistic quantum field theory. We use a lattice duality transformation to map the transition onto that of a Bose gas in $d=1$, involving the onset of a single boson condensate in the background of a higher-dimensional $N$-boson condensate. We present a renormalization group analysis of the strongly coupled field theory for the Bose gas transition in an expansion in $2-d$, with $4-N$ chosen to be of order $2-d$. At two-loop order, we find a regime of parameters
with a renormalization group fixed point which can describe a direct phase transition. We also present numerical
density-matrix renormalization group studies of lattice chiral clock and Bose gas models for $N=3$, finding good evidence for a direct phase transition, and obtain estimates for $z$ and the correlation length exponent $\nu$.
\end{abstract}

\maketitle
\tableofcontents
\section{Introduction}

Recent experiments on one-dimensional chains of Rb atoms excited to Rydberg states by \citet{bernien2017probing} have displayed quantum transitions to ordered states with a period of $N$ sites, with $N \geq 2$. 
This phase transition is described by a model  of  hard-core  bosons  proposed  by  Fendley {\it et al.} \cite{fendley2004competing}.
Such phase transitions are in the universality class of the $\mathbb{Z}_N$ clock model with couplings which preserve both time-reversal and spatial inversion symmetries. For $N \geq 3$, the required clock models must be {\it chiral} \cite{huse1982domain,ostlund1981incommensurate}: domain walls have distinct energies depending upon whether the clock rotates clockwise or counterclockwise upon crossing the wall while moving to the right. 

There has been much theoretical and numerical work on $\mathbb{Z}_N$ chiral clock models,
both as quantum models in one spatial dimension ($d$), and as classical models in two spatial dimensions \cite{huse1981simple,ostlund1981incommensurate,huse1982domain,huse1983melting,haldane1983phase,howes1983quantum,au1987commuting,cardy1993critical,fendley2012parafermionic,ZCTH,dai2017entanglement,SCPLS18,au1997many,baxter2006challenge,au1987commuting,albertini1989commensurate, mccoy1990excitation}. These models exhibit a complex phase diagram with 3 types of phases: \\
({\it i\/}) a gapped phase with long-range $\mathbb{Z}_N$ order (this phase was referred to as `topological' in a parafermionic formulation \cite{fendley2012parafermionic,ZCTH}),\\ ({\it ii\/}) a gapped phase with no broken symmetry and exponentially decaying $\mathbb{Z}_N$ correlations, and\\ ({\it iii\/}) a gapless phase with incommensurate $\mathbb{Z}_N$ correlations decaying as a power-law.\\ 
It is important to note, however, that many of the previous studies are under conditions in which the Hamiltonian does not preserve time-reversal and/or spatial inversion symmetries. Imposing time-reversal and spatial inversion symmetries will be crucial for our theoretical analysis, and indeed, such symmetries are present in the Rydberg atom realization \cite{bernien2017probing}.
With these symmetries imposed, we will examine the {\em direct\/} transition between the two gapped phases noted above, without an intermediate incommensurate phase. The possibility of such a direct transition was already noted in early work \cite{huse1981simple}, but was questioned subsequently \cite{haldane1983phase} (see Appendix~\ref{app:luttinger}). However, numerical evidence for a direct transition for $N=3$ has emerged in recent work \cite{ZCTH,SCPLS18}. This paper will provide a field-theoretic renormalization group analysis
of the direct transition, along with additional numerical density-matrix renormalization group (DMRG) results. 
Our main theoretical tool will be a duality mapping of the chiral clock model transition in $d=1$ onto that of a Bose gas, involving the onset of a single boson condensate in the background of a higher-dimensional $N$-boson condensate \cite{DS96}.

\begin{figure}[!htb]
    \centering
    \includegraphics[width=3.9in]{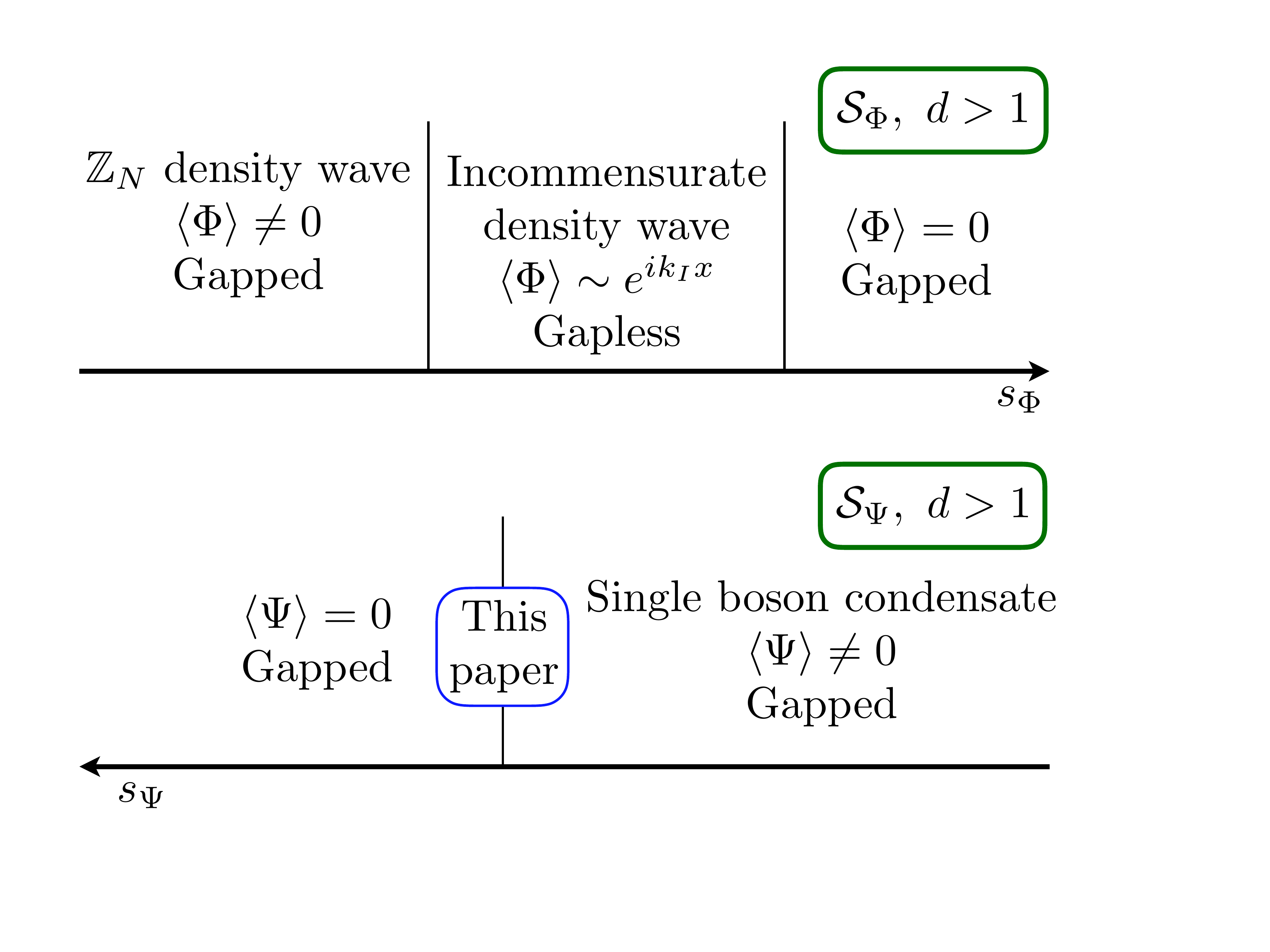}
    \caption{Zero temperature phase diagrams of $\mathcal{S}_\Phi$ (Eq.~(\ref{SPhi})) and $\mathcal{S}_\Psi$ (Eq.~(\ref{SPsi})) in spatial dimensions $d>1$. This paper studies the transition in $\mathcal{S}_\Psi$ in an expansion in $(2-d)$. In $d>1$, $\mathcal{S}_\Phi$ and $\mathcal{S}_\Phi$ describe distinct physical phenomena, and are expected to have different phase diagrams and transitions.}
    \label{fig:pdiag1}
\end{figure}
\begin{figure}[!htb]
    \centering
    \includegraphics[width=3.9in]{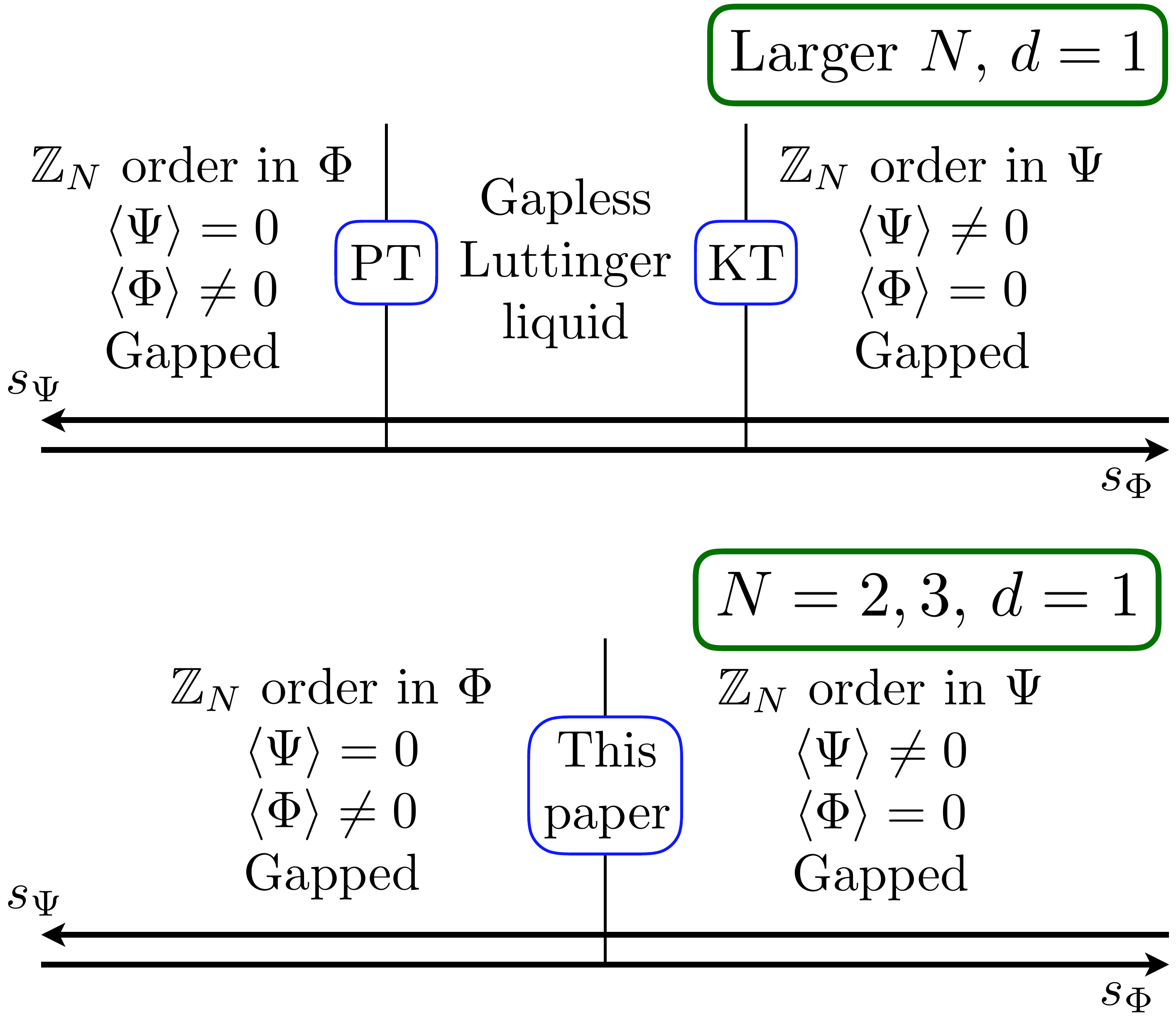}
    \caption{The common zero temperature phase diagrams of $\mathcal{S}_\Phi$ (Eq.~(\ref{SPhi})) and $\mathcal{S}_\Psi$ (Eq.~(\ref{SPsi})) in spatial dimension $d=1$. There is a Kramers-Wannier duality between $\mathcal{S}_\Phi$ and $\mathcal{S}_\Psi$ in $d=1$, and so the two actions have the same phases and transitions. For larger $N$ (possibly all $N \geq 4$) there is an intermediate gapless phase, while for $N=2,3$ there can be a direct transition between gapped phases. This paper describes the direct transition between gapped phases for $N=3$. The transitions out of the gapless
    Luttinger liquid are in the Kosterlitz-Thouless \cite{kosterlitz1973ordering} (KT) and Pokrovsky-Talapov \cite{pokrovsky1978phase,pokrovsky1979ground} (PT) classes.}
    \label{fig:pdiag2}
\end{figure}Let us begin by writing down a possible field theory for period-$N$ ordering \cite{fendley2004competing}.
Let $\Phi$ be the density wave order parameter, so that $\Phi \rightarrow e^{2 \pi i n/N}\Phi $ 
under translation by $n$ lattice spacings, where $n$ is a positive or negative integer. Using translational and time-reversal symmetries (described in more detail below), we obtain an action defined on continuous $d=1$ space ($x$)
and imaginary time ($\tau$):
\beq
\mathcal{S}_\Phi = \int dx \, d\tau \Big[ |\partial_\tau \Phi|^2 + |\partial_x \Phi|^2 + i \alpha_x \Phi^\ast \partial_x \Phi + s_\Phi |\Phi|^2 + u |\Phi|^4 + \lambda \left(\Phi^N + (\Phi^\ast)^N \right) \Big]
\label{SPhi}
\eeq
We show the phase diagram of $\mathcal{S}_\Phi$ in $d>1$ in Fig.~\ref{fig:pdiag1}, and in $d=1$ in Fig.~\ref{fig:pdiag2}.
The field theory $\mathcal{S}_\Phi$ also applies to the chiral clock model with order parameter $\Phi$, in which case $\Phi \rightarrow e^{2 \pi i n/N}\Phi $ is an internal symmetry of the clock model, without combining with spatial translations. So in the clock model, a state with $\langle \Phi \rangle \neq 0$ has a spatially uniform condensate, while this state has period $N$ ordering in the boson model of Ref.~\onlinecite{fendley2004competing}.
The term proportional to the real number $\alpha_x$ is crucial, and is responsible for the chirality in both models. A nonzero $\alpha_x$ yields an inverse propagator for $\Phi$ which has a minimum at a nonzero wavevector $k_I = \alpha_x/2$, and hence induces incommensurate order parameter correlations. When treated perturbatively in $u$ and $\lambda$, $\mathcal{S}_\Phi$ will lead to condensation of $\Phi$ at $k_I$, and hence a to state with long-range incommensurate order. 
Taking $\Phi \sim e^{i k_I x}$, we see that the phase-locking term proportional to 
$\lambda$ spatially averages to zero. Consequently, although $\mathcal{S}_\Phi$ has only a discrete $\mathbb{Z}_N$ symmetry, the low-energy theory of the incommensurate state has an emergent U(1) symmetry which leads to a gapless `phason' mode \cite{Gruner88} (note that this argument applies also in spatial dimensions $d>1$, as illustrated in Fig.~\ref{fig:pdiag1}).
This is the reason for the difficulty in obtaining a theory for the direct transition in the chiral model from a gapped disordered phase, to a commensurate $\mathbb{Z}_N$-ordered phase: the perturbative analysis of the field theory in Eq.~(\ref{SPhi}) implies that such a direct transition does not exist, and there is an intermediate gapless incommensurate phase. On the other hand, there is ample evidence from numerical studies for the existence of a direct transition \cite{ZCTH,SCPLS18} in $d=1$.

One of our main results will be an exact duality between 
models described by $\mathcal{S}_\Phi$ in $d=1$, and a theory of the condensation of a nonrelativistic Bose gas in $d=1$. Specifically, we consider a Bose gas, with Bose field $\Psi$, which undergoes a condensation transition in the presence of a higher-dimensional background condensate of a `molecule' of $N$ bosons. This implies that we always have $\langle \Psi^N \rangle \neq 0$.
The continuum theory for the onset of a single boson condensate in the presence of a $N$-boson condensate is \cite{DS96}
\beq
\mathcal{S}_\Psi = \int dx \, d\tau \Big[ |\partial_\tau \Psi|^2 + |\partial_x \Psi|^2 +  \alpha_\tau \Psi^\ast \partial_{\tau} \Psi + s_\Psi |\Psi|^2 + u |\Psi|^4 + \lambda (\Psi^N + (\Psi^\ast)^N ) \Big]\,,
\label{SPsi}
\eeq
where $\alpha_\tau$ (and all other couplings) are real; note that there is no direct relationship between the values of $s_{\Psi,\Phi},u,\lambda$ between $\mathcal{S}_\Phi$ and $\mathcal{S}_\Psi$.
At first glance, it might appear that the relationship between $\mathcal{S}_\Phi$ and $\mathcal{S}_\Psi$ is trivial, and they are related simply by a Wick rotation which exchanges space ($x$) and imaginary time ($\tau$). However, that is {\it not} the case. There is a crucial difference in a factor of $i$ between the first-order derivative terms in Eqs.~(\ref{SPhi}) and (\ref{SPsi}), and this difference is required by the unitarity of both theories. 
A Wick rotation relationship would imply that the dynamical critical exponent $z$ of $\mathcal{S}_\Phi$ is the inverse of the $z$ of $\mathcal{S}_\Psi$, and that the scaling dimensions of $\Phi$ and $\Psi$ are equal. The actual relationship between the theories is a Kramers-Wannier type duality between the $\Phi$ and $\Psi$ fields, and one is the `disorder' field of the other. Furthermore, unlike the $N=2$ Ising case, the duality is not a self-duality for $N > 2$; consequently the scaling dimensions of $\Phi$ and $\Psi$ are {\it not equal\/} to each other for $N \neq 2$. 
Finally, because the duality does not actually involve a Wick rotation, the values of $z$ of the theories $\mathcal{S}_\Phi$ and $\mathcal{S}_\Psi$ are {\it equal\/} to each other, as are the values of their correlation length exponents $\nu$.

The advantage of working with $\mathcal{S}_\Psi$ is that it allows a perturbative study (near two spatial dimensions) of a direct transition between a phase with $\langle \Psi \rangle = 0$, to a phase with a uniform condensate $\langle \Psi \rangle \neq 0$ (see Fig.~\ref{fig:pdiag1}). Under the duality mapping in $d=1$, these phases correspond to clock model states
with $\langle \Phi \rangle \neq 0$ (and spatially uniform) and $\langle \Phi \rangle = 0$, respectively (see Fig.~\ref{fig:pdiag2}).. Note that with a spatially uniform $\Psi$ condensate, the $\lambda$ term in Eq.~(\ref{SPsi}) does {\it not\/} average to zero, and so there is no emergent U(1) symmetry and the $\Psi$-condensed phase is also gapped. A gapless phase of $\mathcal{S}_\Psi$ can appear only in $d=1$, and requires nonperturbative effects which are special to $d=1$ \cite{jkkn}.
The Bose gas formulation of $\mathcal{S}_\Psi$ is naturally set up to provide a perturbative theory of a transition between two gapped phases, without an intermediate gapless phase. 
We shall present a renormalization group analysis of such a transition, building upon the analysis in Ref.~\onlinecite{DS96}, which examined $\mathcal{S}_\Psi$ in an expansion in $2-d$. 

We reiterate (as illustrated in Figs.~\ref{fig:pdiag1} and \ref{fig:pdiag2}) that the duality mapping between $\mathcal{S}_\Phi$ and $\mathcal{S}_\Psi$ applies only in $d=1$, and their global phase diagrams are expected to coincide only in $d=1$; $\mathcal{S}_\Phi$ can have a gapless incommensurate phase for $d\geq 1$, while $\mathcal{S}_\Psi$ has no gapless phase for $d>1$. We will use the direct transition between gapped phases of $\mathcal{S}_\Psi$, present for $d \geq 1$, to obtain a $2-d$ expansion for the transition in $d=1$; then, we will employ duality to map it onto the direct transition of $\mathcal{S}_\Phi$ in $d=1$.

The outline of the paper is as follows. Section~\ref{sec:ccm} defines the lattice chiral clock model, and analyzes its symmetries
and duality properties. Section~\ref{sec:qft} contains further discussion of the duality in the context of models which can be connected
to the continuum quantum field theories; we also present a general discussion on the nature of phases and phase transitions for different values of $N$. Our main results on the renormalization group analysis of $\mathcal{S}_\Psi$ in a $(2-d)$ expansion appear in Section~\ref{sec:rgdbg}. Section~\ref{sec:num} contains our numerical density-matrix renormalization group (DMRG) results. We extend the numerical results of Ref.~\onlinecite{SCPLS18} on the chiral clock model, and compare critical exponents between the transitions of the $\theta \neq 0$, $\phi=0$ and $\theta=0$, $\phi \neq 0$ models, which are related by the the duality transition. We also examine a lattice discretization of the Bose gas model $\mathcal{S}_\Psi$, and determine the exponents of its transition.
The Appendices contain various technical details. Appendix~\ref{sec:P=T} presents DMRG results on the chiral clock model along the
self-dual line $\phi = \theta$; note that this case does not have the parity and time-reversal symmetries to be the models studied in the
body of the paper. Here, we find numerical evidence for an intermediate incommensurate phase.

\section{Chiral clock model}
\label{sec:ccm}

\subsection{Definition of the model}

The quantum chiral clock model (CCM) in $d=1$ spatial dimension may be defined on an open chain of $M$ sites by \cite{PF2012}
\beq
H = - f \sum_{j=1}^M \left( \tau_j e^{i \phi} + \tau_j^{\dagger} e^{- i \phi} \right) - J \sum_{j=1}^{M-1} \left( \sigma_j \sigma^{\dagger}_{j+1} e^{i \theta} + \sigma^{\dagger}_j \sigma_{j+1} e^{-i \theta} \right),
\label{ccmham}
\eeq
where we conventionally take $f,J > 0$. Here, the operators $\tau$ and $\sigma$ commute on each site and obey the algebra
\begin{equation}
\tau^N = \sigma^N = \mathbb{I},\quad
\tau^{\dagger} = \tau^{-1},\quad
\sigma^{\dagger} = \sigma^{-1},\quad 
\sigma\, \tau = \omega\, \tau\, \sigma, 
\end{equation}
where $\omega = e^{2 \pi i/N}$. 
An explicit matrix representation of these operators is
\renewcommand{\arraystretch}{0.85}
\beq
\sigma = 
\begin{pmatrix}
1 & 0 & 0 & \cdots & 0 \\
0 & \omega & 0& \cdots & 0 \\
0 & 0 & \omega^2 & \cdots & 0 \\
\vdots & \vdots & \vdots & \ddots & \vdots \\
0 & 0 & 0 & \cdots & \omega^{N-1}
\end{pmatrix},
\qquad
\tau = 
\begin{pmatrix}
0 & 0 & 0 & \cdots & 0 & 1 \\
1 & 0 & 0 & \cdots & 0 & 0 \\
0 & 1 & 0 & \cdots & 0 & 0 \\
\vdots &\vdots &\vdots &\ddots & \vdots & \vdots \\
0 & 0 & 0 & \cdots & 1 & 0
\end{pmatrix}.
\label{eq:basis}
\eeq
Here, $\sigma$ measures where on the ``clock'' the state is, while $\tau$ rotates the state clockwise through the discrete angle $2 \pi/N$. 
\begin{figure}
    \centering
    \includegraphics[trim=0cm 1cm 0cm 4cm,clip, width=0.6\linewidth]{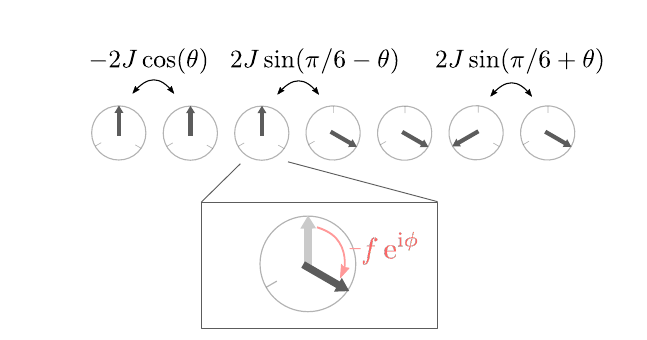}
    \caption{A visual representation of the interactions in the Hamiltonian given in Eq.~(\ref{ccmham}) for the case $N=3$. The state on each site can be represented as a linear combination of the three different states on the clock, and the interaction strength between two adjacent states is given by the factors shown.}
    \label{fig:ccmfig}
\end{figure}
This model has a global $\mathbb{Z}_N$ symmetry implemented by the unitary operator
\beq
G = \prod_{j=1}^M \tau_j ,
\eeq
which acts as $G^{\dagger} \sigma_j G = \omega \sigma_j$ and $G^{\dagger} \tau_j G = \tau_j$.

For generic values of $\theta$, $\phi$, and $N$, the phase diagram of this model is very intricate and not well-understood. 
For small values of $\theta$ and $\phi$, where the interaction is always ferromagnetic, there will be a disordered ground state with $\langle \sigma_i \rangle = 0$ in the $J \ll f$ limit and a commensurate ordered ground state with $\langle \sigma_i \rangle \neq 0$ in the $J \gg f$ limit. 
For intermediate values of $J/f$, then depending on the precise values of $\theta$, $\phi$, and $N$, these two phases may be separated by an intermediate gapless phase or a direct continuous transition 
\footnote{For this particular model there are no first-order transitions, but in models with the same symmetry and different microscopic interactions, a first-order transition between the ordered and disordered phases is possible \cite{cardydiscreteplanar}.}. 
In addition, for large enough angles $\theta$ and $\phi$, this model has incommensurate gapless phases persisting for the entire region $0\leq J/f \leq\infty$.

In this paper, we will largely be interested in the cases $(\theta \neq 0, \phi = 0)$ and $(\theta = 0, \phi \neq 0)$, where the model has both time-reversal and spatial inversion symmetries.
Our goal will be to exploit the duality in this microscopic model (reviewed below) to map out the critical theories for these transitions, and thus gain a better understanding of criticality in all systems in the same universality class.
To this end, we review the important symmetries of this model.

\subsection{Discrete symmetries}
We now introduce the operators $C$, $P$, and $T$ \cite{ZCTH}. 
Charge conjugation is a unitary operator defined by the relations
\beq
C \sigma_j C = \sigma^{\dagger}_j, \qquad C \tau_j C = \tau^{\dagger}_j, \qquad C^2 = \mathbb{I}.
\eeq
The operator $C$ can be explicitly represented in the basis of Eq.~(\ref{eq:basis}) as $C = \prod_j C_j$, where $C_j$ acts at each site as
\beq
C = \begin{pmatrix} 1 & 0 & 0 & \cdots & 0 & 0 \\ 
0 & 0 & 0 & \cdots & 0 & 1 \\ 
0 & 0 & 0 & \cdots & 1 & 0 \\
\vdots & \vdots & \vdots & \ddots &\vdots & \vdots \\
0 & 1 & 0 & \cdots & 0 & 0
\end{pmatrix}.
\eeq
Parity is a unitary operator defined as
\beq
P \sigma_j P = \sigma_{-j}, \qquad P \tau_j P = \tau_{-j}, \qquad P^2 = 1,
\eeq
and time-reversal is an anti-unitary defined as
\beq
T \sigma_j T = \sigma^{\dagger}_j, \qquad T \tau_j T = \tau_j, \qquad T^2 = \mathbb{I}.
\eeq
In the particular basis (\ref{eq:basis}), we have $T = K$ where $K$ is complex conjugation. 

Considering all three of these transformations, we see that our Hamiltonian in Eq.~\eqref{ccmham} transforms as
\beq
C H(\phi,\theta) C = H(-\phi,-\theta), \quad P H(\phi, \theta) P = H(\phi,-\theta), \quad T H(\phi, \theta) T = H(-\phi,\theta).
\eeq
For $\theta = \phi = 0$, all three of these discrete transformations are symmetries, and this is the usual (achiral) clock model. If $\theta = 0$, we have the discrete symmetries
\beq
\left(CT\right)^{-1} H(\phi,0) CT = H(\phi,0), \qquad P^{-1} H(\phi,0) P = H(\phi,0),
\eeq
while if $\phi = 0$, we have
\beq
T^{-1} H(0,\theta) T = H(0,\theta), \qquad \left(CP\right)^{-1} H(0,\theta) C P = H(0,\theta).
\eeq
Therefore, both Hamiltonians $H(\phi,0)$ and $H(0,\theta)$ have separate time-reversal and parity symmetries, though their explicit definitions are different because they must be combined with $C$ in different ways. In contrast, for both $\phi \neq 0$ and $\theta \neq 0$, it is not possible to define separate $T$ and $P$ symmetries. The only discrete spacetime symmetry is $CPT$: 
\beq
\left( CPT \right)^{-1} H(\phi,\theta) CPT = H(\phi,\theta),
\eeq
which involves a simultaneous reversal of space and time.


\subsection{Duality}
\label{sec:duality}

The reason for our specific choice of the microscopic Hamiltonian of Eq.~(\ref{ccmham}) is the existence of an exact microscopic duality in the thermodynamic limit. Similar to the Kramers-Wannier duality in the one-dimensional transverse-field Ising model \cite{kogutreview}, the duality transformation proceeds by defining a ``disorder operator'' $\tilde{\sigma}$ which creates domain walls. Explicitly, we define a set of operators on the links of the chain by
\beq
\tilde{\tau}_{j+1/2} = \sigma_{j} \sigma^{\dagger}_{j+1}, \qquad \tilde{\tau}_{M+1/2} = \sigma_M, \qquad \tilde{\sigma}_{j + 1/2} = \prod_{k=1}^j \tau_k^{\dagger},
\label{eq:duality}
\eeq
which satisfy the same algebra as the original $\tau$ and $\sigma$. Here, we parametrize the position of the dual variable $\tilde{\sigma}$ on the link connecting sites $j$ and $j+1$ by $\tilde{\sigma}_{j+1/2}$, and similarly for $\tilde{\tau}$.

From Eq.~(\ref{eq:duality}), the operator $\tilde{\sigma}_{j+1/2}$ twists all of the states from $j = 1, ..., j$ counterclockwise by an angle $2 \pi/N$, creating a domain wall at the link $j+1/2$.
In terms of these operators, the Hamiltonian takes the form
\bea
H &=& - J \sum_{j=1}^{M-1} \left( \tilde{\tau}_{j+1/2} e^{i \theta} + \tilde{\tau}_{j+1/2}^{\dagger} e^{- i \theta} \right) - f \sum_{j=1}^{M-1} \left( \tilde{\sigma}_{j+1/2} \tilde{\sigma}^{\dagger}_{j+3/2} e^{i \phi} + \tilde{\sigma}^{\dagger}_{j+1/2} \tilde{\sigma}_{j+3/2} e^{-i \phi} \right) \nn
&& - \ f \left( \tilde{\sigma}_{3/2}^{\dagger} e^{i \phi} + \tilde{\sigma}_{3/2} e^{-i \phi} \right).
\eea
This is the same form as our original Hamiltonian, except that the first term does not include the operator $\tilde{\tau}_{M+1/2}$, and the last term acts as an external field acting on the first link of the chain. Ignoring these boundary effects, the bulk part of this model is dual under the simultaneous transformation
\beq
\phi \longleftrightarrow \theta, \qquad J \longleftrightarrow f.
\eeq
This duality will be used in the following sections to obtain quantum field theories for the critical chiral clock model where the fundamental continuum field represents the disorder variable $\tilde{\sigma}$, which will give new insight into these transitions.

We note that the order and disorder operators, $\sigma$ and $\tilde{\sigma}$, are both local and will have nontrivial scaling dimensions at criticality. Since the mapping between them is highly nonlocal, there is no simple relation between the scaling dimensions of these operators for generic $\theta$ and $\phi$. An exception is along the self-dual line $\theta = \phi$, where the two operators must have identical anomalous dimensions.

\section{Chiral clock duality in quantum field theory}
\label{sec:qft}

This section is split into two parts. 
In Section \ref{sec:gendual}, we give a heuristic argument for the duality described in the introduction and then give an overview of the duality in several cases of interest. 
Section \ref{sec:qcderivation} contains an explicit derivation of the duality directly from the microscopic chiral clock model for $N=3$ by mapping to a Euclidean lattice field theory. The more complicated construction for general $N$ is presented in Appendix \ref{app:qcderivation}, which also fills in some other technical details in constructing the critical continuum field theory.

\subsection{General discussion of the duality}
\label{sec:gendual}

We motivate the field-theoretic statement of the duality by considering a family of anisotropic quantum rotor models whose phase transitions and critical field theories are well-understood. 
The derivation of these models from the microscopic Hamiltonian Eq.~(\ref{ccmham}) will contain some suspect arguments, but the final critical theories can be rigorously related to the microscopic model using the methods of Section \ref{sec:qcderivation} and Appendix \ref{app:qcderivation}.
This simpler setting is intended to give an intuitive outline of the duality, after which we give some general statements and conjectures about the critical behavior of the CCM for various values of $\theta$, $\phi$, and $N$.

We begin by softening the discrete nature of the order parameter, replacing the Hilbert space of each site by a rotor degree of freedom:
\beq
\sigma_j | \zeta_i \rangle = e^{i \zeta_j} |\zeta_j \rangle,
\eeq
where the eigenvalue of $\zeta_i$ can be any real number, but with a $2 \pi$ redundancy.
With this alteration, the spatial part of the CCM is unchanged,
\beq
\sigma_j \sigma^{\dagger}_{j+1} e^{i \theta} + \mathrm{h.c.} = \cos\left( \zeta_j - \zeta_{j+1} + \theta \right).
\eeq
Now recall that $\tau$ rotates the eigenvalue of $\sigma_j$ by an angle $2 \pi/N$. 
Therefore, we can write
\beq
\tau_j = e^{- \frac{2 \pi i}{N} L_j},
\eeq
where $L_j$ generates infinitesimal rotations of $\zeta_j$; e.g., in the $\zeta$ basis,
\beq
L_{j} = - i \frac{\partial}{\partial \zeta_j}.
\eeq
Then, the remainder of the CCM can be written as
\beq
\tau_j e^{i \phi} + \mathrm{h.c.} = \cos\left( - \frac{2 \pi}{N} L_j + \phi \right).
\eeq
At this point, we replace this operator by the first term in its power series:
\beq
\cos\left( - \frac{2 \pi}{N} L_j + \phi \right) \longrightarrow \mathrm{const.} - \frac{(2 \pi/N)^2}{2} L_j^2 + h_{\phi} L_j,
\eeq
where $h_{\phi} \propto \phi$. 
One may attempt to justify this step by appealing to a large $N$ limit where $N \phi$ remains small, although we will consider all $N\geq 3$ below.
Alternatively, one may argue that the term on the right-hand side will have the same disordering effect on the $\zeta$ field.
In either case, we may always appeal to the more technical derivation below to justify our conclusions.

Our final rotor representation for the CCM will be
\begin{equation}
H = \frac{f'}{2} \sum_j L_j^2 - J \sum_j \cos\left( \zeta_j - \zeta_{j+1} + \theta \right) + h_{\phi} \sum_j L_j - h_N \sum_j \cos\left( N \zeta_j \right).
\label{eq:rotorccm}
\end{equation}
Here, in addition to the terms described above, we have also added an anisotropic external field proportional to $h_N$ which breaks the U(1) symmetry of the model back down to $\mathbb{Z}_N$. 
The statement of duality is that the critical properties of this Hamiltonian for $(\phi = 0, \theta \neq 0)$ and $(\phi \neq 0, \theta = 0)$ map onto each other. 
Such a duality can only be valid at $h_N > 0$, where a phase with $\mathbb{Z}_N$ order exists and a corresponding disorder operator can be defined. 
The action of the discrete symmetries of the CCM are implemented as
\begin{alignat}{2}
G: \qquad & \zeta_j \rightarrow \zeta_j + \frac{2 \pi}{N}, \qquad &&L_j \rightarrow L_j, \nn
C: \qquad &\zeta_j \rightarrow - \zeta_j, \qquad &&L_j \rightarrow -L_j, \nn
P: \qquad &\zeta_j \rightarrow \zeta_{-j}, \qquad &&L_j \rightarrow L_{-j}, \nn
T: \qquad &\zeta_j \rightarrow \zeta_j, \qquad &&L_j \rightarrow -L_j,
\end{alignat}
and $T$ is still anti-unitary, $T \, i \, T = - i$.

Our reason for introducing this model is to utilize the well-known results mapping critical quantum rotors to quantum field theories \cite{sachdev2011quantum}. 
For $\theta = \phi = h_N = 0$, we have a U(1) rotor, whose critical field theory is the Lorentz and U(1)-invariant theory for a complex field $\Phi$:
\beq
\mathcal{S}_{1} = \int dx \, d\tau \Big[ |\partial_\tau \Phi|^2 + |\partial_x \Phi|^2 + s_\Phi |\Phi|^2 + u |\Phi|^4 + \cdots \Big],
\eeq
where the ellipsis denotes all other allowed terms. 
The main effect of the anisotropic coupling $h_N$ will be to break the U(1) symmetry down to $\mathbb{Z}_N$, so we expect the critical theory will be altered to
\beq
\mathcal{S}_{2} = \int dx \, d\tau \Big[ |\partial_\tau \Phi|^2 + |\partial_x \Phi|^2 + s_\Phi |\Phi|^2 + u |\Phi|^4 + \lambda \left(\Phi^N + (\Phi^\ast)^N \right) + \cdots \Big],
\eeq
where the ellipsis now also includes all real terms invariant under $\Phi \rightarrow e^{2 \pi i/N} \Phi$. 
The actions $\mathcal{S}_1$ and $\mathcal{S}_2$ are not very useful starting points for studying the critical U(1) rotor model and achiral clock models in one dimension, where we expect these field theories to be very strongly coupled, but the critical points of these models have been understood using other methods (their behavior is reviewed below).

If we add a nonzero $\theta$, we only change the couplings in the spatial direction. In particular, since $\theta \neq 0$ breaks parity symmetry, we expect that we should add odd spatial derivative terms \cite{fendley2004competing}:
\beq
\mathcal{S}_{\Phi} = \int dx \, d\tau \Big[ |\partial_\tau \Phi|^2 + |\partial_x \Phi|^2 + i \alpha_x \Phi^\ast \partial_x \Phi + s_\Phi |\Phi|^2 + u |\Phi|^4 + \lambda \left(\Phi^N + (\Phi^\ast)^N \right) + \cdots \Big].
\label{SPhi2}
\eeq
Finally, we consider the effect of $\phi \neq 0$ and $\theta = 0$, which we have argued should describe the critical field theory for the disorder operator $\Psi$ that is dual to the clock order parameter $\Phi$ experiencing the chiral transition in Eq.~(\ref{SPhi2}). 
From Eq.~(\ref{eq:rotorccm}), this is equivalent to adding an external field coupled to the conserved charge $Q = \sum_j L_j$ associated with U(1) rotations of the theory. 
We do not need to perform a detailed derivation of the field theory in this case; we simply need to find the corresponding U(1) Noether charge of the action $\mathcal{S}_1$ to find the operator which couples to $h_{\phi}$.
This is equivalent to replacing $\partial_{\tau} \longrightarrow \partial_{\tau} + h_{\phi}$, and we obtain the theory
\beq
\mathcal{S}_{\Psi} = \int dx \, d\tau \Big[ |\partial_\tau \Psi|^2 + |\partial_x \Psi|^2 +  \alpha_\tau \Psi^\ast \partial_{\tau} \Psi + s_\Psi |\Psi|^2 + u |\Psi|^4 + \lambda (\Psi^N + (\Psi^\ast)^N ) + \cdots \Big]
\eeq
as claimed in the introduction. 


\subsubsection{Achiral clock model: $\phi = \theta = 0$}

Setting $\phi = \theta = 0$ in either Eqs.~(\ref{ccmham}) or (\ref{eq:rotorccm}), we expect the phase diagram to have the same general structure as the phases mapped out for the discrete planar models studied in Refs.~\onlinecite{elitzur1979} and \onlinecite{cardydiscreteplanar}, which also discuss duality in these and related models. 
We review these results, which are useful for what follows.
Here we describe the possible behavior for all models with the same symmetries as the specific models we gave above.

We first recall the properties of the U(1)-symmetric limit, $h_N \rightarrow 0$, which has relevance for the critical properties of the large-$N$ clock models.
At $h_N = 0$, the system undergoes a Kosterlitz-Thouless (KT) transition between a disordered gapped phase and a gapless critical phase (which we call a KT phase) \cite{kosterlitz1973ordering}. 
The critical phase may be described by a free bosonic field theory, and the scaling dimensions of physical operators vary continuously with the couplings.
For small $h_N$, the gapless critical phase is always unstable to a gapped phase with $\mathbb{Z}_N$ order for small enough $f'/J$. 
The critical behavior for various values of $N$ are as follows:

\paragraph{$N=3:$} The U(1)-symmetric fixed point is unstable to the perturbation $h_3$ \cite{jkkn}. The $\mathbb{Z}_3$ clock model is identical to the three-state Potts model, which realizes the full $\mathbb{S}_3$ permutation symmetry, and there is a direct continuous transition between the disordered and ordered phases. The critical point is described by the $c = \frac{4}{5}$ CFT, and the scaling dimension of every operator is known exactly \cite{DMS97}.

\paragraph{$N=4:$} There is a direct transition between the $\mathbb{Z}_4$-ordered phase and the disordered phase.
The critical points are in the Ashkin-Teller (AT) universality class, which is really a family of universality classes. 
The AT model may be defined as two copies of the Ising $\left(c=\frac{1}{2}\right)$ CFT which are coupled together by their energy operators.
This coupling is exactly marginal, and the resulting AT model describes a line of fixed points with central charge $c=1$. 
This line passes through the four-state Potts fixed point, and eventually meets the U(1)-symmetric Kosterlitz-Thouless point at $h_4 = 0$ \cite{kadanoffAT,jkkn}.

\paragraph{$N \geq 5:$} In this case, the phase transition will either be first-order or there will be an intermediate KT phase. 
For these larger values of $N$, we may use intuition from the U(1)-symmetric limit when describing transitions into this KT phase, where thermodynamic variables diverge with an essential singularity.
We note that the scaling dimension of the order parameter at the KT transitions will not coincide with the U(1) value $\Delta_{\sigma} = 1/8$.

\subsubsection{Chiral clock model: $\theta \neq 0$}

We now consider the expected critical theories for $\phi = 0$ and $\theta \neq 0$, along with the dual formulations obtained by applying the $\theta \leftrightarrow \phi$ duality.

We first consider the $h_N \rightarrow 0$ limit. 
Although the duality between these theories only holds at finite $N$, for $N$ large enough, one expects that the leading operators $\Phi^N + \mathrm{c.c.}$ will be irrelevant compared to the U(1) invariant part of the action, leading to enlarged U(1) symmetry in the critical regime (we discuss this point further below).
We note that the $N \geq 5$ CCM is expected to always have an incommensurate KT phase separating the ordered regions of the phase diagram \cite{ostlund1981incommensurate}.



Using the $\phi \leftrightarrow \theta$ duality, we expect that the phase transition from the commensurate $\mathbb{Z}_N$-ordered phase to the incommensurate phase (the C--IC transition) can be described by
\beq
\mathcal{S}_{\Psi,\mathrm{U(1)}} = \int dx \, d\tau \Big[ |\partial_\tau \Psi|^2 + |\partial_x \Psi|^2 +  \alpha_\tau \Psi^\ast \partial_{\tau} \Psi + s_\Psi |\Psi|^2 + u |\Psi|^4 + \cdots \Big],
\label{eq:u1bosegas}
\eeq
where $\Psi$ creates a domain wall in the $\mathbb{Z}_N$-ordered state. 
Interestingly, the disorder operator condenses at zero momentum.
This is the same critical theory which describes the Bose superfluid--Mott insulator transition at variable density \cite{ssbook}.

In fact, the identification of $\mathcal{S}_{\Psi,\mathrm{U(1)}}$ as a description of the C--IC transition could be argued along completely different lines. 
\citet{schulz1980} has shown that the critical Pokrovsky-Talapov (PT) theory of the C--IC transition can be mapped to that of a one-dimensional spinless fermion at the bottom of a quadratic band, undergoing an insulator--Luttinger liquid transition \cite{pokrovsky1978phase,pokrovsky1979ground}.
The equivalence of this theory with the critical theory $\mathcal{S}_{\Psi,\mathrm{U(1)}}$ has also been well-established \cite{SSS94}.
Our present derivation of the relation between $\mathcal{S}_{\Psi}$ and the PT transition completes this circle of dualities, and gives an interesting interpretation of the field $\Psi$ as the operator which creates domain walls in the commensurate phase.

For $N=3$, it is known that $\theta$ couples to an operator with scaling dimension $9/5$, so it is a relevant perturbation. 
Recent numerical progress \cite{ZCTH,SCPLS18} has provided evidence that the $N=3$ theory flows to a new fixed point where there is a direct continuous transition between the two phases for intermediate $\theta$; for other small values of $N$, less is known in the literature. 
As with the achiral models, it would be interesting to establish a critical value $N_c$ above which these models cross over to U(1)-symmetric behavior.

As stressed in the introduction, the theory of Eq.~(\ref{SPhi2}) cannot perturbatively describe the onset of $\mathbb{Z}_N$ order at zero momentum, while the dual theory $\mathcal{S}_{\Psi}$ admits a  perturbative expansion in $2 - d$ and $4 - N$.
In doing so, we envision a scenario where the U(1)-symmetric transition of Eq.~\eqref{eq:u1bosegas} in $d < 2$ dimensions is unstable to the addition of the operator $\Psi^N + \mathrm{c.c.}$, and the theory flows to a fixed point which is smoothly connected to the CCM in $d=1$.
However, the precise nature of the renormalization group flow of these field theories for general $N$ at $d =1$ cannot be addressed by our methods; we cannot rule out the possibility that our perturbative fixed point is unstable, and the CCM fixed point of interest does not smoothly connect to the $d=2$, $N=4$ case where we apply perturbation theory.

If we assume that the critical CCM is smoothly connected to the perturbative fixed point considered below, we may make some predictions for the value of $N_c$ where the critical CCM crosses over to a U(1)-symmetric theory.
\citet{DS96} computed the scaling dimension of the operator $\Psi^N + \mathrm{c.c.}$ at the U(1) symmetric fixed point of Eq.~(\ref{eq:u1bosegas}) in an expansion in $2 - d$.
Extrapolating their results to $d=1$ gave the unusual result that the operator was relevant for $N \lesssim 2.6$ and $N \gtrsim 5.4$, and irrelevant for $N$ between these values.
As pointed out in that work, the predictions for large $N$ are certainly an artifact of the expansion (in particular, the expansion predicts unphysical negative scaling dimensions for $N\geq 6$).
In Section \ref{sec:rgdbg}, we obtain equivalent results to those in Ref.~\onlinecite{DS96} truncated at small $4 - N$.
There, we find that the operator $\Psi^N + \mathrm{c.c.}$ is relevant for $N<N_c$ and irrelevant for $N>N_c$, where
\beq
N_c \approx 3.6.
\eeq
Here, we have extrapolated to $d=1$ and arbitrary $N$, but $N_c$ is close enough to $N=4$ that this may be a quantitatively accurate estimate.

Throughout this paper, we have ignored the $N=2$ case.
From Eq.~\eqref{ccmham}, it is clear that the lattice CCM reduces to the transverse-field Ising model in this case, and that nonzero angles $\theta$ and $\phi$ are simply redefinitions of the constants $f$ and $J$.
At the level of our field theory duality, we can see this by the fact that our order parameter can be chosen to be real, so that the couplings $\Phi \partial_x \Phi$ and $\Psi \partial_{\tau} \Psi$ are total derivatives and do not contribute to the action.
Then, the duality between $\mathcal{S}_{\Phi}$ and $\mathcal{S}_{\Psi}$ reduces to the ordinary Kramers-Wannier self-duality of the Ising model \cite{kogutreview}.
The exact computations of Ref.~\onlinecite{DS96} show that $\mathcal{S}_{\Psi}$ flows to the Ising fixed point for the $N=2$ case, which serves as an added verification of our assumption for the RG flows of these models.


\subsection{Explicit derivation of the duality for $N=3$}
\label{sec:qcderivation}

We now consider the explicit mapping of the one-dimensional quantum model (\ref{ccmham}) to a Euclidean lattice field theory using transfer matrix methods \cite{kogutreview}. 
In this section we treat the simplest case, $N=3$, and leave the more technically complicated but conceptually similar $N>3$ case for Appendix \ref{app:qcderivation}. 
We write the partition function as
\beq
\mathcal{Z} = \Tr \exp\left( - \beta H \right) = \lim_{a \rightarrow 0} \lim_{M_{\tau} \rightarrow \infty} \Tr \left( e^{- a H} \right)^{M_{\tau}},
\eeq
where $a M_{\tau} = \beta$. This represents $M_{\tau}$ products of a $3^M \times 3^M$ transfer matrix $e^{-a H}$. 
We first decompose this into a product,
\beq
e^{- a H} = T_1 T_2 + O(a^2 ),
\eeq
where
\bea
T_1 &=& \exp\left( a \beta f \sum_j^M \tau_j e^{i \phi} + \mathrm{h.c.} \right), \quad T_2 = \exp\left( a \beta J \sum_{j}^M \sigma_j \sigma^{\dagger}_{j+1} e^{i \theta} + \mathrm{h.c.} \right).
\eea
We now insert a complete set of states between each factor of the transfer matrix. 
For the $3^M$-dimensional Hilbert space defined in the problem, we use the basis $| n_j \rangle$ where
\beq
\sigma_{j} | n_j \rangle = e^{2 \pi i n_j/3}| n_j \rangle
\eeq
with possible eigenvalues $n_j = 0, 1, ..., 2$. The partition function becomes
\beq
\mathcal{Z} = \sum_{\{ n_j(\ell) \}} \prod_{\ell = 1}^{M_{\tau}} \left\langle \{ n_j(\ell) \} | T_1 T_2 | \{ n_{j}(\ell + 1) \} \right\rangle .
\label{eq:part1}
\eeq
The sum is over the $3^{M M_{\tau}}$ values of $n_j(\ell)$. The matrix elements of $T_2$ are trivial, 
\beq
T_2 | \{ n_{j}(\ell) \} \rangle = \exp\left( 2 a \beta J \sum_j^{M} \cos\left[ \frac{2 \pi}{3}\Big(n_j(\ell) - n_{j+1}(\ell)\Big) + \theta \right] \right) | \{ n_{j}(\ell) \} \rangle,
\eeq
and it remains to evaluate the matrix elements
\beq
T_1(n,n') \equiv \left\langle n | T_1 | n' \right\rangle .
\eeq
For this, we write the eigenbasis $| n \rangle $ in terms of the eigenbasis of $\tau$:
\beq
\tau | \omega \rangle = e^{2 \pi i \omega/3} | \omega \rangle,
\eeq
where $\omega = 0, 1, ..., 2$. These bases are related by
\beq
| n \rangle = \frac{1}{\sqrt{3}} \sum_{\omega = 0}^{2} e^{2 \pi i \omega n/3} | \omega \rangle.
\eeq
Using the above equations, we can evaluate the matrix elements in Eq.~(\ref{eq:part1}), obtaining an expression resembling a classical partition function defined on a 2D lattice:
\bea
\mathcal{Z} &=& \frac{1}{3^{M_{\tau}}}\sum_{\{ n_j(\ell) \}}  \exp\left( 2 a \beta J \sum_{\ell = 1}^{M_{\tau}} \sum_{j=1}^{M} \cos\left[ \frac{2 \pi}{3}\Big(n_j(\ell) - n_{j+1}(\ell)\Big) + \theta \right] \right) \nn
&& \ \times \prod_{\ell = 1}^{M_{\tau}} \prod_{j=1}^{M} \sum_{\omega = 0}^{2} \exp\left( 2 a \beta f \cos\left[ \frac{2 \pi}{N} \omega + \phi \right] \right) \nn
&& \ \times \exp\left( - \frac{2 \pi i \omega}{3} \Big(n_j(\ell) - n_j(\ell + 1)\Big) \right).
\label{eq:beforesum}
\eea

Our next step is to evaluate the sum over the $\omega$, and then rewrite the resulting terms as a single exponential.
Explicitly, we can write
\bea
S_3(\Delta n) &\equiv& \sum_{\omega = 0}^{2} \exp\left( 2 a \beta f \cos\left[ \frac{2 \pi}{N} \omega + \phi \right] - \frac{2 \pi i \omega}{3} \Big(n_j(\ell) - n_j(\ell + 1)\Big) \right) \nn[.5cm]
&=& A \exp\left[ B(\phi) \cos \left( \frac{2 \pi}{3} \Delta n \right) \right] \exp\left[ i \varphi(\phi) \frac{2}{\sqrt{3}} \sin\left( \frac{2 \pi}{3} \Delta n \right) \right]
\eea
with the definitions
\bea
A &=& \left[ e^{2 a \beta f \cos \phi} + 2 e^{-a \beta f \cos \phi} \cosh\left( \sqrt{3} a \beta f \sin \phi \right) \right] e^{- B(\phi)}, \nonumber \\[14pt]
B(\phi) &=& \frac{1}{3} \log \left[ \frac{\left( e^{3 a \beta f \cos \phi} + 2 \cosh \left( \sqrt{3} a \beta f \sin \phi \right) \right)^2}{e^{6 a \beta f \cos \phi} - 2 e^{3 a \beta f \cos \phi} \cosh\left( \sqrt{3} a \beta f \sin \phi \right) + 2 \cosh\left( 2 \sqrt{3} a \beta f \sin \phi \right) - 1 } \right],  \nonumber \\[14pt]
\tan \varphi(\phi) &=& \frac{\sqrt{3} \sinh \left( \sqrt{3} a \beta f \sin \phi \right)}{e^{3 a \beta f \cos \phi} - \cosh\left( \sqrt{3} a \beta f \sin \phi \right)}. 
\label{quantend}
\eea
For small $a$, $\varphi(\phi) = \phi$; we take this as a strict equality from now on. We can also show that
\beq
B(\phi) \approx - \frac{2}{3}\log a
\eeq
for small $a$, so the $\phi$-dependence disappears from the $B(\phi)$ term.

From this analysis, we expect that the critical behavior of the quantum model is equivalent to the Euclidean lattice field theory obtained by the partition function
\beq
\mathcal{Z} = C \sum_{\{ n_{x,\tau} \}} e^{-\mathcal{S}}
\eeq
with the action
\bea
- \mathcal{S} &=& K_x \sum_{x,\tau} \cos\left[ \frac{2 \pi}{3}\left( n_{x,\tau} - n_{x+1,\tau} \right) + \theta \right] + K_{\tau} \sum_{x,\tau} \cos\left[ \frac{2 \pi}{3}\left( n_{x,\tau} - n_{x,\tau+1} \right) \right] \nn
& +& \ \frac{2 i \phi}{\sqrt{3}} \sum_{x,\tau} \sin\left[ \frac{2 \pi}{3}\left( n_{x,\tau} - n_{x,\tau+1} \right) \right].
\label{classicalHthree}
\eea
Here, the quantum model is obtained in the limit $K_x \rightarrow 0$, $K_y \rightarrow \infty$ such that $K_{x} e^{3 K_y/2}$ is finite and tuned to the phase transition. We have $K_x \sim a \beta J$ and $K_y \sim B = B(a,\beta f)$, but choose the $K_{x}$ and $K_y$ as tuning parameters instead of $J$ and $f$.

For $\phi = 0$, the action (\ref{classicalHthree}) is equivalent to the two-dimensional classical chiral clock model \cite{huse1982domain,ostlund1981incommensurate}, and this mapping has been known for a long time \cite{centen1982non,howes1983quantum,howes1983commensurate}.
For $\phi \neq 0$, there is a purely quantum term proportional to $\phi$ contributing complex Boltzmann weights. 
This term was noticed in Ref.~\onlinecite{howes1983quantum}, but the proper interpretation of the term as describing the Euclidean path integral of a \emph{quantum} field theory was overlooked. 
The coefficient of $\phi$ is such that each term in the partition sum contributes phases of $1$ and $e^{\pm i \phi}$, so the model still has an exact $2 \pi$ periodicity in $\phi$ as required. The original $\theta \leftrightarrow \phi$ duality of the microscopic model is invisible here; it is a nontrivial infrared self-duality of the theory which also involves some nontrivial transformation on $K_x$ and $K_{\tau}$. 
The global symmetries of the original quantum model are now implemented as
\bea
&G:& \quad \quad n_{x,\tau} \rightarrow n_{x,\tau} + 1, \nn
&T:& \quad \quad n_{x,\tau} \rightarrow n_{x,-\tau}, \nn
&C:& \quad \quad n_{x,\tau} \rightarrow -n_{x,\tau}, \nn
&P:& \quad \quad n_{x,\tau} \rightarrow n_{-x,\tau}.
\eea

In Appendix \ref{app:qcderivation} we derive an equivalent field theory for this model in the scaling limit in terms of a complex field $\Phi(x,\tau)$, which acts as an order parameter of the spins $\sigma$.
Our final field theory in terms of this continuum complex field is
\bea
\mathcal{S}' = \int d \tau \, dx \bigg( &i& \alpha_x \Phi^{\ast} \partial_x \Phi + \alpha_{xx} |\partial_x \Phi|^2 + \alpha_{\tau} \Phi^{\ast} \partial_{\tau} \Phi + \alpha_{\tau\tau} |\partial_{\tau} \Phi|^2 \nn
&& + \ s_\Phi |\Phi|^2 + \lambda \left( \Phi^3 + \Phi^{\ast 3} \right) + u |\Phi|^4 + \cdots \bigg),
\label{eq:Sprime}
\eea
where $\alpha_x$ goes to zero for $\theta = 0$, and $\alpha_{\tau}$ goes to zero for $\phi = 0$.
The symmetries of the original model are now implemented by
\bea
&G:& \quad \quad \Phi(x,\tau) \rightarrow e^{2 \pi i/3} \Phi(x,\tau), \nn
&T:& \quad \quad \Phi(x,\tau) \rightarrow \Phi(x,-\tau), \nn
&C:& \quad \quad \Phi(x,\tau), \rightarrow \Phi^{\ast}(x,\tau) \nn
&P:& \quad \quad \Phi(x,\tau) \rightarrow \Phi(-x,\tau).
\eea
From Eq.~\eqref{eq:Sprime}, we can see that specializing to the case $(\theta \neq 0,\phi = 0)$, gives the action $\mathcal{S}_{\Phi}$ of Eq.~\eqref{SPhi}. 
Then, after applying the duality of Section \ref{sec:duality}, and following the same steps for the ``disorder parameter'' $\Psi \sim \tilde{\sigma}$, we obtain the dual action $\mathcal{S}_{\Psi}$ of Eq.~\eqref{SPsi}, completing our proof.

\section{Renormalization group analysis of the $\mathbb{Z}_N$ dilute Bose gas}
\label{sec:rgdbg}

\begin{figure}
\includegraphics[width=10cm]{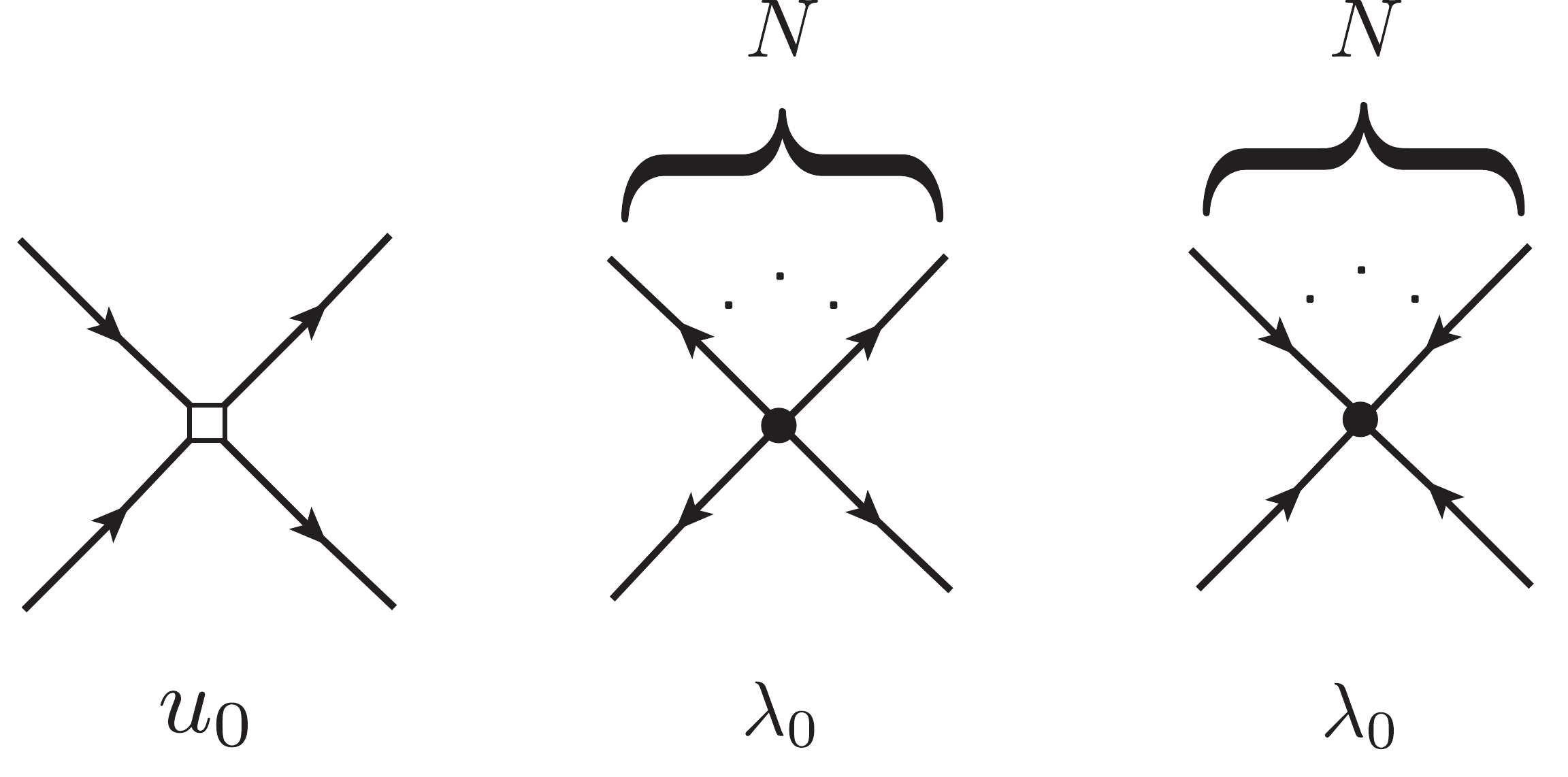}
\caption{The interaction vertices in the diagrammatic expansion of the $\mathbb{Z}_N$ DBG. Both of the $\lambda_0$ vertices have a total of $N$ propagators attached to them.}
\label{fig:vertices}
\end{figure}

In this section, we will study the renormalization group (RG) properties of the $\mathbb{Z}_N$ dilute Bose gas (DBG) starting from the action
\beq
\label{eq:DBG}
\mathcal{S}_{B} = \int d\tau \, d^d x \left[ \Psi^{\ast} \partial_\tau \Psi + \left| \nabla \Psi \right|^2 + s |\Psi|^2 + \frac{u}{2} \left| \Psi \right|^4 + \frac{\lambda_0}{N!} \left( \Psi^N + \Psi^{\dagger N} \right) \right].
\eeq
Here, we have generalized the action $\mathcal{S}_{\Psi}$ to $d$ spatial dimensions, and truncated the action to the most relevant terms.
We drop the subscript on the `mass' term, $s_\Psi$, in the remainder of this section.
The units in the space and imaginary time directions have been chosen such that the coefficients of the first two terms are unity. At the free theory, $u = \lambda_0 = 0$, the dynamical critical exponent is given by $z = 2$, and the scaling dimensions of the couplings in units of momentum or inverse length are
\bea
\mathrm{dim} (s) &=& 2, \nn
\mathrm{dim}( u ) &=& 2 - d, \nn
\mathrm{dim}( \lambda_0 ) &=& 2 + d - Nd/2.
\eea
Close to the free theory, $s$ is always relevant, identifying it as the coupling which tunes through the phase transition.
We will hereafter always assume this coupling is tuned to criticality, and define it to vanish at this value: $s = s_c=0$.
The couplings $u$ and $\lambda$ are both marginal for $d = 2$ and $N = 4$, and there are no additional relevant or marginal operators allowed by symmetry. 
This suggests an expansion in both $\epsilon = 2 -d$ and $\delta = 4 - N$, so that we may exhibit a flow to an interacting fixed point which remains perturbatively accessible. 
We may then perform a diagrammatic expansion on this model, where the free propagator is
\beq
G(\omega,k) = \frac{1}{- i \omega + k^2},
\eeq
and the interaction vertices are pictured in Figure \ref{fig:vertices}.

We now define renormalized fields and couplings,
\bea
\tau &=& \frac{Z_{\tau} \mu^{-2}}{Z}\tau_R, \nn
\psi &=& Z^{1/2} \psi_R, \nn
u &=& \frac{Z_{g} \mu^{2 - d}}{ Z Z_{\tau} S_{d}} g, \nn
\lambda_0 &=& \frac{Z_{\lambda} \mu^{2 + d - Nd/2}}{Z^{N/2-1} Z_{\tau} S_d^{N/2 - 1} } \lambda,
\label{Eq:renorm}
\eea
where $S_d = (4 \pi)^{-d/2}$ is a dimensional factor defined to simplify future expressions.
We have also introduced an arbitrary momentum scale, $\mu$, which renders the renormalized couplings dimensionless. 
We renormalize the theory by first computing correlation functions in bare perturbation theory using the action $\mathcal{S}_B$ for arbitrary $\epsilon$ and $\delta$.
These correlation functions will be divergent in some dimension-one manifold of the $\epsilon-\delta$ plane, including at the point $\epsilon = \delta = 0$.
We then express these correlation functions in terms of the renormalized fields and couplings specified in Eq.~(\ref{Eq:renorm}), and choose the renormalization constants $Z_i$ such that the correlation functions of the renormalized fields are regular in a finite neighborhood of the origin of the $\epsilon-\delta$ plane when expressed in terms of the renormalized couplings.

In particular, if we consider the one-particle irreducible (1PI) vertex of $n$ fields in momentum and frequency space, the bare and renormalized quantities are related by
\beq
\Gamma^{(n)}_R(\{ \omega_{Ri},k_i \},g,\lambda,\mu) = Z^{n/2}\left( \frac{Z_{\tau}}{Z} \right) \Gamma^{(n)}(\{ \omega_{i},k_i  \},u_0,\lambda_0),
\label{eq:1pirenorm}
\eeq 
where we have defined $\omega_R = Z_{\tau} \mu^{-2}\omega/Z$ in congruence with Eq.~\eqref{Eq:renorm}, and these vertex functions are defined without overall delta functions enforcing momentum and frequency conservation. 
By the renormalizability of our theory, the constants $Z_i$ may be specified by computing the three 1PI vertices displayed in Figure \ref{fig:gamma2}.

Once we have obtained the renormalization constants, we may consider the dependence of the interaction couplings on our arbitrary momentum scale $\mu$ by defining the usual beta functions,
\bea
\beta_g = \mu \frac{d g}{d \mu}, \qquad \beta_{\lambda} = \mu \frac{d \lambda}{d \mu}.
\eea
These may be computed directly from the definitions of $g$ and $\lambda$ in Eq.~\eqref{Eq:renorm}. 
Introducing the convenient shorthand
\bea
\mathcal{Z}_g &\equiv& \log \left( \frac{Z_g}{Z Z_{\tau}} \right), \nn
\mathcal{Z}_\lambda &\equiv& \log \left( \frac{Z_\lambda}{Z^{N/2-1} Z_{\tau}} \right), \nn
h_g &\equiv& 2 - d = \epsilon, \nn
h_\lambda &\equiv& 2 + d - Nd/2 = \epsilon + \delta - \epsilon \delta/2,
\label{eq:defs}
\eea
we can write the beta functions as
\bea
\beta_g &=& \frac{- h_g g - h_g g \lambda \frac{d \mathcal{Z}_{\lambda}}{d \lambda} + h_{\lambda} g \lambda \frac{d \mathcal{Z}_g}{d \lambda}}{1 + g \frac{d \mathcal{Z}_g}{dg} + \lambda \frac{d \mathcal{Z}_\lambda}{d \lambda} + g \lambda \frac{d \mathcal{Z}_g}{d g} \frac{d \mathcal{Z}_\lambda}{d\lambda} - g \lambda \frac{d \mathcal{Z}_g}{d \lambda} \frac{d \mathcal{Z}_\lambda}{dg} }, \nn[.5cm]
\beta_\lambda &=& \frac{- h_\lambda \lambda - h_\lambda g \lambda \frac{d \mathcal{Z}_{g}}{d g} + h_{g} g \lambda \frac{d \mathcal{Z}_\lambda}{d g}}{1 + g \frac{d \mathcal{Z}_g}{dg} + \lambda \frac{d \mathcal{Z}_\lambda}{d \lambda} + g \lambda \frac{d \mathcal{Z}_g}{d g} \frac{d \mathcal{Z}_\lambda}{d\lambda} - g \lambda \frac{d \mathcal{Z}_g}{d \lambda} \frac{d \mathcal{Z}_\lambda}{dg} }.
\label{eq:betaSW}
\eea
The critical points of the system are given by solving $\beta_g = \beta_{\lambda} = 0$.

Once we obtain a fixed point, we compute critical exponents. 
For example, the scaling of the dimensionless renormalized coupling $\tau_R$ determines the scaling of the time dimension with respect to momentum, which gives the dynamical critical exponent $z$:
\bea
\mu \frac{d \tau_R}{d \mu} &\equiv& z \tau_R \nn
\Rightarrow \  z &=& 2 - \beta_g \frac{d}{dg} \log\left( \frac{Z_{\tau}}{Z} \right) - \beta_{\lambda} \frac{d}{d\lambda} \log\left( \frac{Z_{\tau}}{Z} \right).
\label{eq:zdef}
\eea
All other critical exponents are related to the scaling dimensions of operators.
For example, by renormalizing the two-point function, we have effectively computed the scaling dimension $\Delta_{\Psi}$ associated with the operator $\Psi$:
\beq
\Delta_{\Psi} = \frac{d}{2} + \frac{1}{2} \beta_g \frac{d}{dg} \log Z + \frac{1}{2}\beta_{\lambda} \frac{d}{d \lambda} \log Z.
\label{eq:p1def}
\eeq
Similarly, by renormalizing the interaction vertices, we have effectively computed the scaling dimensions of the operators $\Psi^N + \mathrm{c.c.}$ and $|\Psi|^4$. 
We will find below that these operators will generically mix at the interacting fixed point, as they have the same scaling dimension at $\epsilon = \delta = 0$, and they transform identically under the symmetries of $\mathcal{S}_B$ when $\lambda \neq 0$. 
The eigenoperators under dilatations will have scaling dimensions given by
\beq
\Delta_{\Psi^4_{\pm}} = 2d + z + \omega_{\pm}
\eeq
with
\beq
\det \left[  \begin{pmatrix} \frac{\partial \beta_g}{\partial g} & \frac{\partial \beta_g}{\partial \lambda} \\ \frac{\partial \beta_{\lambda}}{\partial g} & \frac{\partial \beta_{\lambda}}{\partial \lambda} \end{pmatrix} - \omega_{\pm} \, \mathbb{I} \right] = 0
\label{eq:omegapmdef}
\eeq
i.e., the two numbers $\omega_{\pm}$ are the two eigenvalues associated with the matrix formed by linearizing the beta functions at the critical couplings.
The eigenvectors of this matrix determine the precise nature of the operator mixing.

The last operator we are interested in is the leading relevant operator, $|\Psi|^2$. 
Since this does not appear in our action at criticality, we need to define a new renormalization constant,
\beq
| \Psi |^2 = \frac{Z_2}{Z_{\tau}} \left( | \Psi |^2 \right)_R,
\eeq
where $Z_2$ is chosen to cancel divergences upon insertion of this composite operator. 
With this particular definition, the 1PI vertex with $n$ insertions of $\Psi$ or $\Psi^{\ast}$ and $m$ insertions of $|\Psi|^2$ is renormalized as
\beq
\Gamma^{(n,m)}_R(\{ \omega_{Ri},k_i \},g,\lambda,\mu) = Z^{n/2 - 1} Z_{\tau}^{1-m} Z_2^m \, \Gamma^{(n,m)}(\{ \omega_{i},k_i  \},u_0,\lambda_0).
\label{eq:renormp2}
\eeq
We will calculate $Z_2$ by renormalizing the vertex $\Gamma^{(2,1)}$, pictured in Figure \ref{fig:gamma12}.
With this definition, the scaling dimension of $|\Psi|^2$ is
\beq
\Delta_{|\Psi|^2} \equiv d + \beta_g \frac{d}{dg} \log\left( \frac{Z_2}{Z_{\tau}} \right) + \beta_{\lambda} \frac{d}{d\lambda} \log\left( \frac{Z_2}{Z_{\tau}} \right).
\label{eq:p2def}
\eeq

After computing these scaling dimensions, we will obtain the critical exponents of $\mathcal{S}_B$.
However, because the field $\Psi$ is the disorder operator of the CCM, many of the critical exponents will not have a simple relation to the critical exponents associated with the CCM order parameter $\Phi$, which is a nonlocal operator in this theory.
We do expect that $|\Psi|^2$, as the lone relevant operator allowed by symmetry at the critical point, will map to the corresponding relevant operator in the CCM transition.
This implies that the critical exponent $\nu$ will coincide at the $\mathbb{Z}_N$ CCM and DBG critical points.
Furthermore, the dynamical critical exponent $z$ is a property of the exact low-energy dispersion of the quantum critical point rather than any particular operator, and therefore it should also be the same in both theories.

\subsection{Diagrammatic expansion}

\begin{figure}
\includegraphics[width=16cm]{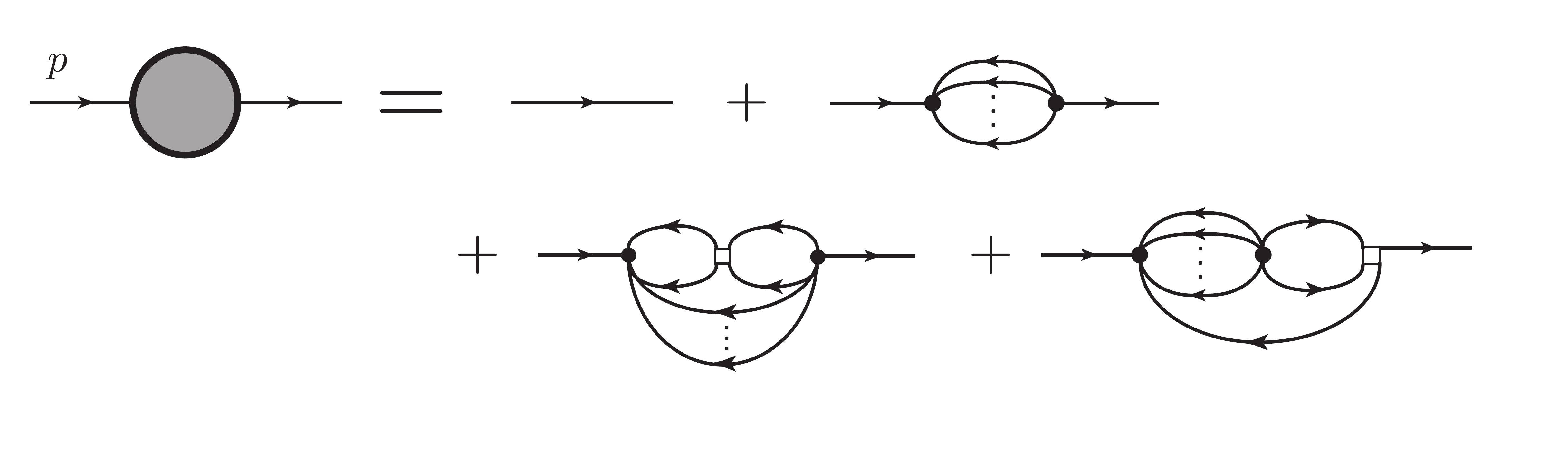}
\includegraphics[width=16cm]{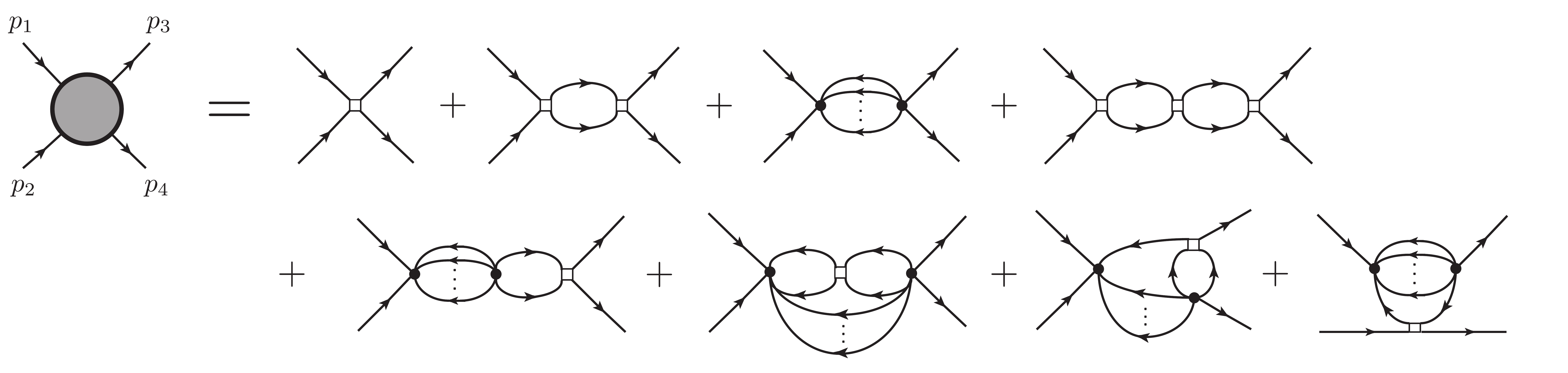}
\includegraphics[width=16cm]{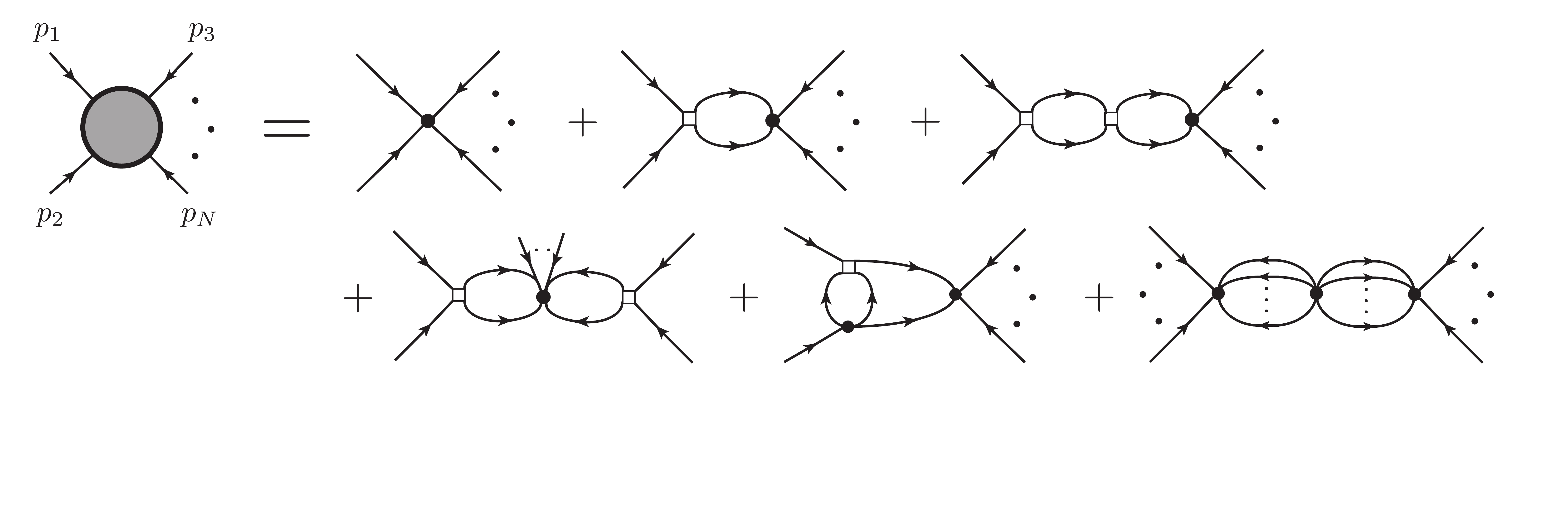}
\caption{Diagrams contributing to the 1PI vertices (top) $\Gamma^{(2)}$, (middle) $\Gamma^{(4)}$, (bottom) $\Gamma^{(N)}$. The ellipses represent the insertion of propagators required so that each $\lambda_0$ vertex has a total of $N$ lines attached.}
\label{fig:gamma2}
\end{figure}
We now outline the diagrammatic expansion for renormalizing $\mathcal{S}_B$.
The diagrams needed to renormalize the interactions are pictured in Figure \ref{fig:gamma2}.
We can immediately identify the main technical challenge, which is that the loop diagrams are only defined for integer $N$.
For example, the first correction to $\Gamma^{(2)}$ pictured in Figure \ref{fig:gamma2} is an $(N-1)$-loop diagram, and a given loop diagram only makes sense for an integer number of loops.
However, we are interested in an analytic expansion in the theory in small $\delta = 4 - N$.
This requires finding an expression for this diagrams for all integers $N$, analytically continuing this expression to arbitrary values of $N$, and then performing the expansion in $N = 4 - \delta$.
The method by which we compute and analytically continue these diagrams is outlined in Appendix \ref{app:mloop}, which also contains derivations of the integrals needed.

\begin{figure}
\includegraphics[width=16cm]{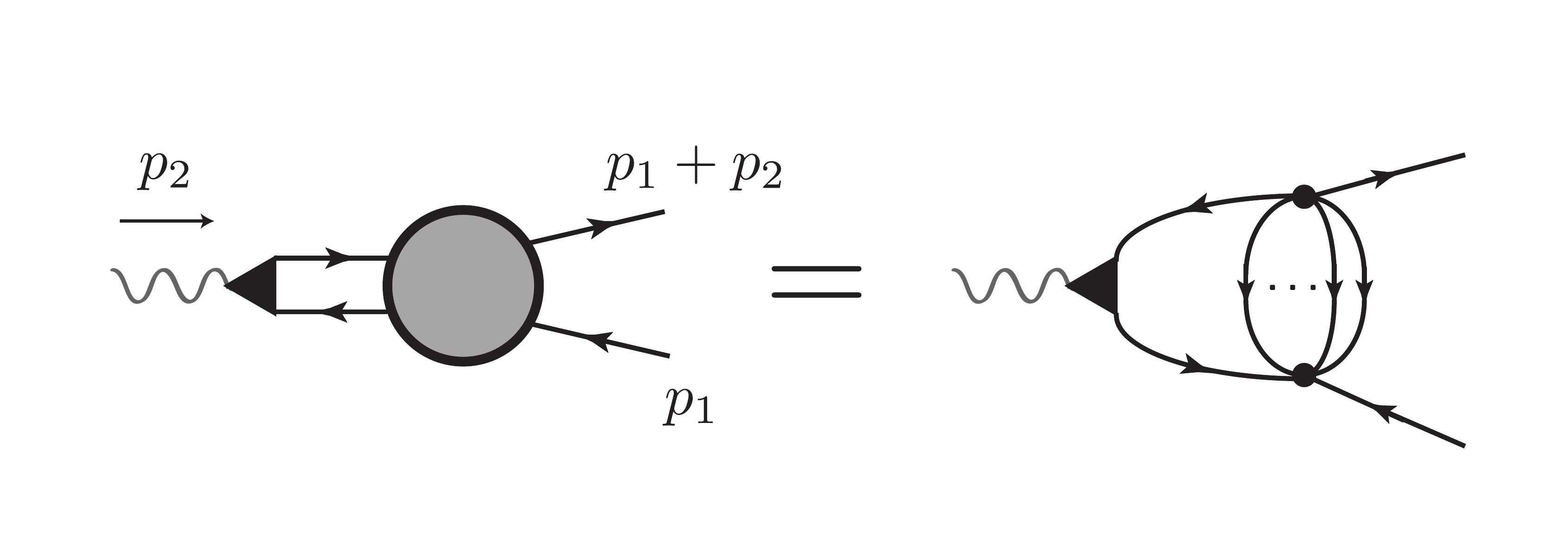}
\caption{Diagrams contributing to the renormalization of $\Gamma^{(2,1)}$.}
\label{fig:gamma12}
\end{figure}

Using the expressions for $I^{(M)}_{1-4}$ given in Appendix \ref{app:mloop}, the bare 1PI vertices pictured in Figures \ref{fig:gamma2} and \ref{fig:gamma12} are given by
\begin{alignat}{2}
\Gamma^{(2)}(\omega,k) &= &&- i \omega + k^2 - \frac{\lambda_0^2}{\Gamma(N)} I_1^{(N-2)}(\omega,k) + \frac{u \lambda_0^2}{2 \Gamma(N - 2)} I_2^{(N - 1)}(\omega,k) \nn[.25cm]
& && + \  \frac{2 u \lambda_0^2}{\Gamma(N-1)} I_3^{(N - 1)}(\omega,k),\\
\label{eq:g2bare}
\Gamma^{(4)} &= &&- 2 u  + 2 u^2 I_1^{(1)}(p_1+p_2) + \frac{\lambda_0^2}{\Gamma(N - 1)}  I_1^{(N-3)}(p_1+p_2) \nn
& && - \ 2 u^3 I_1^{(1)}(p_1+p_2)^2 \nn
& && - \ \frac{2 u \lambda_0^2}{\Gamma(N - 1)} \bigg[ I_1^{(1)}(p_1+p_2) I^{(N-3)}_1(p_1+p_2) \bigg] \nn
& && - \ \frac{u \lambda_0^2}{2 \Gamma(N - 3)} I_2^{(N-2)}(p_1+p_2) \nn
& && - \frac{u \lambda_0^2}{\Gamma\left(N - 2 \right)} \left[ I_3^{(N-2)}(p_1 + p_2,p_3) + 3 \ \mathrm{perms.} \right] \nn
& && - \frac{2 u \lambda_0^2}{\Gamma(N - 1)} \bigg[ I_4^{(N-2)}(p_1,p_3) + 3 \ \mathrm{perms.}  \bigg],\\
\Gamma^{(N)}(\omega_i,k_i)  &= &&- \lambda_0 + \lambda_0 u \sum_{i < j}^N I_1^{(1)}\left( p_i + p_j \right) - \lambda_0 u^2 \sum_{i < j}^N I_1^{(1)}\left( p_i + p_j \right)^2 \nn
& && - \ \lambda_0 u^2 \sum_{i < j < k < \ell} \Big[ I_1^{(1)}(p_i + p_j) I_1^{(1)}(p_k + p_{\ell}) + I_1^{(1)}(p_i + p_j) I_1^{(1)}(p_k + p_{\ell}) \nn
& && \qquad \qquad \qquad \qquad \qquad \qquad + \ I_1^{(1)}(p_i + p_j) I_1^{(1)}(p_k + p_{\ell}) \Big] \nn
& && - \ 2\lambda_0 u^2 (N-2) \sum_{i < j}^N I_3^{(2)}\left( p_i + p_j \right) \nn
& && - \ \frac{\lambda^2}{N!} \sum_{n = 2}^{\left \lfloor{\frac{N}{2}}\right \rfloor } \binom{N}{n} I_1^{(n-1)}(p_i) I_1^{(N-n-1)}(-p_i),\\
\Gamma^{(2,1)}(\omega_i,k_i) &= && 1 + \frac{\lambda_0^2}{\Gamma\left( N - 1 \right)} I_4^{(N-2)}(p_1,p_1 + p_2).
\end{alignat}
Here, the terms labelled ``$\mathrm{perm.}$'' denote the same integrals with permutations of the labelled external momenta.
We then apply the renormalization conditions of Eqs.~\eqref{eq:1pirenorm} and \eqref{eq:renormp2}:
\bea
\Gamma^{(2)}_R(\omega,k) &=& Z_{\tau} \, \Gamma^{(2)}(\omega,k), \nn
\Gamma^{(4)}_R(\omega,k) &=& Z \, Z_{\tau} \, \Gamma^{(4)}(\omega,k), \nn
\Gamma^{(N)}_R(\omega,k) &=& Z^{N/2 - 1} \, Z_{\tau} \, \Gamma^{(N)}(\omega,k), \nn
\Gamma^{(2,1)}_R(\omega,k) &=& Z_2 \, \Gamma^{(2,1)}(\omega,k).
\label{eq:grenormed}
\eea
Finally, we express the bare couplings appearing on the right-hand side of the equation in terms of the renormalized 1PI couplings defined in Eq.~\eqref{Eq:renorm}, and then choose the renormalization constants to render these functions finite near $\epsilon = \delta = 0$.

Using the above expressions, we will renormalize the theory to second order in the couplings.
We have also obtained the renormalized vertex $\Gamma^{(2)}$ to third order in the couplings; because this vertex does not have a contribution at first order, knowing the fixed point to second order is sufficient to obtain $Z_{\tau}$ and $Z$ (and therefore $z$ and $\Delta_{\psi}$) to third order.
The third-order calculation does not involve any extra difficulty because the diagrams appearing in $\Gamma^{(2)}$ at third order involve no new integrals compared to those needed to renormalize the interaction vertices at second order.

From the expressions for $I_{1-4}$, the bare 1PI vertices have divergences in the form of poles in the $\epsilon-\delta$ plane.
For example, in the 1PI four-point function we find simple poles of the form
\beq
\Gamma^{(4)}_R = Z_g + \frac{f_1(\epsilon,\delta)}{\epsilon} + \frac{f_2(\epsilon,\delta)}{\epsilon/2 + \delta - \epsilon \delta/2} + \mathrm{reg.}
\label{eq:g4renorm}
\eeq
at leading order, where the $f_i(\epsilon,\delta)$ are complicated functions and ``$\mathrm{reg.}$'' indicates finite contributions.
As usual, there is a very large ambiguity in defining the $Z_g$ to subtract these poles, and this ambiguity will not affect universal quantities at the critical point.
If we only had simple poles in $\epsilon$, a common choice is to subtract the poles in $\epsilon$ without subtracting any finite part of the bare vertices.
This is the modified minimal subtraction scheme, $\overline{\mathrm{MS}}$, where ``modified'' refers to the extra factors of $S_d$ inserted in the definitions of our couplings in Eq.~\eqref{Eq:renorm}.
However, is it not possible to subtract the pole $1/(\epsilon/2 + \delta - \epsilon \delta/2)$ without retaining some of the $\epsilon$ or $\delta$ dependence in the numerator. 
One may choose to subtract the pole with the entire function $f_2(\epsilon,\delta)$ in the numerator, but instead we have chosen
\beq
Z_g = 1 -  \frac{f_1(0,0)}{\epsilon} - \frac{f_2(\epsilon_1(\delta),\delta)}{\epsilon/2 + \delta - \epsilon \delta/2},
\eeq
where
\beq
\epsilon_1(\delta) \equiv -\frac{2\delta}{1 - \delta}.
\eeq
One may check that this choice renders Eq.~\eqref{eq:g4renorm} a regular function of $\epsilon$ and $\delta$ wherever the $f_i$ remain regular.
Our reason for this choice is that it reduces exactly to the $\overline{\mathrm{MS}}$ scheme in the $\delta = 0$ limit, allowing us to check the renormalization constants against those of the $N=4$ theory in the $\overline{\mathrm{MS}}$ scheme, which are much easier to obtain.

At second order, we find a more complicated pole structure in the $\epsilon-\delta$ plane, but we continue to renormalize the theory by demanding that our renormalization scheme reduces to $\overline{\mathrm{MS}}$ for $\delta = 0$. 
For example, we find a contribution of the form
\beq
Z_g + \frac{f_3(\epsilon,\delta)}{\epsilon (\epsilon/2 + \delta - \epsilon \delta/2)}.
\eeq
We renormalize this by choosing
\beq
Z_g = 1 - \frac{f_3(0,\delta)}{\epsilon \delta} + \frac{f_3(\epsilon_1(\delta),\delta)}{\delta (\epsilon - \epsilon_1(\delta))},
\eeq
which satisfies the two conditions that (1) the resulting expression is regular for all $\epsilon$ and $\delta$, and (2) the $\delta \rightarrow 0$ limit of $Z_g$ reduces to the $\overline{\mathrm{MS}}$ scheme if we work exactly at $N = 4$.

\subsection{RG results and critical exponents}

The renormalization constants are tabulated in Appendix \ref{sec:rgconstants}.
One may check that with these choices, the $\Gamma_R$ are finite functions of external frequency and momentum for small $\epsilon$ and $\delta$ and arbitrary $\epsilon/\delta$.
We can obtain the beta functions to second order in the couplings using Eqs.~(\ref{eq:defs})-(\ref{eq:betaSW}), and then truncate the resulting expressions by assuming that $\epsilon$ and $\delta$ are of the same order as $g$ and $\lambda$. 
The resulting beta functions are
\bea
\beta_g &=& - \epsilon g + g^2 + \frac{\lambda^2}{4}\left( 1 + \alpha_1 \delta\right) - g \lambda^2 \left( \frac{38}{27} + \ln \frac{4}{3} \right), \nn[.5cm]
\beta_{\lambda} &=& -\left( \epsilon + \delta - \delta \epsilon/2 \right) \lambda + 6 \lambda g (1 - 7 \delta/12) - 12 g^2 \lambda \ln \frac{4}{3} - \frac{2}{27}\lambda^3.
\eea
Here, $\alpha_1 = 2 - \gamma_E - \ln 2$. 
We may obtain higher-order dependence on $\epsilon$ and $\delta$ using our obtained renormalization constants, but this will not contribute to the perturbative fixed point to the order that we are working. 

Similarly, we can calculate the quantities $z$, $\Delta_{\Psi}$, and $\Delta_{|\Psi|^2}$ directly from Eqs.~\eqref{eq:zdef}, \eqref{eq:p1def}, and \eqref{eq:p2def}, obtaining
\begin{alignat}{1}
z &= 2 - \frac{4 \lambda^2}{27} \left( 1 + \alpha_z \delta \right) + \frac{4 \ln 2}{9} g \lambda^2,\\
\Delta_{\Psi} &= \frac{d}{2} - \frac{\lambda^2}{18} \left( 1 + \alpha_{\Psi} \delta \right) + \frac{\ln 2}{6} g \lambda^2,\\
\Delta_{|\Psi|^2} &= d + \frac{8 \lambda^2}{27} ,
\end{alignat}
where $\alpha_z = \frac{9}{4} - \gamma_E - \frac{1}{2} \ln 3$ and $\alpha_{\Psi} = \frac{13}{6} - \gamma_E - \frac{1}{2} \ln 3$.

We now consider solutions to the equations $\beta_g = \beta_{\lambda} = 0$. We find the $U(1)$-symmetric fixed point at $(g^{\ast},\lambda^{\ast}) = (\epsilon,0)$. Computing the stability of this fixed point to $\lambda$ perturbations, we have:
\beq
\dim(\lambda) = \delta - 5 \epsilon + 3 \epsilon \delta + 12 \log(4/3) \epsilon^2, \qquad (g,\lambda) = (\epsilon,0).
\eeq
This equation also follows from Eq.~(4) in Ref.~\onlinecite{DS96}.
The $U(1)$ fixed point is stable for small values of $\epsilon$ and $\delta$, but it becomes unstable if we tune them to be large enough. 
The properties of this fixed point have been studied in detail, see {\it e.g.,} Ref.~\onlinecite{ssbook}.

\begin{figure}
\includegraphics[width=0.5\linewidth]{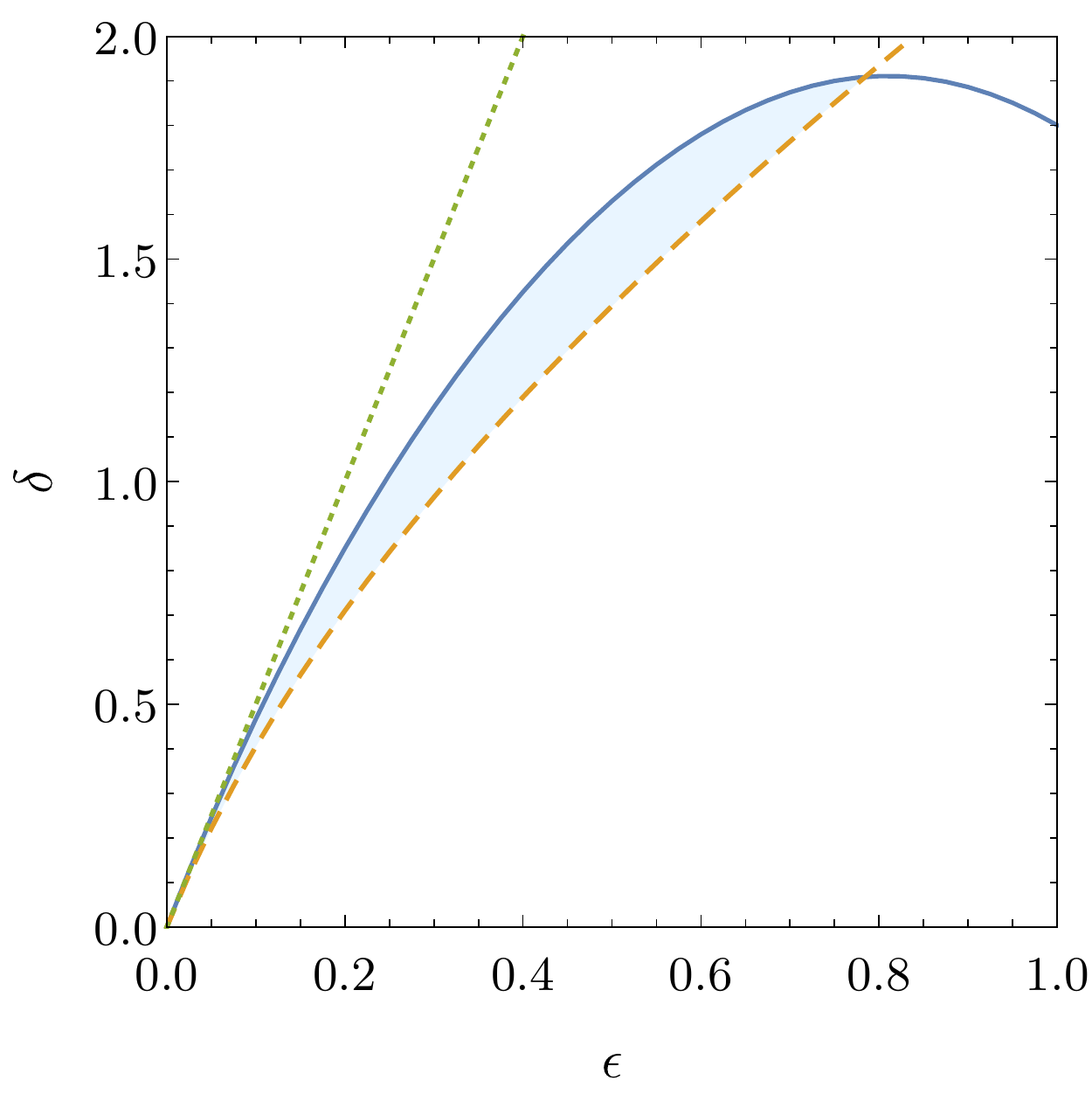}
\caption{Region in $\epsilon-\delta$ space where the $\mathbb{Z}_{N}$ fixed point is stable, evaluated to second order in an expansion in $\epsilon = 2 - d$ and $\delta = 4 - N$. The eigenvalue $\omega_+$ is zero along the full line, and the eigenvalue $\omega_-$ is zero along the dashed line. The shaded region between these two lines is the region where both eigenvalues are positive, representing the region of stability of the fixed point. The dotted line is $\delta = 5 \epsilon$; the fixed point only exists in the region $\delta < 5 \epsilon$.}
\label{fig:eigs}
\end{figure}

For $\lambda \neq 0$, we have the following weak-coupling fixed point:
\bea
g^{\ast} &=& \frac{1}{6} \left( \epsilon + \delta \right) + \frac{40 \epsilon^2 + 559 \delta^2 + 113 \epsilon \delta + 324 \ln(4/3) (\epsilon + \delta)^2}{5832}, \nn[.5cm]
\lambda^{\ast} &=& \frac{1}{3} \sqrt{ \left( \epsilon + \delta \right) \left( 5 \epsilon - \delta \right)} + \frac{2360 \epsilon^3 - 1015 \delta^3 + 4290 \delta \epsilon^2 + 2373 \delta^2 \epsilon + 324(7 \epsilon - 2 \delta)(\delta + \epsilon)^2 \ln(4/3)}{2916 \sqrt{ \left( \epsilon + \delta \right) \left( 5 \epsilon - \delta \right)}}\nn[.5cm]
&& \qquad - \ \frac{\alpha_1\delta}{6} \sqrt{\left( \epsilon + \delta \right) \left( 5 \epsilon - \delta \right)}.
\eea
This fixed point only exists when $\delta < 5 \epsilon$. 
We first discuss the stability of this fixed point, and the possibility that it is a stable endpoint from the free or $U(1)$-symmetric fixed points. 
The condition for stability is that $\omega_+$ and $\omega_-$, defined by Eq.~\eqref{eq:omegapmdef}, are both positive.
Computing these at the fixed point, we find the two eigenvalues to be
\bea
\omega_{\pm} &=& \frac{1}{6} \left( \delta - 2 \epsilon \right) \pm \frac{1}{6} \sqrt{64 \epsilon^2 - 11 \delta^2 + 44 \epsilon \delta} \nn[.5cm]
&& + \ \frac{1063 \delta^2 - 2480 \epsilon^2 - 1903 \epsilon \delta - 648(2 \epsilon - \delta)(\epsilon + \delta) \log \frac{4}{3}}{5832} \nn[.5cm]
&& \pm \ \frac{4960 \epsilon^3 - 2339 \delta^3 - 13740 \delta \epsilon^2 - 5001 \epsilon \delta^2 - 648 (\epsilon + \delta)(14 \epsilon^2 -  \delta^2 + 22 \epsilon \delta) \log\frac{4}{3}}{5832 \sqrt{64 \epsilon^2 - 11 \delta^2 + 44 \epsilon \delta}}.
\eea
To leading order, we find that there is no region where this fixed point exists and is stable. This is the primary reason we have carried out the present computation to second order.
Using the full expression, we find that there is a region where where the fixed point is stable and both eigenvalues $\omega_{\pm}$ are positive, although this region does not include $\epsilon = \delta = 1$; see Figure \ref{fig:eigs}. 
However, we note that the exact results of Ref.~\onlinecite{DS96} imply that for $\epsilon = 1$, $\delta = 2$, the U(1)-symmetric fixed point flows to the transverse-field Ising fixed point, so this region of stability presumably continues to increase at higher orders.
Therefore, this prediction of an unstable fixed point may be a failure of our expansion in obtaining quantitatively accurate values of the $\omega_{\pm}$.
In what follows, we assume that the region of stability extends to $\epsilon = \delta = 1$, the primary case of interest.

Evaluating $z$, $\Delta_{\Psi}$, and $\Delta_{|\Psi|^2}$ at this fixed point, we find
\bea
z &=& 2 - \frac{4(\epsilon + \delta)(5 \epsilon - \delta)}{243} \nn[.5cm]
&& + \ \frac{\left( -4720 - 4536 \log \frac{4}{3} + 2430 \log 2 \right) \epsilon^3 + \left( 2273 + 1782 \log \frac{4}{3} - 486 \log 2 \right) \delta^3 }{59049} \nn[.5cm]
&& + \ \frac{ \left( -3265 - 3402 \log \frac{4}{3} + 1458 \log 2 \right) \delta \epsilon^2- \left( 1906 + 1296 \log \frac{4}{3} - 486 \log 2 \right) \delta^2 \epsilon}{19683},
\eea
\bea
\Delta_{\psi} &=& \frac{2 - \epsilon}{2} - \frac{(\epsilon + \delta)(5 \epsilon - \delta)}{162} \nn[.5cm]
&& + \ \frac{\left( -2360 - 2268 \log \frac{4}{3} + 1215 \log 2 \right) \epsilon^3 + \left( 1096 + 891 \log \frac{4}{3} - 243 \log 2 \right) \delta^3 }{78732} \nn[.5cm]
&& + \ \frac{ \left( -1565 - 1701 \log \frac{4}{3} + 729 \log 2 \right) \delta \epsilon^2 - \left( 899 + 648 \log \frac{4}{3} - 243 \log 2 \right) \delta^2 \epsilon}{26244},
\eea
\beq
\Delta_{|\psi|^2} = 2 - \epsilon + \frac{8 (\epsilon + \delta)(5 \epsilon - \delta)}{243}.
\eeq
These expressions are simply related to the critical exponents of the theory. 
As noted above, the exponent $\nu$, defined as the exponent characterizing the divergence of the correlation length, will coincide with the exponent $\nu$ in the CCM in one dimension.
This exponent is given by
\beq
\nu^{-1} = d + z - \Delta_{|\Psi|^2}.
\eeq
We also give the anomalous dimension of the field $\Psi$, defined as
\beq
\eta = 2 \Delta_{\Psi} + 2 - d - z.
\eeq
This anomalous dimension will characterize the correlations of the \emph{domain walls} of the CCM rather than the order parameter.
\begin{table}
\begin{tabular}{| c | c | c |}
\hline
Exponent & LO & NLO \\
\hline
$z$ & 1.87 & 1.57 \\
\hline
$\nu $ & 0.60 & N/A  \\
\hline
$\eta$ & 0.03 & 0.11 \\
\hline
\end{tabular}
\caption{Critical exponents predicted for the $\mathbb{Z}_3$ DBG in one dimension to leading order (LO) and next-to-leading order (NLO). The exponents $z$ and $\nu$ in the one-dimensional DBG coincide with those in the one-dimensional CCM.}
\label{tab:critexp}
\end{table}

We note that all of these exponents lie between those for the one-dimensional 3-state Potts model and the U(1)-symmetric DBG model. 
In those cases, the exponents are known exactly:
\begin{alignat}{2}
(z,\nu,\eta) &= (2,1/2,0) \qquad \qquad &&\mathrm{U(1) \ DBG}, \nn
(z,\nu,\eta) &= (1,5/6,4/15) \qquad && 3\operatorname{-}\mathrm{state \ Potts},
\end{alignat}
(since the one-dimensional 3-state Potts model is self-dual, the exponent $\eta$ for $\Phi$ and $\Psi$ coincide).
The second-order correction to the exponents $z$ and $\eta$ is rather large, which may indicate that the series is already diverging and may require resummation at higher order.

Finally, we may also compare these results with those recently obtained using exact diagonalization on a lattice boson model in the same universality class as the $\mathbb{Z}_3$ CCM \cite{SCPLS18},
\beq
z \approx 1.33, \qquad \nu \approx 0.71.
\eeq
Our field-theoretic results do not give precise quantitative matches with these results, but we do find that the exponents shift in the correct direction for agreement with the CCM.

\section{Numerical results}
\label{sec:num}

In this section, we numerically investigate the $\mathbb{Z}_3$-symmetry-breaking QPT in the context of both the chiral clock (Sec.~\ref{sec:num_dual}) and dilute Bose gas (Sec.~\ref{sec:lattice}) models. The critical exponents of interest to us in characterizing the nature of this transition are the dynamical critical exponent $z$ and the correlation length exponent $\nu$, which are defined by \cite{sachdev2011quantum}
\begin{alignat}{1}
\label{eq:def}
\Delta &\sim \left \lvert g - g_c \right \rvert^{z\,\nu};\qquad
\xi \sim \left \lvert g - g_c \right \rvert^{-\,\nu}, 
\end{alignat}
where $g$ is some tuning parameter, $\Delta$ denotes the mass gap, and $\xi$ is the correlation length. For the purpose of numerically evaluating these exponents, we resort to finite-size scaling (FSS) \cite{fisher1972scaling, hamer1980finite} as sketched below.

The FSS approach employs the relation between the divergence of a thermodynamic quantity $\mathcal{K}\, (g)$ in the bulk system, as $\mathcal{K} \,(g) \sim \lvert g - g_c \rvert^{-\kappa}$ when $g \rightarrow g_c$, and its scaling at criticality, as $\mathcal{K}\, (g_c) \sim L^{\kappa /\nu}$, on a lattice of $L$ sites. The exponent $\kappa/\nu$ can thus be estimated by plotting $\mathcal{K}$ against the system size, where $\mathcal{K}$ is to be chosen appropriately. For instance, near the quantum critical point (QCP), one can assume that the gap obeys a scaling ansatz of the functional form
\begin{equation}
\label{eq:ansatz}
\Delta = L^{-z} \, \mathcal{F} \left( L^{1/\nu}\,(g-g_c) \right),
\end{equation}
with $\mathcal{F}$ some universal scaling function. Additionally, in this regard, it is also useful to consider the Callan-Symanzik $\beta$ function \cite{hamer1979strong} defined as
\begin{equation}
\label{eq:beta}
\beta\, (g) = \dfrac{\Delta\, (g)}{\Delta\, (g)  - 2\, {\displaystyle \dfrac{\partial\,\Delta (g) \vphantom{\huge e^{x^x}}} {\partial\, \ln g}} }.
\end{equation}
From Eqs.~\eqref{eq:ansatz} and \eqref{eq:beta}, it follows that these two quantities scale as $-z$ and $-1/\nu$, respectively, at the QCP, thus giving us access to the required exponents.

Our numerical calculations in this section are based on the density-matrix renormalization group (DMRG) algorithm \cite{white1992density, white1993density,ostlund1995thermodynamic, rommer1997class, dukelsky1998equivalence,peschel1999density}. We use finite-system DMRG \cite{schollwock2005density, schollwock2011density} with a bond dimension $m = 150$ for a chain of up to $L=100$ sites with open boundary conditions; the first and second energy levels are individually targeted.  After three sweeps, the energy eigenvalues were found to be suitably converged to an accuracy of one part in $10^{10}$.

\subsection{The $\phi \leftrightarrow \theta$ duality}
\label{sec:num_dual}

The $\phi \leftrightarrow \theta$ duality, introduced in Sec.~\ref{sec:duality}, maps the Hamiltonian of the chiral clock model onto itself under the simultaneous interchange of both $f \leftrightarrow J$ and $\phi \leftrightarrow \theta$. Despite this mapping, the two sides of the phase diagram are \textit{not} the same in that the energy levels are not identical owing to boundary effects. 

The critical exponents of the chiral $\mathbb{Z}_3$ transition were recently studied for $\phi =0$, $\theta \ne 0$ by Ref.~\onlinecite{SCPLS18}; here, we consider its dual case with $\theta = 0$ and $0 \le \phi < \pi/6$, varied in steps of $\pi/48$, in the subspace $J = 1 - f$. The precise location of the QPT can be ascertained by plotting $L^z \, \Delta_L$ against the tuning parameter $f$ for various lattice sizes (ranging from $L=60$ to $L=100$) and values of $z$. Eq.~\eqref{eq:ansatz} asserts that the quantity $L^z\Delta$ is independent of the length of the system $L$ right at the QCP $f=f_c$. This, in turn, implies that, with the correct choice of $z$, all the curves for $L^z\Delta$ should cross at $f_c$ for different values of $L$, thereby allowing us to determine both $f_c$ and $z$ simultaneously. Following this prescription, we are able to determine the intersection point of the curves for different lengths to an accuracy of $10^{-4}$ by scanning progressively finer intervals. The variation of the crossing points with $\phi$ (for $\theta = 0$) is noted in Table~\ref{Table:z}, along with the corresponding values for $ \phi = 0$, $\theta \ne 0$ (from Ref.~\onlinecite{SCPLS18}). Although the QCPs in the two cases are obtained separately, it is easy to observe that $f_c \vert_{\phi = 0} = 1- f_c \vert_{\theta= 0}$, as predicted by the duality.

\bgroup
\def\arraystretch{1.5}
\begin{table}[htb]
\centering
{
\bgroup
\setlength{\tabcolsep}{11.25pt}
\begin{ruledtabular}
\begin{tabular}{l l l l | l l l l} 
\multicolumn{1}{c}{$\theta$} &\multicolumn{1}{c}{$f_c \vert_{\phi = 0}$} &\multicolumn{1}{c}{\large $z$}  &\multicolumn{1}{c |}{$\mbox{\large $z$}_{c}$} &\multicolumn{1}{c}{$\phi$} &\multicolumn{1}{c}{$f_c \vert_{\theta = 0}$} &\multicolumn{1}{c}{\large $\bar{z}$}  &\multicolumn{1}{c}{$\mbox{\large $\bar{z}$}_{c}$}  \\ \hline
$\pi/48$  & $0.4990$ & $1.003$ &$1.00(9)$ &$\pi/48$  & $0.5010$ & $1.000$ &$1.00(7)$ \\
$\pi/24$  & $0.4961$ & $1.021$ &$1.029(6)$ &$\pi/24$  & $0.5040$ & $1.013$ &$1.01(4)$  \\
$\pi/16$  & $0.4913$ & $1.022$ &$1.02(3)$ & $\pi/16$  & $0.5090$ & $1.047$ &$1.04(8)$\\
$\pi/12$  & $0.4842$ & $1.078$ &$1.078(2)$ &$\pi/12$  & $0.5161$ & $1.118$ &$1.119(0)$\\
$5 \pi/48$  & $0.4748$ & $1.135$ &$1.132(7)$ & $5 \pi/48$  & $0.5257$ & $1.155$ &$1.150(6)$\\
$\pi/8$  & $0.4627$ & $1.229$ & $1.226(8)$ &$\pi/8$  & $0.5379$ & $1.224$ & $1.216(4)$\\
$7\pi/48$ &$0.4475$ & $1.368$ &$1.366(1)$ & $7\pi/48$ &$0.5531$ & $1.331$ &$1.324(6)$\\
\end{tabular}
\end{ruledtabular}
}
\egroup
\caption{\label{Table:z}Calculated dynamical critical exponents for $\phi = 0$, $\theta \ne 0$ ($z$) and $\phi \ne 0$, $\theta = 0$ ($\bar{z}$). Two independent sets of values of $z$ are distinguished: the first series is our estimate from the crossing points whereas the second (designated by the subscript $c$) is for the values  determined from fitting $\Delta$ to $c \,L^{-z}$. }
\end{table}

The values of $z$ obtained in this fashion can be independently corroborated in order to check for any dependence (or lack thereof) on the particular system sizes over which FSS is applied. While our former approach relied on considering $\Delta$ as a function of $f$, one can alternatively study the scaling of $\Delta$ as a function of $L$ instead, at $f= f_c$. Using the ansatz  $\Delta \,(L) = c \, L^{-z}$, we obtain the best functional fit for the gap, treating the coefficient $c$ and the exponent $z$ as free parameters. Table~\ref{Table:z} lists the values of $z$ thus obtained, together with those for $\phi =0$, $\theta \ne 0$. The exponents in these two cases while close, are not exactly the same since they are essentially determined from curve-fitting. Nonetheless, the good agreement between the two serves as a highly nontrivial check of the duality.

Another interesting aspect of the duality is the physics along the self-dual line of the CCM: when $f = J$ and $\phi = \theta$, the model is self-dual. Unlike previously, in this case, we find that there is no direct order--disorder phase transition and a sliver of the gapless incommensurate phase always intervenes. Further numerical details pertaining to the self-dual phase boundary are presented in Appendix~\ref{sec:P=T}.

\subsection{Lattice model for the dilute Bose gas}
\label{sec:lattice}
The renormalization group analysis of the $\mathbb{Z}_N$ dilute Bose gas \eqref{eq:DBG} in Sec.~\ref{sec:rgdbg} is complemented by our numerics in this section, which focus exclusively on the case $N = 3$. To this end, we study a lattice Hamiltonian, which is a variation upon the usual Bose-Hubbard model, described by a hopping strength $t$, a chemical potential $\mu$, and an onsite repulsion $U$. The only addition is a point-split perturbation which breaks the symmetry down to $\mathbb{Z}_3$. Explicitly,
\begin{equation*}
H = - t \sum_{\langle i, j \rangle} \left( b_i^\dagger \,b_j + b_i \,b_j^\dagger \right) - \mu\, \sum_i b_i^\dagger \,b_i + U \sum_i b_i^\dagger \,b_i \left(b_i^\dagger \,b_i - 1 \right) + \lambda \sum_i \left (b_i \, b_{i+1}\, b_{i+2} + b_i^\dagger \, b_{i+1}^\dagger\, b_{i+2}^\dagger \right),
\end{equation*}
where $b_i$ and $b_i^\dagger$ are the bosonic annihilation and creation operators, respectively. Taking the limit $U \rightarrow \infty$ imposes a hard-boson constraint i.e., each site can be occupied by no more than one boson. With this constraint implicit hereafter, the Hamiltonian reduces to
\begin{equation}
H = - t \sum_{\langle i, j \rangle} \left( b_i^\dagger \,b_j + b_i \,b_j^\dagger \right) - \mu\, \sum_i n_i + \lambda \sum_i \left (b_i \, b_{i+1}\, b_{i+2} + b_i^\dagger \, b_{i+1}^\dagger\, b_{i+2}^\dagger \right); \quad n_i \le 1,
\label{eq:numham}
\end{equation}
where we have introduced the number operator $n_i = b^\dagger_i \,b_i$ for notational clarity.

With $\lambda =0$, this system exhibits a $U(1)$-symmetry-breaking  QPT: this is the usual Bose gas transition between a  Mott-insulating ground state ($ t \ll \mu$) a superfluid phase ($t \gg \mu$), in which the $U(1)$-symmetry is spontaneously broken. The more interesting case, which we now turn to, is the ground state of the system for $\lambda > 0$; without loss of generality, we set $t=1$ . For $ \lvert \mu/\lambda \rvert \gg 1$, the ground state is unique and tends towards being entirely empty or entirely filled depending on the sign of $\mu$. On the other hand, for $\lvert \mu/\lambda \rvert \ll 1$, one is in the $\mathbb{Z}_3$-ordered phase, where three states with different fillings $\mathrm{mod}\, 3$ (each of which is individually $\mathbb{Z}_3$-symmetric) are degenerate, and the system can break the $\mathbb{Z}_3$ symmetry by choosing a linear combination of the three states. Both the disordered and the $\mathbb{Z}_3$-ordered phase are gapped, provided $\lambda/t$ is sufficiently large.

A convenient metric to characterize the phases and observe the transition between them is the spatial entanglement entropy (EE) \cite{calabrese2004entanglement, levin2006, kitaev2006}. Formally, if a system is partitioned into two regions, say, A and B, then the reduced density matrix of region A is obtained by tracing
over the degrees of freedom of region B as $\rho_\textsc{a} \equiv \mathrm{Tr}_\textsc{b}\, \rho$; the EE is then defined as $S \equiv - \mathrm{Tr} (\rho_\textsc{a} \ln \rho_\textsc{a})$. If there is topological ground-state degeneracy, one expects an EE of order $\sim \ln D$, where $D$ is the degeneracy \cite{pollmann2010entanglement}. Hence, we set $t = 1$ and numerically calculate the EE using DMRG with $m = 100$ to detect the QPT and the (approximate) phase boundaries in the $(\mu,\lambda)$ parameter space.  Figure~\ref{fig:EE} shows a sharp increase in the central-cut EE, which saturates to a value $\sim \mathcal{O}( \ln 3 )$, as $\mu$ is varied across the phase transition. Furthermore, both the EE and the energy gap $\Delta$ are symmetric under $\mu \leftrightarrow - \mu$, which implies that the system undergoes the same phase transition at $(\mu_c, \lambda_c)$ and $(-\mu_c, \lambda_c)$.
\noindent
\begin{figure}[htb]
  \centering
\subfigure[]{\label{fig:EEa}\includegraphics[width=0.475\linewidth]{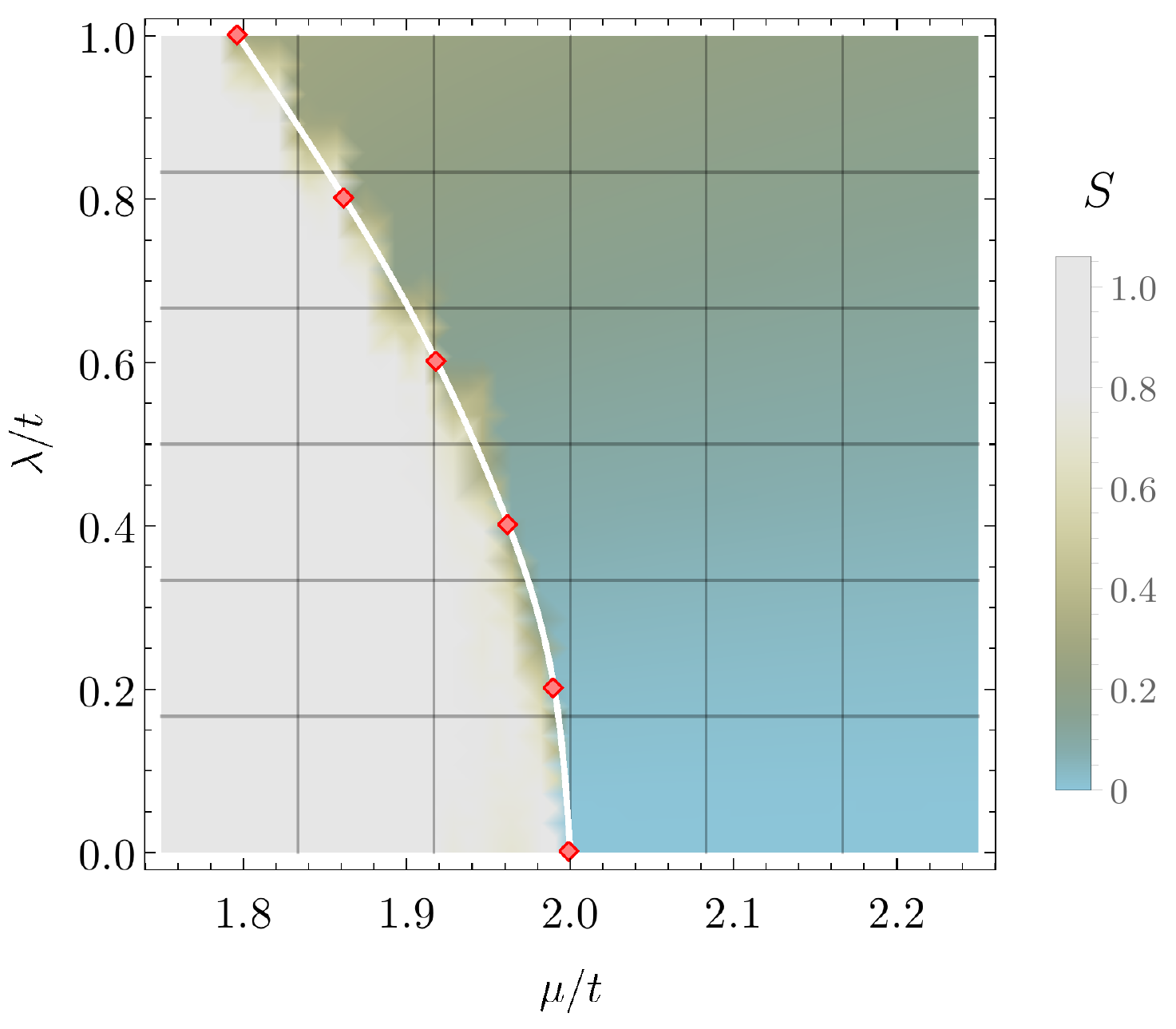}}
\subfigure[]{\label{fig:EEb}\includegraphics[width=0.495\linewidth, trim= 0 -1cm 0 0, clip]{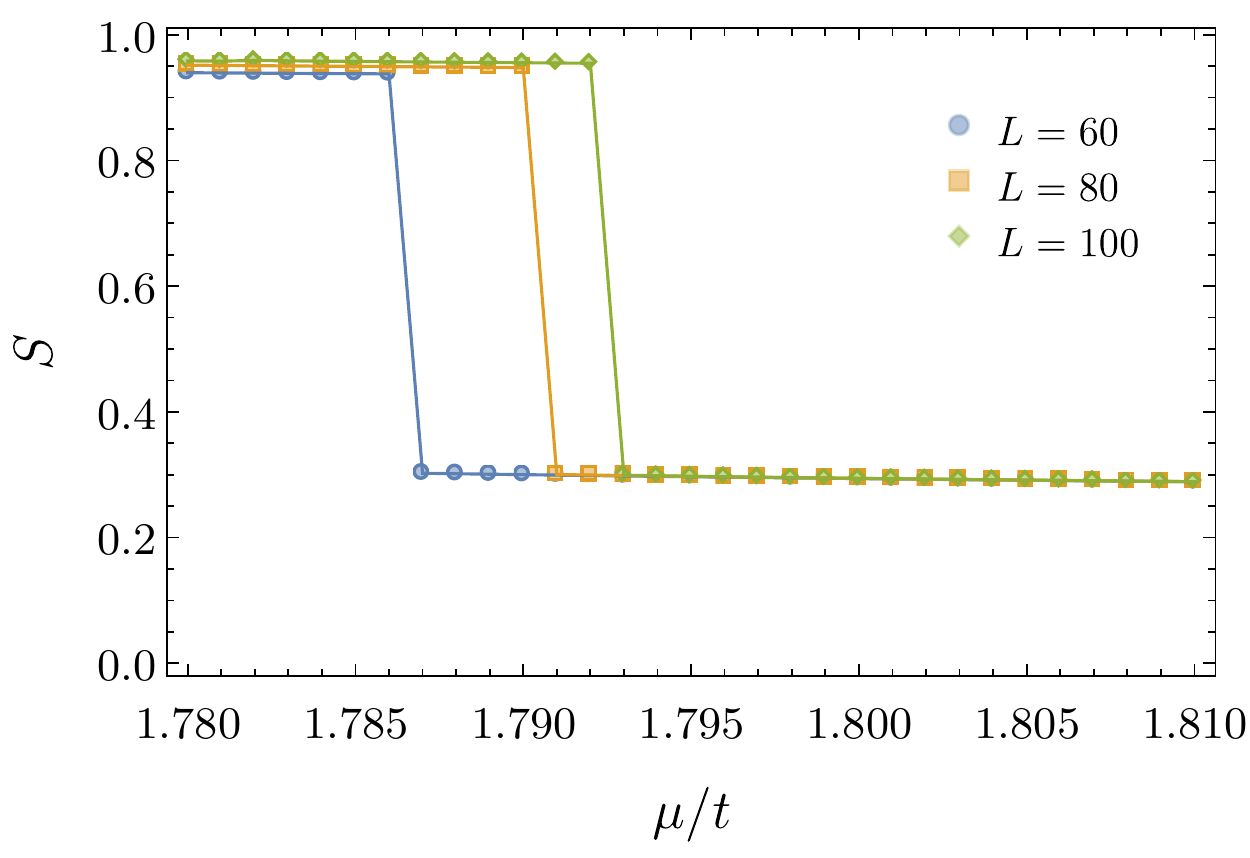}}
  \caption{\label{fig:EE}(a) The central-cut entanglement entropy as a function of $\mu$ and $\lambda$ for a chain of length $L =100$. The EE decreases sharply from $\mathcal{O}(\ln 3)$ to zero on moving across the phase transition from the ordered to the disordered phase. (b) Cross-section of (a) along the line $\lambda/t=1$, greatly zoomed in to the QCP, for three different system sizes. The transition point corresponding to the jump tends to the actual critical point for an infinite lattice as we move to progressively larger systems.}
\end{figure}

In order to precisely pinpoint the QCP, we use the same finite-size scaling considerations as in Sec.~\ref{sec:num_dual}. At each fixed value of $\lambda$, we systematically tune $\mu$ to drive the phase transition and note the intersection point of the curves of $L^z \,\Delta$ for $L$ varying between 60 to 100. For instance, Figure~\ref{fig:Cra} displays an example of this method for $\lambda = 1$, from which we extract the critical point $\mu_c = 1.7970$, and the associated exponent $z = 1.779$. Repeating this procedure over several discrete values of $\lambda$ leads to the phase diagram illustrated in Figure~\ref{fig:PDa}. Similarly, FSS arguments based on the scaling of the Callan-Symanzik $\beta$ function \eqref{eq:beta} can yield the correlation length exponent $\nu$. Accordingly, we fit the $\beta$ function, displayed in Figure~\ref{fig:Beta}, to an ansatz of the form 
\begin{equation}
\beta \,(L) = c_0\, L^{-1/\nu}\, (1 + c_1\, L^{-\zeta} ); \quad \left(\zeta > \frac{1}{\nu}\right),
\end{equation}
allowing for subleading corrections to ensure a more robust fit. The values of $\nu$ thus obtained are compiled in Table \ref{tab:table2} together with the corresponding exponents for $z$.

\begin{figure}[htb]
    \centering
    \begin{minipage}{0.49\textwidth}
        \centering
       \subfigure[]{\label{fig:Cra}\includegraphics[width=\linewidth]{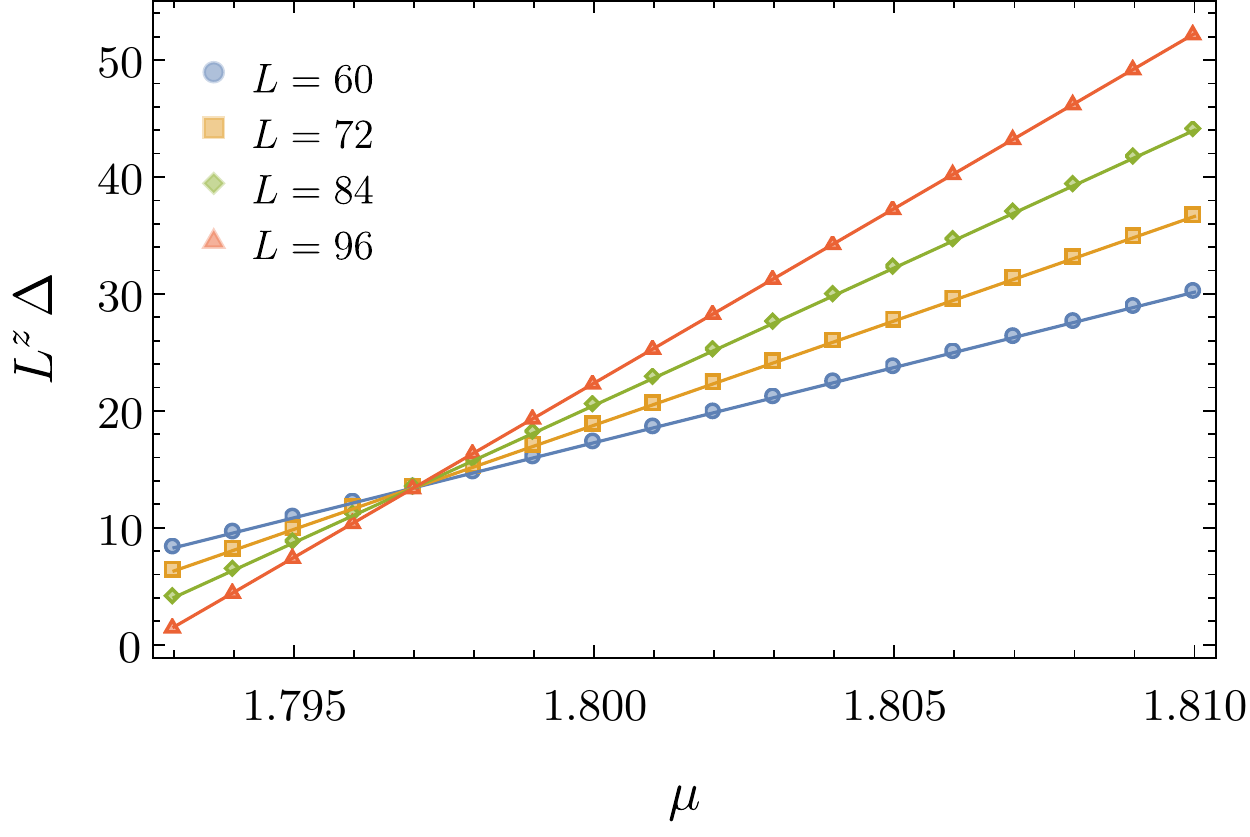}}\\
       \subfigure[]{\label{fig:Crb}\includegraphics[width=0.49\linewidth]{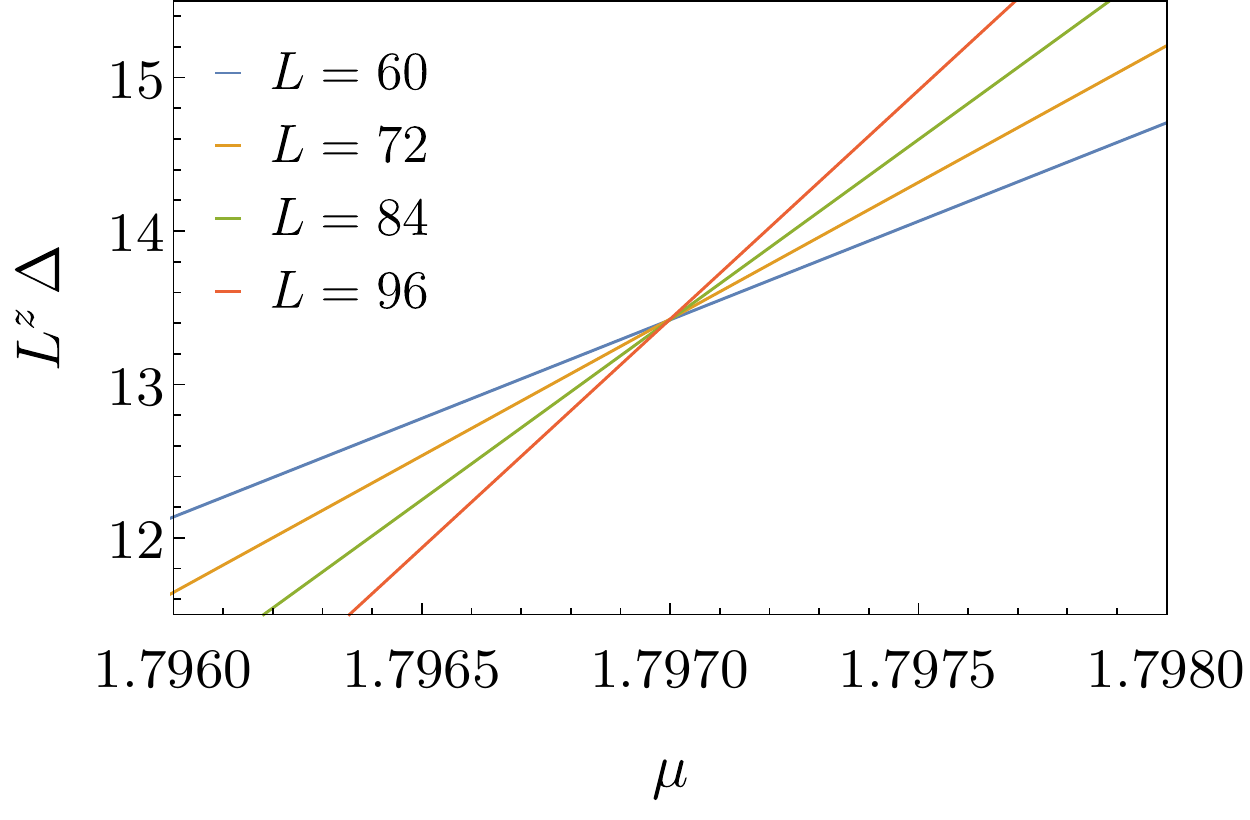}}
       \subfigure[]{\label{fig:Crc}\includegraphics[width=0.49\linewidth]{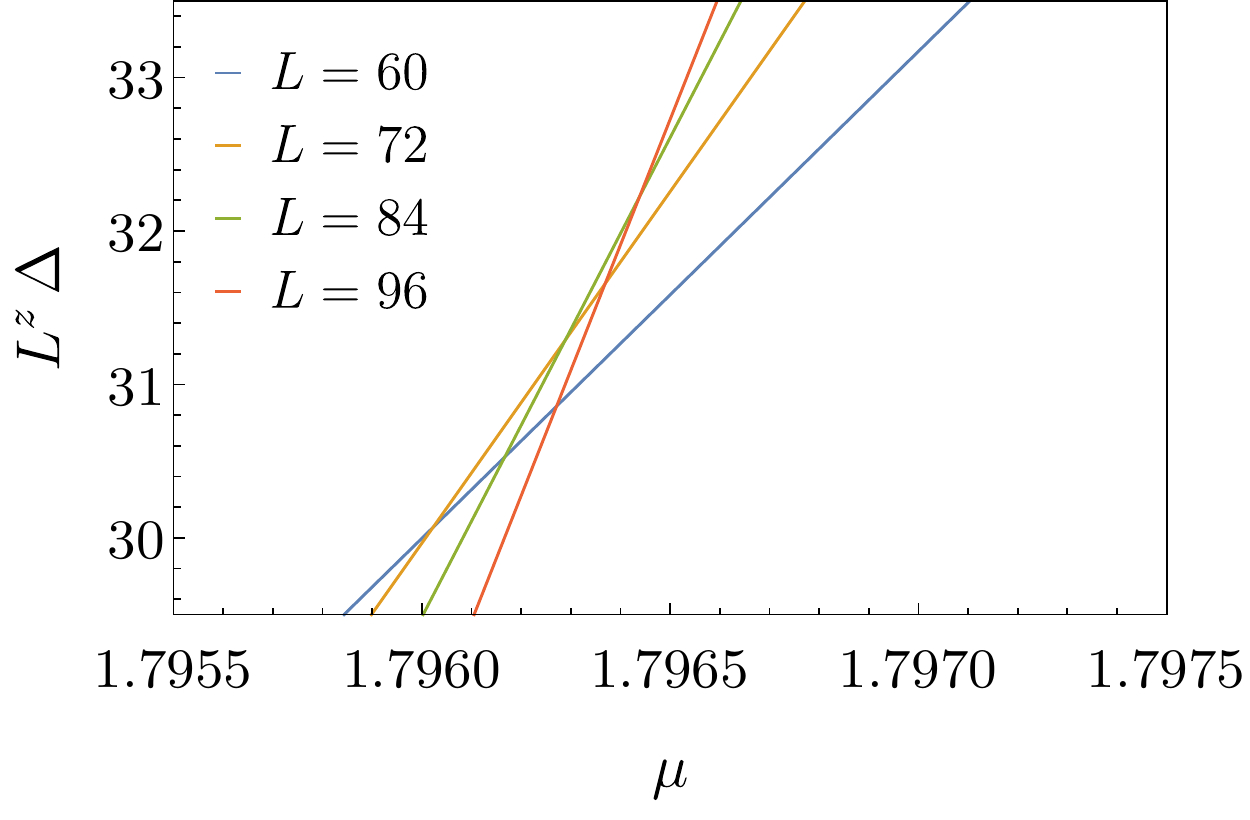}} 
    \end{minipage}
     \begin{minipage}{.5\textwidth}
        \centering
        \subfigure[]{\label{fig:Beta}\includegraphics[width=\linewidth]{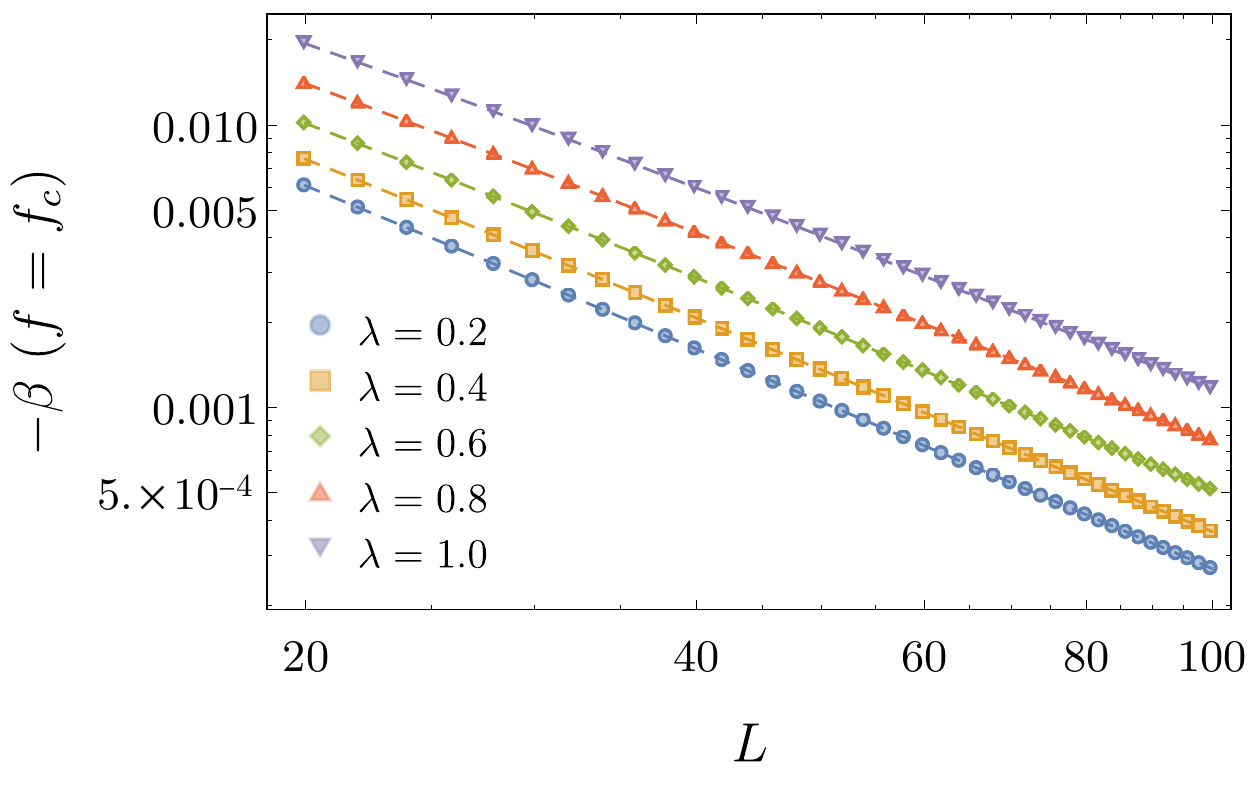}}\
    \end{minipage}%
\caption{\label{fig:Cross}Finite-size scaling analysis of the $\mathbb{Z}_3$ dilute Bose gas. (a) Scaling of the variable $L^z \,\Delta$ as a function of $\mu$ for individual system sizes. With $z = 1.779$, all the curves intersect right at the critical point. The finesse of the crossing depends crucially on the correct choice of $z$: on zooming in, the contrast in sharpness between $z=1.779$ (b) and $z = 2$ (c) is vivid. (d) The Callan-Symanzik $\beta$ function plotted on a logarithmic scale against the system size. The slopes of the curves in the linear region, corresponding to large lattices, give us the respective values of $-1/\nu$.}
\end{figure}

\begin{table}[h]
\centering
\setlength{\tabcolsep}{11.25pt}
\begin{tabular}{ cccl}
\toprule
\multicolumn{1}{c}{$\lambda$} & \multicolumn{1}{c}{$\mu_c$} & \multicolumn{1}{c}{$z$} & \multicolumn{1}{c}{$\nu^{-1}$}\\
\colrule
$0.2$ & $1.9904$ & $1.956 \pm 0.003$ &$1.963(8)$ \\
$0.4$  & $1.9625$ & $1.894 \pm 0.010$ &$1.900(5)$ \\
$0.6$  & $1.9186$ & $1.887 \pm 0.002$ &$1.899(3)$ \\
$0.8$  & $1.8622$ & $1.847 \pm 0.037$ &$1.86(5)$ \\
$1.0$  & $1.7970$ & $1.779 \pm 0.034$ &$1.798(3)$ \\
$1.5$  & $1.6156$ & $1.641 \pm 0.042$ & $1.683(1)$ \\
$2.0$  & $1.4329$ & $1.511 \pm 0.052$ & $1.57(4)$ \\
$2.5$  & $1.2638$ & $1.387 \pm 0.064$ & $1.484(7)$ \\
$3.0$  & $1.1129$ & $1.272 \pm 0.071$ & $1.405(2)$ \\
\botrule
\end{tabular}
\caption{\label{tab:table2}%
Numerically calculated dynamical and correlation length critical exponents for the $\mathbb{Z}_3$ dilute Bose gas with $t=1$. Upon turning on $\lambda$, the exponents start deviating nontrivially from the values of $z = 2$, $\nu = 1/2$ of the $U(1)$-symmetry-breaking transition. The extracted value of $z$ depends slightly on the range of system sizes considered in the FSS procedure. Denoting the exponent obtained from FSS over the interval $L =a$ to $L=b$ by $z_{a,b}$, we take $z \equiv z_{60,100}$, and the uncertainty estimate 
$\varepsilon \equiv \max(\vert z_{60,80}- z\vert, \vert z_{80,100}-z\vert)$.}
\end{table}
We find that the exponents move in the direction of those for the 3-state Potts model (where $z=1$ and $\nu^{-1} = 6/5$), in consistency with our RG results.
Over the range of couplings we have accessed, the critical exponents appear to vary continuously.
A rather unlikely explanation of these results, which we cannot rule out, is that $\lambda$ is exactly marginal, leading to a critical line with varying exponents.
A more generic (and likely) explanation is that the scaling dimension of $\lambda$ is parametrically small, and these results are due to crossover behavior between the U(1) and $\mathbb{Z}_3$ DBG fixed points.
The latter scenario is consistent with the small region of stability found in our RG calculation; this explanation amounts to the claim that the point $\epsilon = \delta = 1$ lies close to the boundary of the region of stability in Figure \ref{fig:eigs} in the exact theory.

\begin{figure}[htb]
    \centering
     \subfigure[]{\label{fig:PDa}\includegraphics[width=0.495\linewidth]{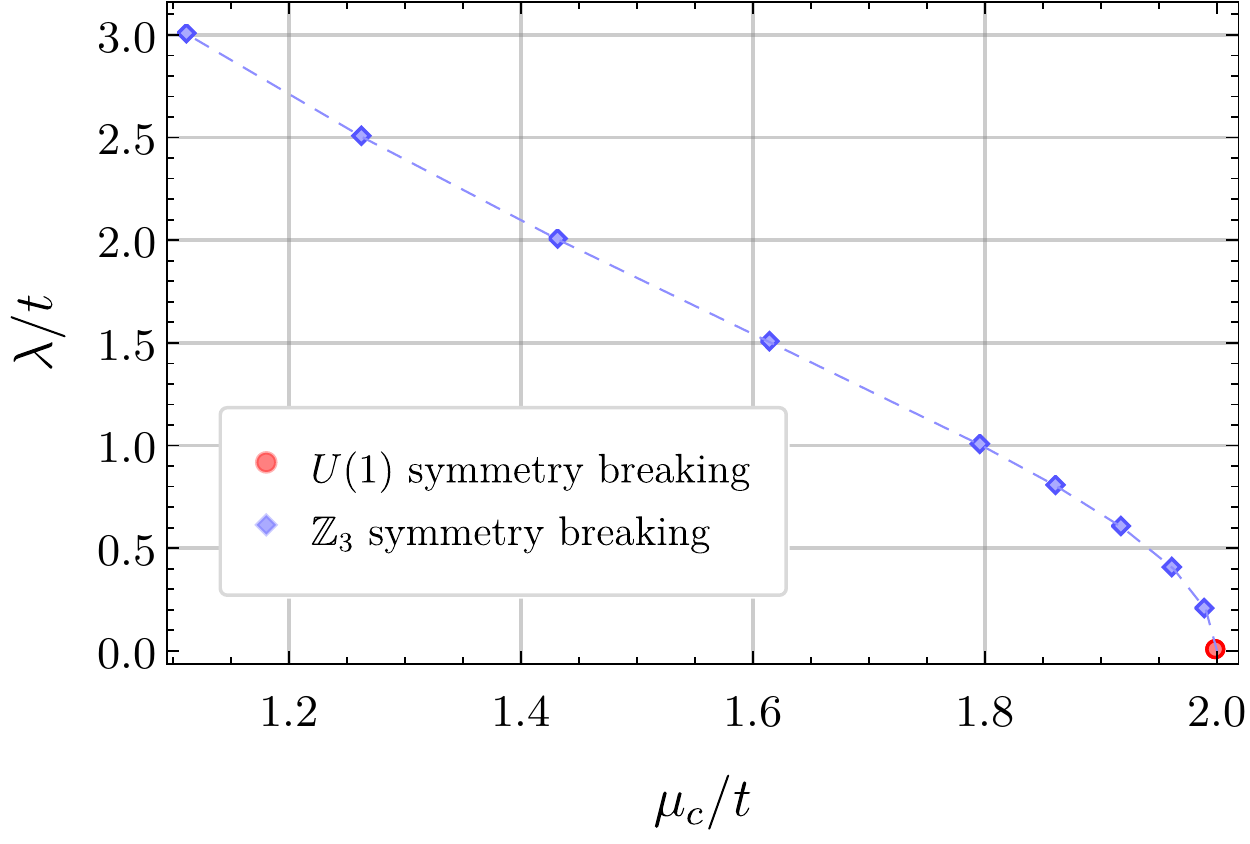}}
     \subfigure[]{\label{fig:4L}\includegraphics[width=0.495\linewidth]{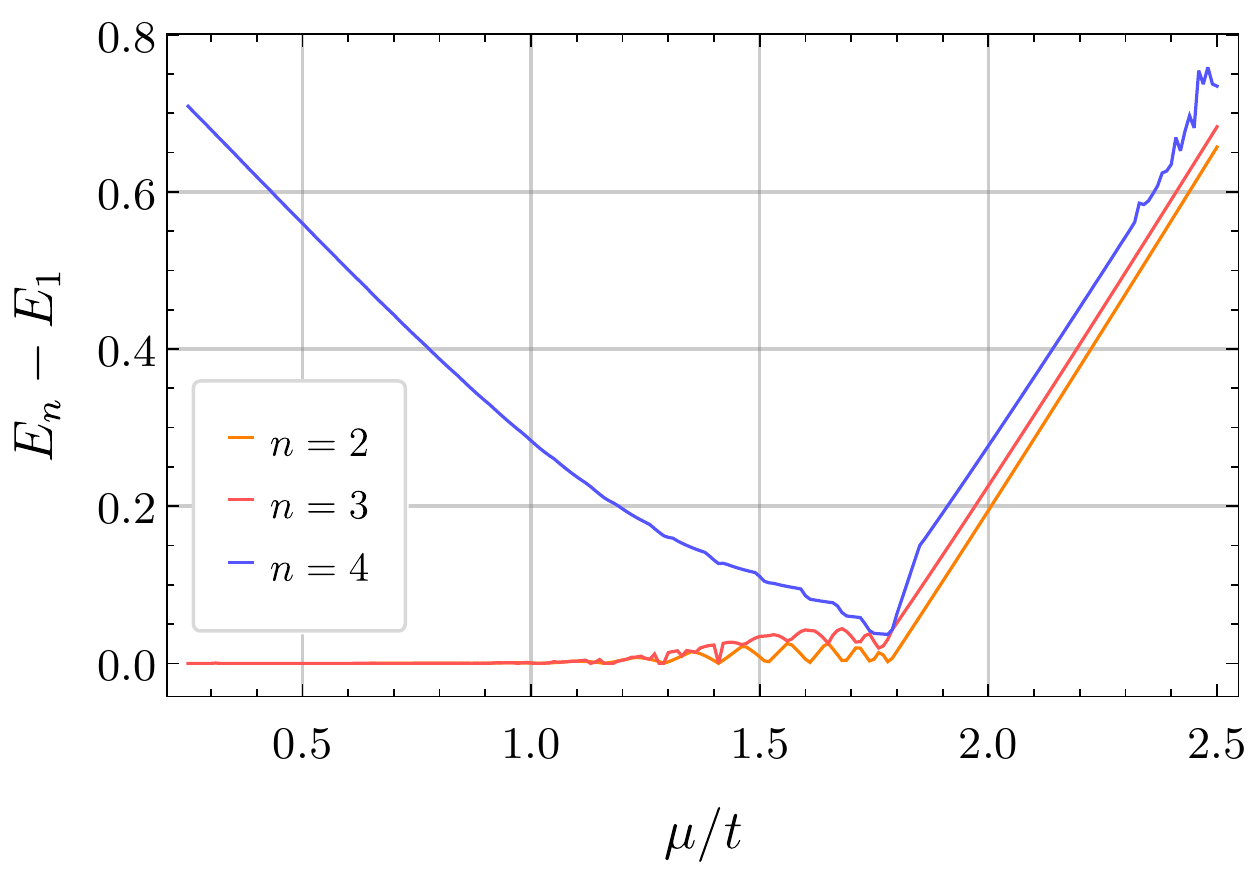}}
    \caption{(a) Schematic phase diagram of the $\mathbb{Z}_3$ dilute Bose gas obtained from DMRG and FSS calculations. For $\lambda = 0$, the QPT involves $U(1)$ symmetry breaking whereas for nonzero $\lambda$, the symmetry broken is $\mathbb{Z}_3$. The phase diagram is symmetric upon reflection about $\mu = 0$. (b) Both phases are gapped. To see this, the difference in energies between the ground state and the first three excited states, are plotted for a system of size $L = 50$ and $\lambda = t = 1$. In the ordered phase, the ground state is threefold degenerate in the infinite-volume limit. However, in a finite chain, there is always a splitting between them and therefore DMRG can probe and distinguish the states within this (ideally) degenerate manifold. Contrarily, on the disordered side, there exists a unique ground state. }
\end{figure}

\section{Conclusions}

The phase transitions of the quantum (classical) $\mathbb{Z}_N$ chiral clock model in one (two) spatial dimensions have been the subject of a number of theoretical and numerical studies \cite{huse1981simple,ostlund1981incommensurate,huse1982domain,huse1983melting,haldane1983phase,howes1983quantum,au1987commuting,cardy1993critical,fendley2012parafermionic,ZCTH,dai2017entanglement,SCPLS18,au1997many,baxter2006challenge,au1987commuting,albertini1989commensurate, mccoy1990excitation}.
The phase structure of this model in the case where $N=3$, and the model has separate time-reversal and spatial inversion symmetry, was debated in the early literature: some studies favored a direct transition between a disordered phase and a gapped phase with broken $\mathbb{Z}_3$ symmetry \cite{ostlund1981incommensurate,huse1983melting}, while others predicted the appearance of a gapless incommensurate phase separating the two gapped ones \cite{haldane1983phase} (we review the arguments of Ref.~\onlinecite{haldane1983phase} in Appendix~\ref{app:luttinger}).
Recent numerical work gives strong evidence for the first scenario \cite{ZCTH,SCPLS18} but a quantum field theory describing the direct transition was lacking. These questions have recently become relevant for experimental
studies of trapped ultracold Rydberg atoms \cite{bernien2017probing}.

In this paper, we have presented a quantum field theory for the critical $\mathbb{Z}_N$ chiral clock model, which is capable of describing the direct transition described above. 
Interestingly, the field theory is written in terms of the `disorder parameter' $\Psi$ (which creates domain walls in the ordered phase) rather than the order parameter of the clock degrees of freedom, $\Phi$. The field theory for $\Psi$ also describes the onset of a condensate in a dilute Bose gas, in the background of a static, higher-dimensional $N$ boson condensate.
We performed a perturbative renormalization group analysis of this field theory for small $\epsilon = 2 - d$ and $\delta =  4 - N$, where we found a weak-coupling fixed point describing a direct transition with $\mathbb{Z}_N$ symmetry breaking.
We give the first analytical predictions for the critical exponents of this transition, obtaining $\nu \approx 0.60$ and $z \approx 1.57$.

We also performed a numerical DMRG study of a lattice boson model the order parameter of which, $\Psi$, is described directly by our field theory. This study contains strong evidence
for a direct transition between a gapped disordered phase and a gapped phase with broken $\mathbb{Z}_3$ symmetry.
The critical exponents obtained numerically do not appear to have fully converged to their universal values in the finite system sizes studied, but their flow is consistent with our field-theoretic results.
We also presented additional DMRG results on the $\mathbb{Z}_3$ clock model, going beyond those in Ref.~\onlinecite{SCPLS18}: these results confirm the duality properties, and yield exponents similar to those for the lattice boson model.

In the future, the field-theoretic advances presented here could be extended to nonequilibrium dynamics, and so could address corresponding experimental studies on Rydberg atoms.

\noindent
{\it Note added:} We learnt of numerical studies \cite{Mila18} on large systems which also find a direct transition without an 
intermediate gapless incommensurate phase.

\section*{Acknowledgements}
We thank Soonwon Choi, Paul Fendley, Mikhail Lukin, Hannes Pichler, and Alex Thomson for valuable discussions. 
This research was supported by the National Science Foundation under Grant No. DMR-1664842. 
Research at Perimeter Institute is supported by the Government of Canada through Industry Canada and by the Province of Ontario through the Ministry of Research and Innovation. SS also acknowledges support from Cenovus Energy at Perimeter Institute. SW acknowledges support from the NIST NRC Postdoctoral Associateship award. The computations in this paper were run on the Odyssey cluster supported by the FAS Division of Science, Research Computing Group at Harvard University. 

\appendix
\section{Derivation of quantum field theory for general $N$}
\label{app:qcderivation}

In Section \ref{sec:qcderivation}, we gave the transfer matrix calculation of the Euclidean lattice field theory for the $N=3$ chiral clock model, and argued for the form of its continuum limit. 
Here, we will generalize this calculation to arbitrary $N$, and give a direct mapping from the Euclidean lattice field theory to the continuum quantum field theory of a complex order parameter field.

Following the derivation in Section \ref{sec:qcderivation}, the steps until Eq.~\eqref{eq:beforesum} generalize to arbitrary $N$ in an obvious way, and we obtain the expression for the partition function
\bea
\mathcal{Z} &=& \frac{1}{N^{M_{\tau}}}\sum_{\{ n_j(\ell) \}}  \exp\left( 2 a \beta J \sum_{\ell = 1}^{M_{\tau}} \sum_{j=1}^{M} \cos\left[ \frac{2 \pi}{N}\Big(n_j(\ell) - n_{j+1}(\ell)\Big) + \theta \right] \right) \nn
&& \ \times \prod_{\ell = 1}^{M_{\tau}} \prod_{j=1}^{M} \sum_{\omega = 0}^{N - 1} \exp\left( 2 a \beta f \cos\left[ \frac{2 \pi}{N} \omega + \phi \right] \right) \nn
&& \ \times \exp\left( - \frac{2 \pi i \omega}{N} \Big(n_j(\ell) - n_j(\ell + 1)\Big) \right).
\label{eq:beforesuma}
\eea
In this expression, the integers $n_j(\ell)$ are defined modulo $N$, and live on a rectangular lattice parametrized by $j = 1, 2, ..., M$ and $\ell = 1, 2, ..., M_{\tau}$. 

We now consider the sum over $\omega$, which has the form
\beq
S_N(\Delta n) = \sum_{\omega = 0}^{N-1} \exp\left( 2 a \beta f \cos\left[ \frac{2 \pi}{N} \omega + \phi \right] - \frac{2 \pi i \Delta n}{N} \omega \right).
\label{eq:sndef}
\eeq
We wish to write this in the form $S_N(\Delta n) \sim e^{f(\Delta n)}$ so that we may combine it with the other exponents in Eq.~(\ref{eq:beforesuma}), and we are only interested in the $a \rightarrow 0$ limit. This sum may be written
\bea
S_N(\Delta n) &=& \exp\left( 2 a \beta f \cos\phi \right) + \delta_{N,2\mathbb{Z}} \exp\left( - 2 a \beta f \cos \phi\right) \cos \left( \pi \Delta n \right) \nn[.25cm]
&& + \ \sum_{k=1}^{\left \lfloor{\frac{N-1}{2}}\right \rfloor } \Bigg\{ \alpha_1(k,N) \cos\left( \frac{2 \pi k \Delta n}{N} \right) + i \alpha_2(k,N) \sin\left( \frac{2 \pi k \Delta n}{N} \right) \Bigg\},
\eea
where $\left \lfloor{x}\right \rfloor$ is the floor function, $\delta_{N,2\mathbb{Z}}$ is zero (one) if $N$ is odd (even), and
\bea
\alpha_1(k,N) &=& 2 \exp\left( 2 a \beta f \cos\left[ \frac{2 \pi k}{N} \right] \cos \phi \right) \cosh\left( 2 a \beta f \sin\left[ \frac{2 \pi k}{N} \right] \sin \phi \right), \nn[.25cm]
\alpha_2(k,N) &=& 2 \exp\left( 2 a \beta f \cos\left[ \frac{2 \pi k}{N} \right] \cos \phi \right) \sinh\left( 2 a \beta f \sin\left[ \frac{2 \pi k}{N} \right] \sin \phi \right).
\eea
In the small $a$ limit, $S_N(0)$ approaches a constant while the other values of $S_N(\Delta n)$ vanish linearly with $a$.

The function $S_N(\Delta n)$ satisfies $S_N(\Delta n + N) = S_N(\Delta n)$ and $S_N(-\Delta n) = S_N(\Delta n)^{\ast}$, so we take the ansatz
\beq
S_N(\Delta n) = A \exp\left( \sum_{m=1}^{\mathcal{M}} K^{(m)}_{\tau} \cos\left( \frac{2 \pi m}{N} \Delta n \right) + i \varphi[\Delta n] \right),
\eeq
where $\varphi[-x] = -\varphi[x]$. By taking the logarithm of the magnitude of both sides of this equation, we obtain a linear system
\beq
\log \left|S_N(\Delta n) \right| = \log A + \sum_{m=1}^{\mathcal{M}} K^{(m)}_{\tau} \cos\left( \frac{2 \pi m}{N} \Delta n \right).
\label{eq:linearset}
\eeq
The left-hand side is still a periodic even function of $\Delta n$, so this clearly justifies our ansatz. Now, if $\Delta n$ were a real number, we would need to take $\mathcal{M} = \infty$ to represent the real function on the left-hand side. However, we only need this equality to match for the finite set of values $\Delta n = 0, 1, 2, ..., \left \lfloor{\frac{N}{2}}\right \rfloor$, so we only need $\mathcal{M}$ large enough so that we can solve the linear set of equations represented in Eq.~(\ref{eq:linearset}). Assuming that the linear system is nonsingular (which we have checked for $N=3,4,5$), we can take $\mathcal{M} = \left \lfloor{\frac{N}{2}}\right \rfloor$.
Additionally, because most of the coefficients $\left|S_N(\Delta n) \right|$ vanish linearly at small $a$, the coefficients $K^{(m)}_{\tau}$ will diverge as $\log a$ in the limit $a \rightarrow 0$.

We now consider the complex phase of $S_N(\Delta n)$. For determining this, we turn back to Eq.~(\ref{eq:sndef}) and write
\bea
S_N &=& \sum_{\omega = 0}^{N-1} \exp\left( 2 a \beta f \cos\left[ \frac{2 \pi}{N} \omega + \phi \right] - \frac{2 \pi i \Delta n}{N} \omega \right) \nn[.25cm]
&=& \sum_{k = 0}^{\infty}\sum_{\omega = 0}^{N-1} \frac{\left( 2 a\beta f\right)^k}{k!}  \cos\left[ \frac{2 \pi}{N} \omega + \phi \right]^k e^{- \frac{2 \pi i \Delta n}{N} \omega } \nn[.25cm]
&=& \mathcal{N} e^{i \varphi[\Delta n]}.
\label{eq:phase}
\eea
Since we are taking the $a \rightarrow 0$ limit, we determine $\varphi[\Delta n]$ from the first nonzero term in the power series in $k$. 
This can be obtained using the identity
\beq
\sum_{\omega = 0}^{N-1} e^{- \frac{2 \pi i}{N} \omega \kappa} = \begin{cases} N & \kappa = 0 \ \mathrm{mod} \ N \\ 0 & \mathrm{else} \end{cases},
\eeq
which can be used to simplify the sum:
\bea
&& \sum_{\omega = 0}^{N-1}\cos\left[ \frac{2 \pi}{N} \omega + \phi \right]^k e^{- \frac{2 \pi i \Delta n}{N} \omega } \nn
&=& 2^k \sum_{\omega = 0}^{N-1} \left( e^{2 \pi i \omega/N + i \phi} + e^{-2 \pi i \omega/N - i \phi} \right)^k e^{- \frac{2 \pi i \Delta n}{N} \omega }.
\eea
Using the symmetry properties of $\varphi[\Delta n]$, it suffices to consider $0 < \Delta n < N/2$. Then for a given $\Delta n$, we need to go to the $k = \Delta n$ term in this sum before we get a nonzero expression after summing over $\omega$, which is given by expanding the above binomial and evaluating the term
\beq
2^{k} \sum_{\omega = 0}^{N-1} e^{2 \pi i k \omega/N + i k \phi} e^{- \frac{2 \pi i \Delta n}{N} \omega } = N 2^{\Delta n} e^{i \phi \Delta n}.
\eeq
Comparing with Eq.~(\ref{eq:phase}), we find the phase in the small $a$ limit to be
\beq
\varphi[\Delta n] = \phi\Delta n, \quad 0 \leq \Delta n < N/2.
\label{eq:vpdef}
\eeq
Together with the relations $\varphi[-\Delta n] = -\varphi[\Delta n]$ and $\varphi[\Delta n + N] = \varphi[\Delta n]$, this completely determines the function $\varphi[\Delta n]$. For the cases $N = 3,4$, we can write
\begin{alignat}{2}
\varphi(\Delta n) &= \frac{2 \phi}{\sqrt{3}} \sin\left( \frac{2 \pi}{3} \Delta n \right), \qquad &&(N=3), \nn[.25cm]
\varphi(\Delta n) &= \phi \sin\left( \frac{2 \pi}{4} \Delta n \right), \qquad \qquad &&(N=4).
\end{alignat}
For larger values of $N$, the periodicity and symmetry of $\varphi[\Delta n]$ will imply that we can write it as a Fourier series, 
\beq
\varphi[\Delta n] = \phi \sum_{m = 1}^{\left \lfloor{\frac{N-1}{2}}\right \rfloor} c_m \sin\left( \frac{2 \pi m}{N} \Delta n \right),
\eeq
where the numerical coefficients $c_m$ can be calculated directly from matching this expression to Eq.~(\ref{eq:vpdef}). 
For $N = 3$ and $4$, we have $c_1 = \frac{2}{\sqrt{3}}$ and $c_1 = 1$ respectively, and all other coefficients vanish.

Combining these results, we may now write the partition function in Eq.~(\ref{eq:beforesuma}) as
\beq
\mathcal{Z} = C \sum_{\{ n_{x,\tau} \}} e^{-\mathcal{S}[n_{x,\tau}]},
\eeq
with the Euclidean action
\bea
- \mathcal{S}[n_{x,\tau}] &=& K_x \sum_{x,\tau} \cos\left[ \frac{2 \pi}{N}\left( n_{x,\tau} - n_{x+1,\tau} \right) + \theta \right] + \sum_{x,\tau} \sum_{m=0}^{\left \lfloor{\frac{N}{2}}\right \rfloor} K^{(m)}_{\tau} \cos\left[ \frac{2 \pi m}{N}\left( n_{x,\tau} - n_{x,\tau+1} \right) \right] \nn
&& + \ i \phi \sum_{x,\tau} \sum_{m = 1}^{\left \lfloor{\frac{N-1}{2}}\right \rfloor} c_m \sin\left[ \frac{2 \pi m}{N} \left( n_{x,\tau} - n_{x,\tau+1} \right) \right].
\label{classicalH}
\eea
Here, the quantum model is obtained in the extreme anisotropic limit $K_x \rightarrow 0$, $K^{(m)}_{\tau} \rightarrow \infty$. 
In particular, if the divergence of the coefficients is $K^{(m)}_{\tau} \rightarrow \eta_m \log a$ at small $a$, we take the limits such that the combinations $K_{x} e^{- K^{(m)}_{\tau}/\eta_m}$ are finite at $a = 0$, and the couplings $J$ and $f$ are tuned to the phase transition. 

Due to universality, we do not expect the details of the nearest-neighbor interactions to change the critical properties of the phase transition provided the interactions have the same symmetry properties and remain short-ranged. 
Thus, for all values of $N$, it should be valid to truncate the interactions proportional to $K_{\tau}$ to a single cosine potential:
\bea
- \mathcal{S}[n_{x,\tau}] &=& K_x \sum_{x,\tau} \cos\left[ \frac{2 \pi}{N}\left( n_{x,\tau} - n_{x+1,\tau} \right) + \theta \right] + K_{\tau} \sum_{x,\tau} \cos\left[ \frac{2 \pi}{N}\left( n_{x,\tau} - n_{x,\tau+1} \right) \right] \nn
&& + \ i \phi \sum_{x,\tau} \sum_{m = 1}^{\left \lfloor{\frac{N-1}{2}}\right \rfloor} c_m \sin\left[ \frac{2 \pi m}{N} \left( n_{x,\tau} - n_{x,\tau+1} \right) \right].
\label{classicalH2}
\eea
At this point, the couplings $K_x$ and $K_{\tau}$ are chosen as tuning parameters instead of $J$ and $f$. 
We also assume that the critical region of the phase diagram extend away from the extreme anisotropic region, so we may take $K_x/K_{\tau} \sim O(1)$.
Since we expect the model to still contain the same phase structure, this is justified provided the transition remains continuous.

We now turn this into a continuum field theory for a complex scalar. 
We first simplify our action (\ref{classicalH2}) by taking $c_{m>1} = 0$, so that we only keep one of the sine terms. 
For $N>4$ this will ruin the periodicity of the model under $\phi \rightarrow \phi + 2 \pi$, while for the important cases $N=3,4$ it is exact; in all cases, the periodicity of $\phi$ will be obscured by our final expressions anyway. 
We rewrite the fields as a unit vector, $\mathbf{v}_{x,\tau} = \left( \cos \left( \frac{2 \pi }{N}n_{x,\tau} \right) ,\sin \left( \frac{2 \pi }{N}n_{x,\tau} \right) \right)$, 
\bea
- \mathcal{S} &=& \sum_{\mathbf{r},\mathbf{r}'} \mathcal{K}_{\mathbf{r},\mathbf{r}'}^{a b} v_{\mathbf{r}}^a v_{\mathbf{r}'}^b \nn
\mathcal{K}_{\mathbf{r},\mathbf{r}'}^{a b} &=& K_x \delta_{\mathbf{r},\mathbf{r}+\hat{x}}\left( \cos \theta \delta^{ab} + \sin \theta \epsilon^{ab} \right) + K_{\tau} \delta_{\mathbf{r},\mathbf{r}+\hat{\tau}} \delta^{ab} - i \, \phi \,  c_1 \delta_{\mathbf{r},\mathbf{r}+\hat{\tau}} \epsilon^{ab}.
\eea
Then, the partition function can be written as
\beq
\mathcal{Z} = \exp\left( \sum_{\mathbf{r},\mathbf{r}'} \mathcal{K}_{\mathbf{r},\mathbf{r}'}^{a b} \frac{\partial^2}{\partial X^a_{\mathbf{r}} \partial X^b_{\mathbf{r}'}} \right) \prod_{\mathbf{r}} \frac{1}{N} \sum_{ n_{\mathbf{r}} = 0}^{N-1} \exp\left( X^1_{\mathbf{r}} \cos\left[ \frac{2 \pi }{N}n_{\mathbf{r}} \right]  + X^2_{\mathbf{r}} \sin\left[ \frac{2 \pi }{N}n_{\mathbf{r}} \right] \right) \Bigg|_{X_{\mathbf{r}} = 0},
\eeq
where we have introduced the auxiliary real fields $X_{\mathbf{r}}^{1,2}$ at each lattice site. 
Subsequently, we work with the complex field $\Phi_{\mathbf{r}} = X^1_{\mathbf{r}} + i X^2_{\mathbf{r}}$. We define the quantities
\bea
G_{\mathbf{r},\mathbf{r}'} &=& K_x \left( \delta_{\mathbf{r},\mathbf{r}'+\hat{x}} + \delta_{\mathbf{r},\mathbf{r}'-\hat{x}} \right) \cos \theta + K_{\tau} \left( \delta_{\mathbf{r},\mathbf{r}+\hat{\tau}} + \delta_{\mathbf{r},\mathbf{r}-\hat{\tau}} \right) \nn
&& + \ i K_x \left( \delta_{\mathbf{r},\mathbf{r}'+\hat{x}} - \delta_{\mathbf{r},\mathbf{r}'-\hat{x}} \right) \sin \theta + \phi \, c_1 \left( \delta_{\mathbf{r},\mathbf{r}'+\hat{\tau}} - \delta_{\mathbf{r},\mathbf{r}'-\hat{\tau}} \right), \nonumber \\[.5cm]
V\left( \Phi,\Phi^{\ast} \right) &=& -\log\left[ \frac{1}{N} \sum_{n = 0}^{N-1} \exp\left(\frac{\Phi}{2} \omega^{n} + \frac{\Phi^{\ast}}{2} \omega^{\ast n} \right) \right],
\eea
where $\omega = e^{2 \pi i /N}$. Then, performing the sum over $n_{\mathbf{r}}$ leads to the partition function
\beq
\mathcal{Z} = \exp\left( 2 \sum_{\mathbf{r},\mathbf{r}'} G_{\mathbf{r},\mathbf{r}'} \frac{\partial^2}{\partial \Phi^{\ast}_{\mathbf{r}} \partial \Phi_{\mathbf{r}'}}  \right) \exp\left( -\sum_{\mathbf{r}} V\left( \Phi_{\mathbf{r}},\Phi^{\ast}_{\mathbf{r}} \right) \right) \Bigg|_{\Phi_{\mathbf{r}},\Phi^{\ast}_{\mathbf{r}} = 0}.
\eeq
At this point, it is noticed that this generates identical diagrams to the theory with interaction potential $V\left( \Phi,\Phi^{\ast} \right)$ and Green's function $G_{\mathbf{r},\mathbf{r}'}$ \cite{poly69}. That is, the partition function is identical to that obtained using the following action for the field $\Phi$:
\beq
\mathcal{S}_{\Phi} = \sum_{\mathbf{r},\mathbf{r}'} \Phi^{\ast}_{\mathbf{r}} \left( G_{\mathbf{r},\mathbf{r}'} \right)^{-1} \Phi_{\mathbf{r}} + \sum_{\mathbf{r}} V\left( \Phi_{\mathbf{r}},\Phi^{\ast}_{\mathbf{r}} \right).
\eeq
The original $\mathbb{Z}_N$ symmetry is still present, taking the form $\Phi \rightarrow e^{2 \pi i/N} \Phi$. 
We will later expand the potential for small $\Phi$, $\Phi^{\dagger}$, and for now it is useful to pull out the quadratic piece, $U\left( \Phi_{\mathbf{r}},\Phi^{\ast}_{\mathbf{r}} \right) \equiv V\left( \Phi_{\mathbf{r}},\Phi^{\ast}_{\mathbf{r}} \right) + \frac{1}{4}|\Phi |^2$, so the potential $U$ only contains nonlinearities in $\Phi$.
We then have
\bea
\mathcal{S}_{\Phi} &=& \sum_{\mathbf{r},\mathbf{r}'} \psi^{\ast}_{\mathbf{r}} \left[ \left( G_{\mathbf{r},\mathbf{r}'} \right)^{-1}  - \frac{1}{4}\delta_{\mathbf{r},\mathbf{r}'} \right] \Phi_{\mathbf{r}} + \sum_{\mathbf{r}} U\left( \Phi_{\mathbf{r}},\Phi^{\ast}_{\mathbf{r}} \right) \nn
&=& \int_{\mathrm{BZ}} \frac{d^2 k}{(2 \pi)^2} \left( \frac{4 - G(\mathbf{k})}{4G(\mathbf{k})} \right) \left| \Phi(\mathbf{k}) \right|^2 + \sum_{\mathbf{r}} U\left( \Phi_{\mathbf{r}},\Phi^{\ast}_{\mathbf{r}} \right),
\eea
where
\beq
G(\mathbf{k}) = 2 K_x \left[ \cos \theta \cos k_x - \sin \theta \sin k_x \right] + 2 K_{\tau} \cos k_{\tau} + 2 i \phi \, c_1 \sin k_{\tau}.
\eeq
This is a formally exact representation of the lattice field theory in Eq.~(\ref{classicalH2}) except for having taken $c_{m>1} = 0$ for $N>4$ (and as promised, the periodicity of our theory under $\phi \rightarrow \pi + 2 \pi$ is opaque even in the cases $N = 3,4$ where we did not make an approximation). 

We now consider the critical regime, where we may expand near $\mathbf{k} \rightarrow 0$.
For the cases of interest, we have
\beq
\left( \frac{4 - G(\mathbf{k})}{4G(\mathbf{k})} \right) = r + \alpha_x k_x + \kappa_{x} k_x^2 + \kappa_{\tau} k_{\tau}^2 \cdots \qquad (\phi = 0,\theta \neq 0),
\eeq
\beq
\left( \frac{4 - G(\mathbf{k})}{4G(\mathbf{k})} \right) = r' + i \alpha'_{\tau} k_{\tau} + \kappa'_{x} k_x^2 + \kappa'_{\tau} k_{\tau}^2 \cdots \qquad (\phi \neq 0,\theta = 0),
\eeq
where the coefficients are complicated but generically nonzero real functions of $K_x$, $K_{\tau}$, and either $\theta$ or $\tau$.
We may also formally write the potential as its power series in $\Phi$ and $\Phi^{\ast}$,
\beq
U\left( \Phi_{\mathbf{r}},\Phi^{\ast}_{\mathbf{r}} \right) = \lambda \left( \Phi^3 + \Phi^{\ast 3} \right) + u |\Phi|^4 + \cdots.
\eeq
Going back to position space, $k_{x,\tau} \rightarrow i \partial_{x,\tau}$, and in the continuum limit, these become the theories claimed in Eqs.~(\ref{SPhi}) and (\ref{SPsi})

\section{$M$-loop integrals}
\label{app:mloop}

In the text we need to compute certain $M$-loop integrals where $M$ needs to be analytically continued to an arbitrary complex number. 
We show how to perform these integrals in this Appendix. 

Before computing specific integrals, we outline the main steps. 
We first perform all frequency integrals; due to the structure of the propagator, this will combine most of the denominators. 
We then combine any remaining denominators, usually by using Feynman parameters.
This will leave us with $M$ integrals over internal momenta of the form
\beq
I = \int \left( \prod_{i = 1}^{M} \frac{d^d k_i}{\left( 2 \pi \right)^{d}} \right) \frac{1}{f(k_1,k_2,...,k_M)^{\gamma}},
\label{eq:standardI}
\eeq
where the function $f$ is at most quadratic in the $k_i$, and there may be additional integrations over Feynman parameters.

We next repeatedly complete the square for all the $k_i$. 
Focusing on a specific $k$, the most general form for $f(k)$ we encounter is
\beq
f(k) = F + \eta k^2 + \alpha \left( k + K_1 \right)^2 + \beta \left( k + K_2 \right)^2
\eeq
for some constants (which may depend on the other momenta) $F$, $K_1$, $K_2$, $\eta$, $\alpha$, $\delta$. After writing
\bea
f(k) &=& F + \left(\eta + \alpha + \beta\right) \left( k + \frac{\alpha K_1 + \beta K_2}{\eta+\alpha + \beta }\right)^2 \nn
&& + \ \frac{\eta}{\eta + \alpha + \beta}(\alpha K_1^2 + \beta K_2^2) + \frac{\alpha \beta}{\eta + \alpha + \beta}(K_1 - K_2)^2,
\label{eq:shift2}
\eea
we may shift $k \rightarrow k  - (\alpha K_1 + \beta K_2)/(\eta+\alpha + \beta)$ and scale $k \rightarrow k/\sqrt{\eta + \alpha + \beta}$. 
Performing this procedure for all $k_i$, our integral eventually takes the form
\beq
I = J \int \left( \prod_{i = 1}^{M} \frac{d^d k_i}{\left( 2 \pi \right)^{d}} \right) \frac{1}{\left( k_1^2 + k_2^2 + \cdots + k_M^2 + \Delta^2  \right)^{\gamma}}
\eeq
for some Jacobian $J$. 
Applying the identity
\beq
\int \frac{d^d k}{\left( 2 \pi \right)^d} \frac{1}{\left( k^2 + m^2 \right)^\gamma} = \frac{\Gamma\left(\gamma - \frac{d}{2} \right) }{\Gamma\left( \gamma \right)} \frac{S_d}{\left(m^2 \right)^{\gamma - d/2}} 
\eeq
iteratively, we have
\beq
I = J S_d^M \frac{\Gamma\left( \gamma - M d/2 \right)}{\Gamma\left(\gamma \right)} \left( \Delta^2 \right)^{Md/2 - \gamma}.
\label{eq:iterate}
\eeq
In this expression, the dependence on $M$ may be analytically continued to arbitrary values.

\subsection{$I_1^{(M)}$}

\begin{figure}
\includegraphics[width=8cm]{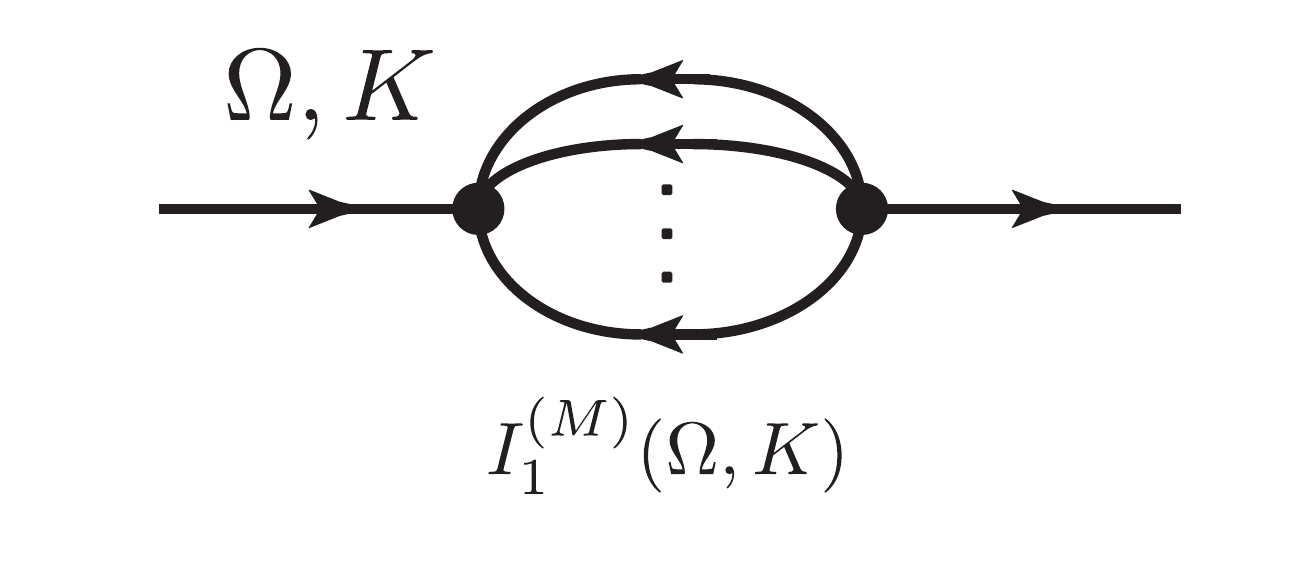}
\caption{Definition of the integral $I_1^{(M)}(\Omega,K)$.}
\label{fig:int1}
\end{figure}
The first integral we encounter is given by the diagram in Figure \ref{fig:int1}. Explicitly, this is
\beq
I_1^{(M)}(\Omega,K) \equiv \int \left( \prod_{i = 1}^{M} \frac{d \omega_i d^d k_i}{\left( 2 \pi \right)^{d+1}} \right)  \frac{1}{\left(- i \omega_1 + k_1^2\right) \cdots \left(- i \omega_{M} + k^2_{M}\right) \left( i \left(\Omega + \sum_i^{M} \omega_i \right) + \left(K + \sum_i^{M} k_i \right)^2 \right)}. \nonumber
\eeq
After performing each $\omega$ integral, all of the denominators combine:
\beq
I_1^{(M)}(\Omega,K) = \int \left( \prod_{i = 1}^{M} \frac{d^d k_i}{\left( 2 \pi \right)^{d}} \right)  \frac{1}{\left( i \Omega + \sum_i^{M} k_i^2 + (K + \sum_i^{M} k_i)^2 \right)},
\eeq
which is of the form Eq~\eqref{eq:standardI}. 
To demonstrate how the steps outlined above look in this simple case, we now apply Eq.~(\ref{eq:shift2}) iteratively:
\bea
&& \int \left( \prod_{i}^{M}\frac{d^d k_i}{(2 \pi)^d}\right) \frac{1}{i \Omega + k_1^2 + k_2^2 + \cdots + k_M^2 + (k_1 + k_2 + \cdots + k_M + K)^2} \nn
&& = \ \int \left( \prod_{i}^{M}\frac{d^d k_i}{(2 \pi)^d}\right) \frac{1}{i \Omega + 2 k_1^2 + k_2^2 + \cdots + k_M^2 + \frac{1}{2} (k_2 + k_3 + \cdots + k_M + K)^2} \nn
&& = \ \int \left( \prod_{i}^{M}\frac{d^d k_i}{(2 \pi)^d}\right) \frac{1}{i \Omega + 2 k_1^2 + \frac{3}{2} k_2^2 + \cdots + k_M^2 + \frac{1}{3} (k_3 + k_4 + \cdots + k_M + K)^2} \nn[.25cm]
&& = \ \cdots \nn[.25cm]
&& = \ \int \left( \prod_{i}^{M}\frac{d^d k_i}{(2 \pi)^d}\right) \frac{1}{i \Omega + 2 k_1^2 + \frac{3}{2} k_2^2 + \cdots + \frac{M+1}{M}k_M^2 + \frac{1}{M+1}K^2} \nn
&& = \ \frac{1}{\left( M + 1 \right)^{d/2}} \int \left( \prod_{i}^{M}\frac{d^d k_i}{(2 \pi)^d}\right) \frac{1}{i \Omega + \frac{1}{M+1} K^2 + \sum_i^{M} k_i^2}.
\label{jacobian}
\eea
Here, we have shifted the integration in each step at will. 
In the last equality, we rescaled the momenta giving the Jacobian $\left[ \left(\frac{1}{2} \right) \left(\frac{2}{3} \right) \cdots \left(\frac{M}{M+1} \right) \right]^{d/2} = \left( M + 1 \right)^{-d/2}$. 
We may now use Eq.~(\ref{eq:iterate}), obtaining the final result
\beq
I_1^{(M)}(\Omega,K)  = S_d^M \frac{\Gamma\left( 1 - M d/2 \right)}{\left( M + 1 \right)^{d/2}} \left(  i \Omega + \frac{1}{M+1}K^2 \right)^{Md/2 - 1}.
\eeq

\subsection{$I_2^{(M)}$}

\begin{figure}
\includegraphics[width=8cm]{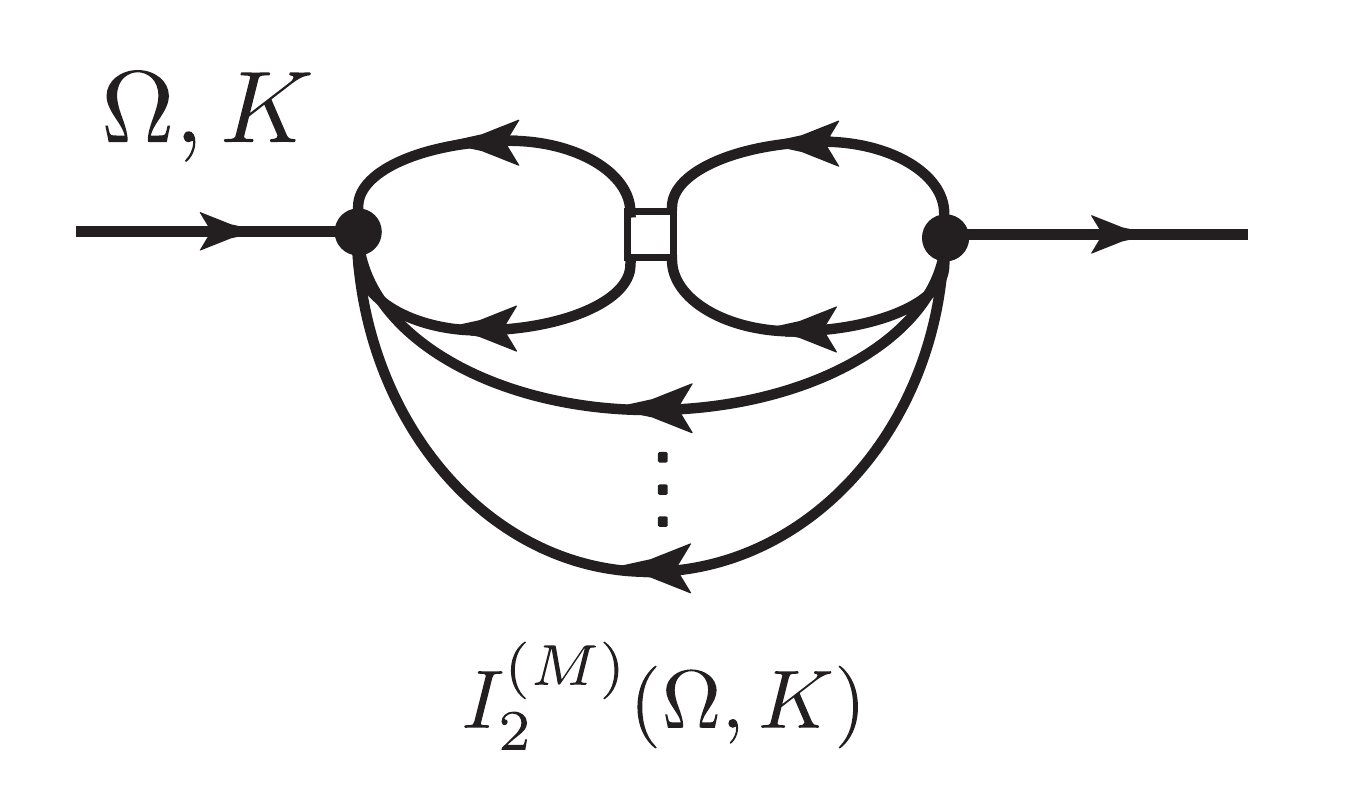}
\caption{Definition of the integral $I_2^{(M)}(\Omega,K)$.}
\label{fig:int2}
\end{figure}
The second integral appears in the diagram in Figure \ref{fig:int2}, which is given by
\bea
I_2^{(M)}(\Omega,K) &\equiv& \int \left( \prod_{i = 1}^{M} \frac{d \omega_i d^d k_i}{\left( 2 \pi \right)^{d+1}} \right)  \frac{1}{\left(- i \omega_1 + k_1^2\right) \cdots \left(- i \omega_{M} + k^2_{M}\right)\left[ i \left(\Omega + \sum_{i=2}^{M} \omega_i \right) + \left(K +\sum_{i=2}^{M} k_i \right)^2 \right]} \nn[.25cm]
&& \bigtimes \ \frac{1}{\left[ i \left(\Omega + \omega_1 + \sum_{i=3}^{M} \omega_i \right) + \left(K + k_1 +\sum_{i=3}^{M} k_i \right)^2 \right]}.
\eea
After integrating over all of the frequencies, this becomes
\bea
I_2^{(M)}(\Omega,K) &\equiv& \int \left( \prod_{i = 1}^{M} \frac{ d^d k_i}{\left( 2 \pi \right)^{d}} \right)  \frac{1}{\left[ i \Omega + \sum_{i=2}^{M} k_i^2 + \left(K + \sum_{i=2}^{M} k_i \right)^2 \right]} \nn[.25cm]
&& \bigtimes \ \frac{1}{\left[ i \Omega + k_1^2 + \sum_{i=3}^{M} k_i^2 + \left(K + k_1 + \sum_{i=3}^{M} k_i \right)^2 \right]}.
\eea
We now integrate over the special momenta $k_1$ and $k_2$, which combines the two denominators:
\bea
I_2^{(M)}(\Omega,K) &=& \frac{S_d^2 \ \Gamma\left(1 - \frac{d}{2}\right)^2}{2^{d}} \int \left( \prod_{i = 3}^{M} \frac{ d^d k_i}{\left( 2 \pi \right)^{d}} \right)  \frac{1}{\left[ i \Omega + \sum_{i=3}^{M} k_i^2 + \frac{1}{2} \left(K + \sum_{i=3}^{M} k_i \right)^2 \right]^{2 - d} }.\quad
\eea
The remaining integral is of the form given in Eq.~(\ref{eq:standardI}), so we can perform the steps as above, obtaining
\beq
I_2^{(M)}(\Omega,K) = \frac{S_d^M}{\left( 2 M \right)^{d/2}} \frac{\Gamma\left(1 - \frac{d}{2}\right)^2\Gamma\left( 2 - Md/2 \right)}{\Gamma\left( 2 - d \right)} \left[ i \Omega + \frac{1}{M} K^2 \right]^{Md/2 - 2}.
\eeq
We can check that $I_2^{(2)}(\Omega,K) = I_1^{(1)}(-\Omega,-K)^2$, as can be seen directly from the diagrams.

\subsection{$I_3^{(M)}$}

\begin{figure}
\includegraphics[width=8cm]{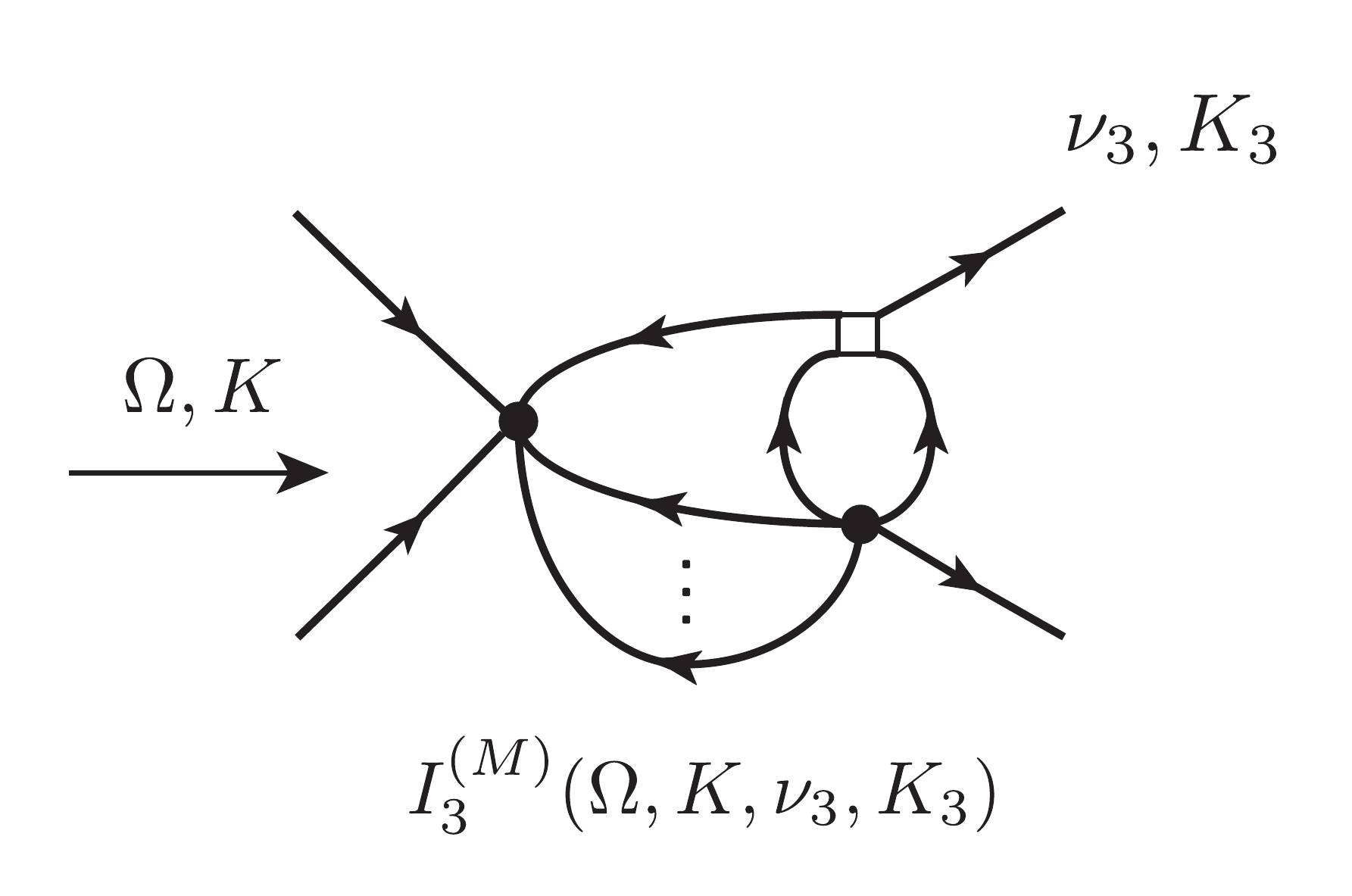}
\caption{Definition of the integral $I_3^{(M)}(\Omega,K,\nu_3,K_3)$. Here, $\Omega$ and $K$ are the total frequency and momentum flowing into the diagram from the left.}
\label{fig:int3}
\end{figure}
The third integral, which is shown in Figure \ref{fig:int3}, is
\bea
I_3^{(M)} &\equiv& \int \left( \prod_{i = 1}^{M} \frac{d \omega_i d^d k_i}{\left( 2 \pi \right)^{d+1}} \right) \frac{1}{\left(- i \omega_1 + k_1^2\right) \cdots \left(- i \omega_{M} + k^2_{M}\right)}\nonumber \\[.5cm]
&& \bigtimes \ \frac{1}{\left[ i \left(\Omega + \sum_{i}^{M-1} \omega_i \right) + \left(K + \sum_{i}^{M-1} k_i \right)^2 \right] \left[ i \left(\nu_3 + \sum_i^{M} \omega_i \right) + \left(K_3 + \sum_i^{M} k_i \right)^2 \right]}. \qquad \qquad
\eea
We first integrate over the frequencies,
\bea
I_3^{(M)} &=& \int \left( \prod_{i = 1}^{M} \frac{d^d k_i}{\left( 2 \pi \right)^{d}} \right) \frac{1}{\left[ i \Omega + \sum_{i}^{M-1} k_i^2 + \left(K + \sum_{i}^{M-1} k_i \right)^2 \right] \left[ i \nu_3 + \sum_i^{M} k_i^2 + \left(K_3 + \sum_i^{M} k_i \right)^2 \right]}, \qquad \qquad
\eea
and then combine the denominators using Feynman parametrization,
\bea
I_3^{(M)} &=& \int_0^1 du \int \left( \prod_{i = 1}^{M} \frac{d^d k_i}{\left( 2 \pi \right)^{d}} \right) \Bigg[ ui \Omega + (1-u)i\nu_3 + \sum_{i}^{M-1} k_i^2 + (1-u) k_M^2 \nn[.5cm]
&& \qquad \qquad + \ u\left(K + \sum_{i}^{M-1} k_i \right)^2 + (1-u) \left(p_3 + \sum_i^M k_i\right)^2\Bigg]^{-2}. 
\eea
We perform the momentum integral using Eq.~(\ref{eq:standardI}), obtaining
\bea
I_3^{(M)} &=& S_d^M \Gamma\left( 2 - Md/2\right) \int_0^1 du (1-u)^{-d/2} \left[ (M+1) + u (M - 1) \right]^{-d/2} \nn[.5cm]
&& \qquad \qquad \qquad \bigtimes \ \Bigg[ u i \Omega + (1-u) i \nu_3 + \frac{u \left[ (M + 1) - u (M - 1)\right]}{(M + 1) + u (M-1)} K^2 \nn[.5cm]
&& \qquad \qquad \qquad + \ \frac{(1-u)\left[ 1 + u (M-1)\right]}{(M + 1) + u (M-1)} K_3^2 - \frac{2 u (1-u) (M-1)}{(M + 1) + u (M-1)} K \cdot K_3 \Bigg]^{Md/2 - 2}.
\eea
We have obtained one pole from the Gamma function sitting out front, but another pole (which we will see is $1/\epsilon$) occurs due to the region near $u \rightarrow 1$ in the integral; in particular, due to the factor of $(1 - u)^{-d/2}$.

To extract the divergent behavior, let us temporarily assume $M = 2 - \delta$, a case which is used in the main text. We write the term in the brackets as
\bea
\bigg[ \bigg]^{Md/2 - 2} &=& \left[ i \Omega + \frac{1}{M} K^2 \right]^{Md/2 - 2} + \left\{ \bigg[ \bigg]^{Md/2 - 2} - \left[ i \Omega + \frac{1}{M} K^2 \right]^{Md/2 - 2} \right\} \nn[.5cm]
&=& \left[ i \Omega + \frac{1}{2-\delta} K^2 \right]^{Md/2 - 2} + \left( \epsilon + \delta - \frac{\epsilon \delta}{2} \right) \log\left[ \frac{\big[ \big]}{i \Omega + \frac{1}{M} K^2} \right] + \cdots,
\eea
where we are exploiting the expansion in $\epsilon = d - 2$ and $\delta = 2 - M$ in the last line. 
Now, the second term has an overall coefficient $\left( \epsilon + \delta - \frac{\epsilon \delta}{2} \right)$ which will cancel the pole in the Gamma function, and near $u \rightarrow 1$, the argument of the logarithm goes to 1, killing the pole in epsilon, so the latter term is finite. 
In the main text, we will also need to evaluate this integral for $M = 3 - \delta$, where identical steps lead to the same cancellation of divergences.

In conclusion, we only need the first term, which only depends on $\Omega$ and $K$. This divergent part of the integral may be written as
\bea
I_3^{(M)} &=& S_d^M \frac{\Gamma\left( 2 - Md/2\right)}{M} \left[ i \Omega + \frac{1}{M} K^2 \right]^{Md/2 - 2} F(d,M),
\eea
with the special function defined as 
\bea
F(d,M) &\equiv& M \int_0^1 du (1-u)^{-d/2} \left[ (M + 1) + u (M - 1) \right]^{-d/2} \nn
&=& \frac{1}{2 - d} + f_{\mathrm{reg}}(d,M).
\eea
Here, $f_{\mathrm{reg}}(d,M)$ is regular for $d \leq 2$, $M > 1$.
Some explicit values we will use are
\bea
f_{\mathrm{reg}}(2,2) &=& \frac{1}{2} \log\left( \frac{16}{3} \right), \nn
f_{\mathrm{reg}}(2,3) &=& \log 3.
\eea

\subsection{$I_4^{(M)}$}

\begin{figure}
\includegraphics[width=8cm]{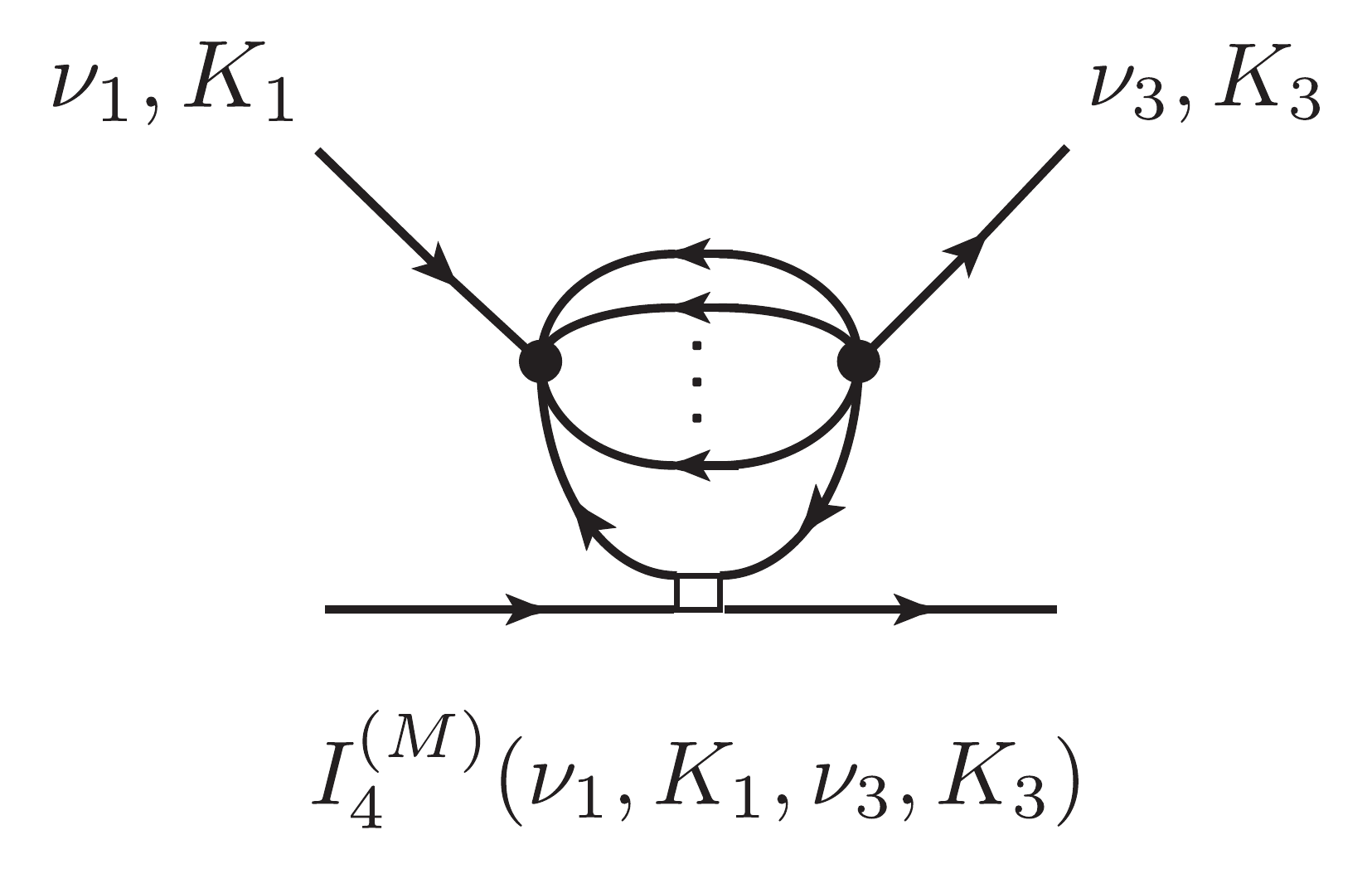}
\caption{Definition of the integral $I_4^{(M)}(\Omega_1,K_1,\nu_3,K_3)$, where $(\Omega_1,K_1)$ enters from the top left and $(\Omega_3,K_3)$ exits from the top right.}
\label{fig:int4}
\end{figure}
This integral, shown in Figure \ref{fig:int4}, is
\bea
I_4^{(M)} = &\equiv& \int \left( \prod_{i = 1}^{M} \frac{d \omega_i d^d k_i}{\left( 2 \pi \right)^{d+1}} \right)  \frac{1}{\left(- i \omega_1 + k_1^2\right) \cdots \left(- i \omega_{M} + k^2_{M}\right)} \nonumber \\[.5cm]
&& \bigtimes \ \frac{1}{\left[ i \left(\nu_1 + \sum_i^{M} \omega_i \right) + \left(K_1 + \sum_i^{M} k_i \right)^2 \right] \left[ i \left(\nu_3 + \sum_i^{M} \omega_i \right) + \left(K_3 + \sum_i^{M} k_i \right)^2 \right]}. \qquad \qquad
\eea
After the frequency integrations, we end up with two denominators.
After combining them with Feynman parameters and performing the momentum integrals, we obtain
\bea
I_4^{(M)}(\Omega,K) &=& \frac{S_d^M}{\left( M+1 \right)^{d/2}} \Gamma\left( 2 - M d/2 \right) \\
&& \ \times\int_0^1 du\left[ u \left( i \nu_1 + \frac{K_1^2}{M+1}  \right) + (1-u) \left( i \nu_3 + \frac{K_3^2}{M+1}  \right) + \frac{Mu (1-u)}{(M+1)}(K_1 - K_3)^2 \right]^{M d/2 - 2}. \nonumber
\eea
We will only need this integral near $M=2$ where the remaining integral is finite, and the divergence is entirely due to the gamma function out front.
Setting $M=2-\delta$ and $d=2-\epsilon$, the divergent part of the integral close to $\epsilon = \delta = 0$ is
\beq
I_4^{(2-\delta)} = \frac{S_d^{2-\delta}}{\left( 3-\delta \right)^{1 - \epsilon/2}} \frac{1}{(\epsilon + \delta - \epsilon \delta/2)}.
\eeq

\section{Renormalization constants for the $\mathbb{Z}_N$ dilute Bose gas}
\label{sec:rgconstants}

In the main text, we outlined our method for computing the renormalization constants for the $\mathbb{Z}_N$ DBG. 
In our renormalization convention, we found it useful to define the auxiliary quantities
\bea
\epsilon_1(\delta) &=& - \frac{2 \delta}{1 - \delta}, \nn
\epsilon_2(\delta) &=& - \frac{\delta}{1 - \delta/2}, \nn
\epsilon_3(\delta) &=& - \frac{2 \delta}{3 - \delta},
\eea
as well as the functions
\bea
\alpha_1(\epsilon,\delta) &=& \frac{4}{\Gamma(3 - \delta) (2 - \delta)^{1 - \epsilon/2}}, \nn
\alpha_2(\epsilon,\delta) &=& \frac{2 \Gamma(\epsilon/2)}{\Gamma(\epsilon) \Gamma(1 - \delta) (4 - 2\delta)^{1 - \epsilon/2}}, \nn
\alpha_3(\epsilon,\delta) &=& \frac{18}{\Gamma(4 - \delta) (3 - \delta)^{1 - \epsilon/2}}, \nn
\alpha_4(\epsilon,\delta) &=& \frac{3 \Gamma(\epsilon/2)^2}{2\Gamma(2 - \delta) (6 - 2 \delta)^{1 - \epsilon/2} \Gamma(\epsilon)^2 }, \nn
\alpha_5(\epsilon,\delta) &=& \frac{6}{\Gamma(3-\delta)(3 - \delta)^{1 - \epsilon/2}}.
\eea
All of these functions have the simple limit $\alpha_i(0,0) = 1$.

In terms of these definitions, the explicit expressions for the renormalization constants are
\bea
Z_g &=& 1 + \frac{g}{\epsilon} + \frac{g^2}{\epsilon^2} + \frac{\alpha_1(\epsilon_1(\delta),\delta) \lambda^2}{8g\left(\delta + \epsilon/2 - \epsilon \delta/2 \right)} - \frac{3 (1 - \delta/4)(1- \delta/3) \alpha_1(\epsilon_1(\delta),\delta) \lambda^2}{2 \delta (\epsilon - \epsilon_1(\delta))} \nn[.25cm]
&& + \ \frac{\alpha_2(\epsilon_2(\delta),\delta) \lambda^2}{4\delta (\epsilon - \epsilon_2(\delta))} + \frac{2 \lambda^2}{\delta (\epsilon - \epsilon_2(\delta)) \Gamma(3 - \delta)} + \frac{\lambda^2 \alpha(\epsilon_1(\delta),\delta)}{4 \epsilon \delta}\nn[.25cm]
&& - \ \frac{\lambda^2}{\delta + \epsilon - \epsilon \delta/2}\Bigg[ \frac{\alpha_2(\epsilon_2,\delta)}{8}\left( \Gamma(\epsilon_2/2) - 2/\epsilon_2\right) + \frac{4}{\Gamma(3 - \delta) (3 - \delta)^{1 - \epsilon_2/2}} \nn[.25cm]
&& \qquad \qquad \qquad \qquad  \qquad \qquad \qquad \qquad + \ \frac{2f_{\mathrm{reg}}(\epsilon_2,2-\delta)}{\Gamma(3 - \delta)} \Bigg]
\eea
\bea
Z_{\lambda} &=& 1 + \frac{6}{\epsilon}g(1 - \delta/4)(1- \delta/3) + \frac{21 g^2}{\epsilon^2}(1 - \delta/4)(1 - \delta/3) (1 - \delta/2 + \delta^2/14) \nn[.25cm]
&& - \ \frac{6 g^2 \log (4/3) }{\epsilon}(1 - \delta/4)(1 - \delta/3)(1 - \delta/2) + \frac{3\lambda^2}{8 \epsilon y_{1}},
\eea
\bea
Z &=& 1 + \frac{\lambda^2 \alpha_3(\epsilon_2,\delta)}{18 y_2} - g \lambda^2 \frac{2 \alpha_3(\epsilon_2,\delta)}{3 \delta (\epsilon - \epsilon_2)} (1 - \delta/4)(1-\delta/3) + g \lambda^2 \frac{\alpha_4(\epsilon_3,\delta)}{3 \delta (\epsilon - \epsilon_3)} \nn[.25cm]
&& + \ g \lambda^2 \frac{\alpha_3(0,\delta) (1 - \delta/3)}{3 \delta (\epsilon - \epsilon_3)} - g \lambda^2 \frac{\alpha_4(\epsilon_3,\delta) [\Gamma_{\mathrm{reg}}(\epsilon_3) - 1/\epsilon_3]}{3 y_3}  \nn
&& - \ g \lambda^2 \frac{\alpha_3(0,\delta) (1 - \delta/3) f_{\mathrm{reg}}(\epsilon_3,3-\delta)}{3 y_3} ,
\eea
\bea
Z_{\tau} &=& 1 - \frac{\lambda^2 \alpha_3(\epsilon_2,\delta)}{54 y_2 (1 - \delta/3)} + g \lambda^2 \frac{2 \alpha_3(\epsilon_2,\delta)}{9 \delta (\epsilon - \epsilon_2)} (1 - \delta/4) - g \lambda^2 \frac{\alpha_4(\epsilon_3,\delta)}{9 \delta (\epsilon - \epsilon_3) (1 - \delta/3)} \nn[.25cm]
&& - \ g \lambda^2 \frac{\alpha_3(0,\delta) }{9 \delta (\epsilon - \epsilon_3)} + g \lambda^2 \frac{\alpha_4(\epsilon_3,\delta) [\Gamma_{\mathrm{reg}}(\epsilon_3) - 1/\epsilon_3]}{9 y_3 (1 - \delta/3)} + g \lambda^2 \frac{\alpha_3(0,\delta) f_{\mathrm{reg}}(\epsilon_3,3-\delta)}{9 y_3} ,
\eea
\beq
Z_2 = 1 - \frac{\lambda^2 \alpha_5(\epsilon_2,\delta)}{6 y_2}.
\eeq
Plugging these definitions into Eqs.~\eqref{eq:g2bare}--\eqref{eq:grenormed}, one can check that the resulting renormalized 1PI vertices are finite for arbitrary external frequency and momentum.
One may also check that the $\delta \rightarrow 0$ limit of these reduce to simple poles in $\epsilon$ with no finite part, which was our defined renormalization scheme.

\section{The self-dual phase boundary in the chiral clock model}
\label{sec:P=T}

A second example of a trivial--topological phase transition can be found in the three-state chiral clock for $f=J$ and $\phi = \theta < \pi/6$. The model is self-dual along this line, which culminates in a tricritical Lifshitz point at $\phi = \theta = \pi/6$. Our first line of investigation is to look at how the gap closes as a function of the detuning from criticality. Figure~\ref{fig:GapP=T} demonstrates that the scaling expected from Eq.~\eqref{eq:def} is reasonably well-satisfied for $\theta \le \pi/12$; however, for larger $\theta$ (and $\phi$), this relation clearly breaks down. The marked distinction between these two regimes can be understood based on the onset of Lifshitz oscillations \cite{rodney2013scaling} as one approaches the tricritical point.

\begin{figure}[htb]
\subfigure[]{\label{fig:GapP=T}\includegraphics[width= 0.485\linewidth]{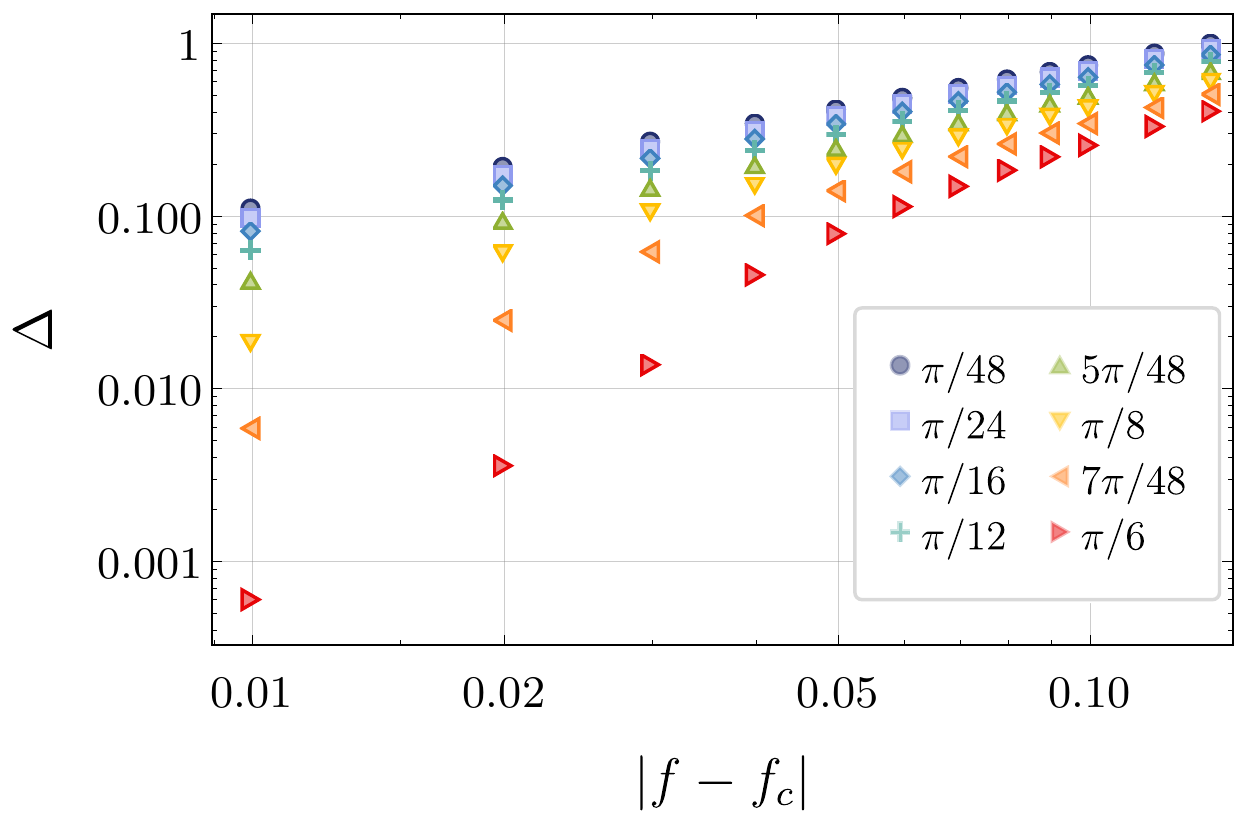}}\,\,\,\,
\subfigure[]{\label{fig:Lifshitz1}\includegraphics[width= 0.485\linewidth]{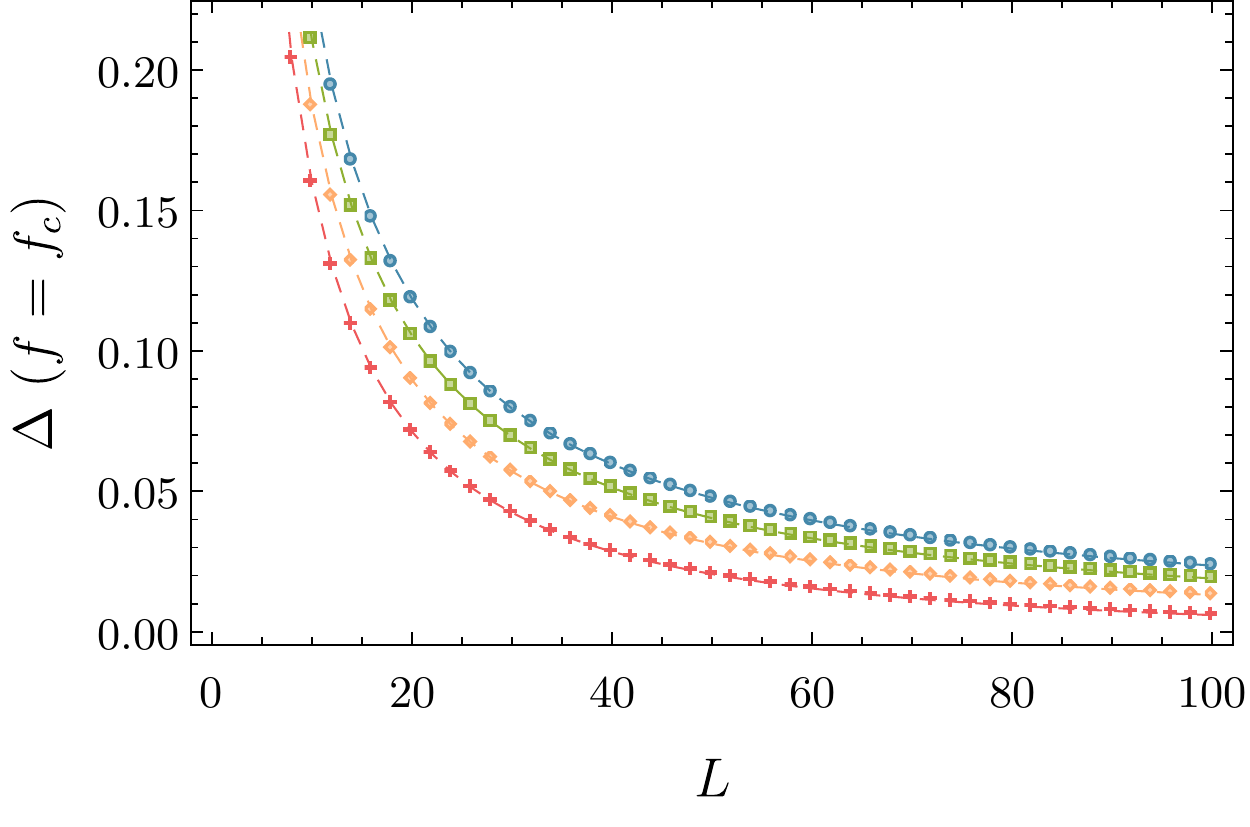}
\begin{picture}(0,0)
\put(-49,21){\includegraphics[height=3cm]{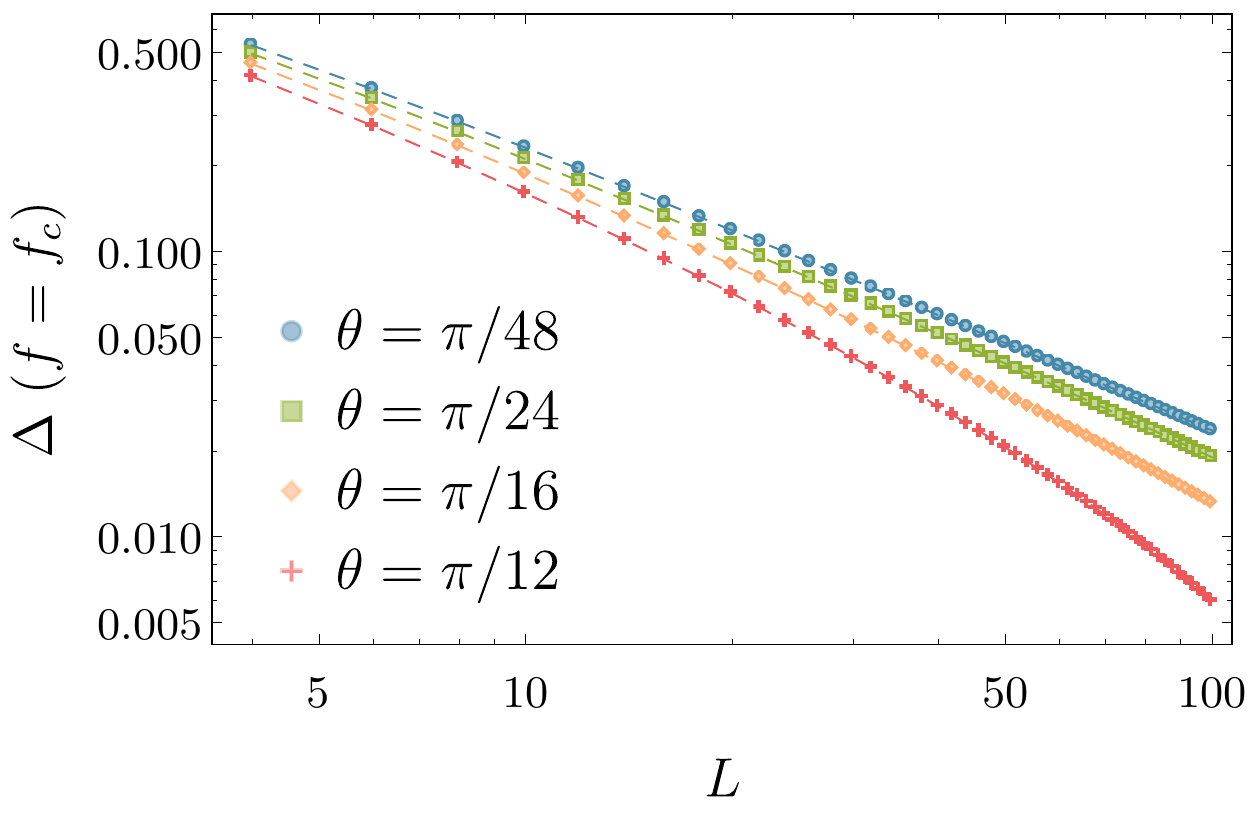}}
\end{picture}}
\subfigure[]{\label{fig:Lifshitz2}\includegraphics[width= 0.485\linewidth]{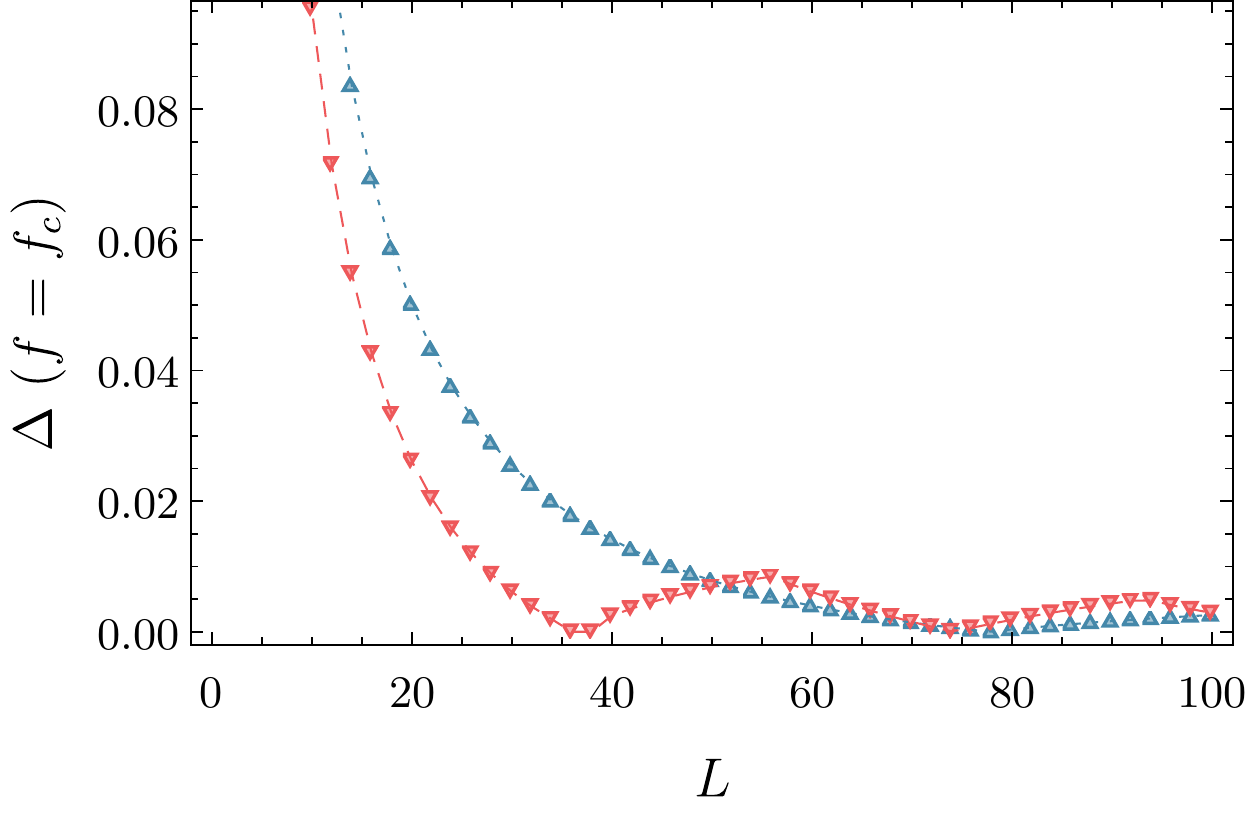}
\begin{picture}(0,0)
\put(-49,21){\includegraphics[height=3cm]{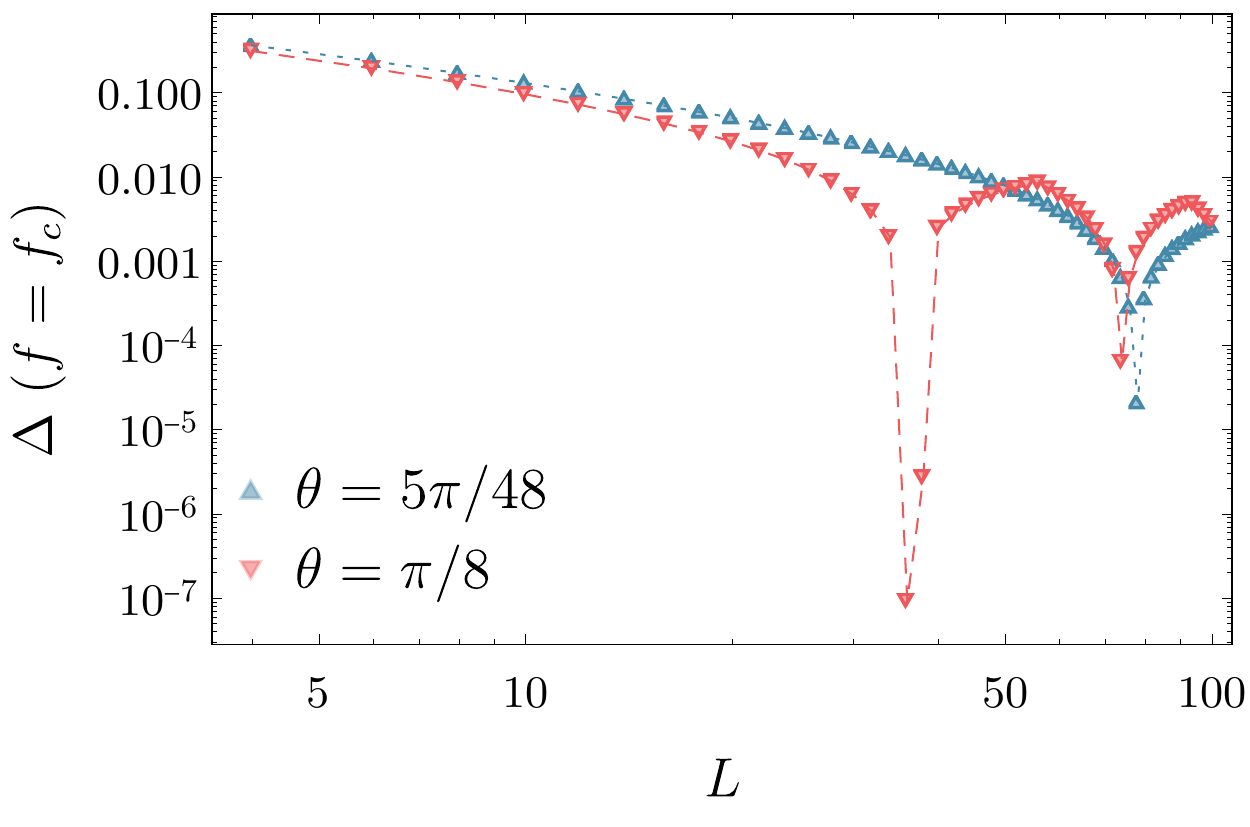}}
\end{picture}}
\subfigure[]{\label{fig:Lifshitz3}\includegraphics[width= 0.485\linewidth]{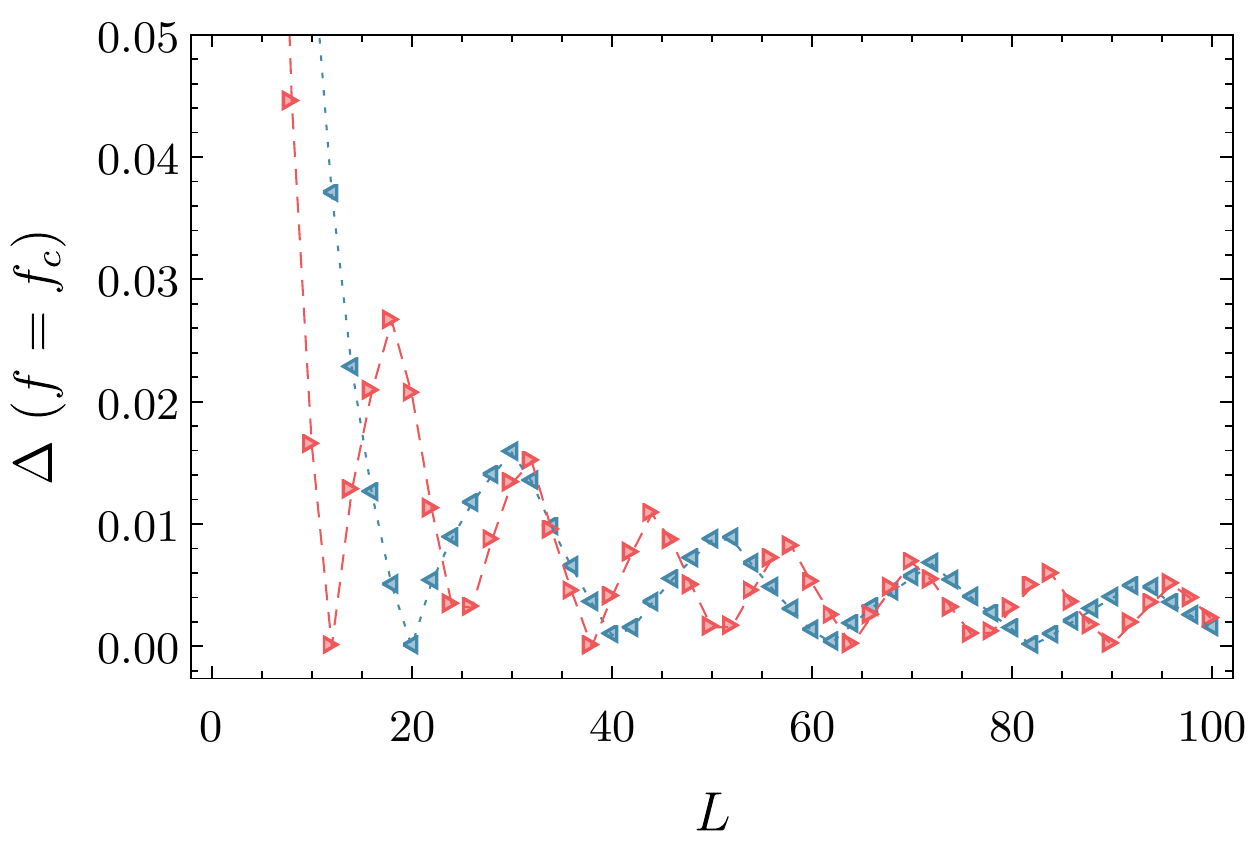}
\begin{picture}(0,0)
\put(-49,21){\includegraphics[height=3cm]{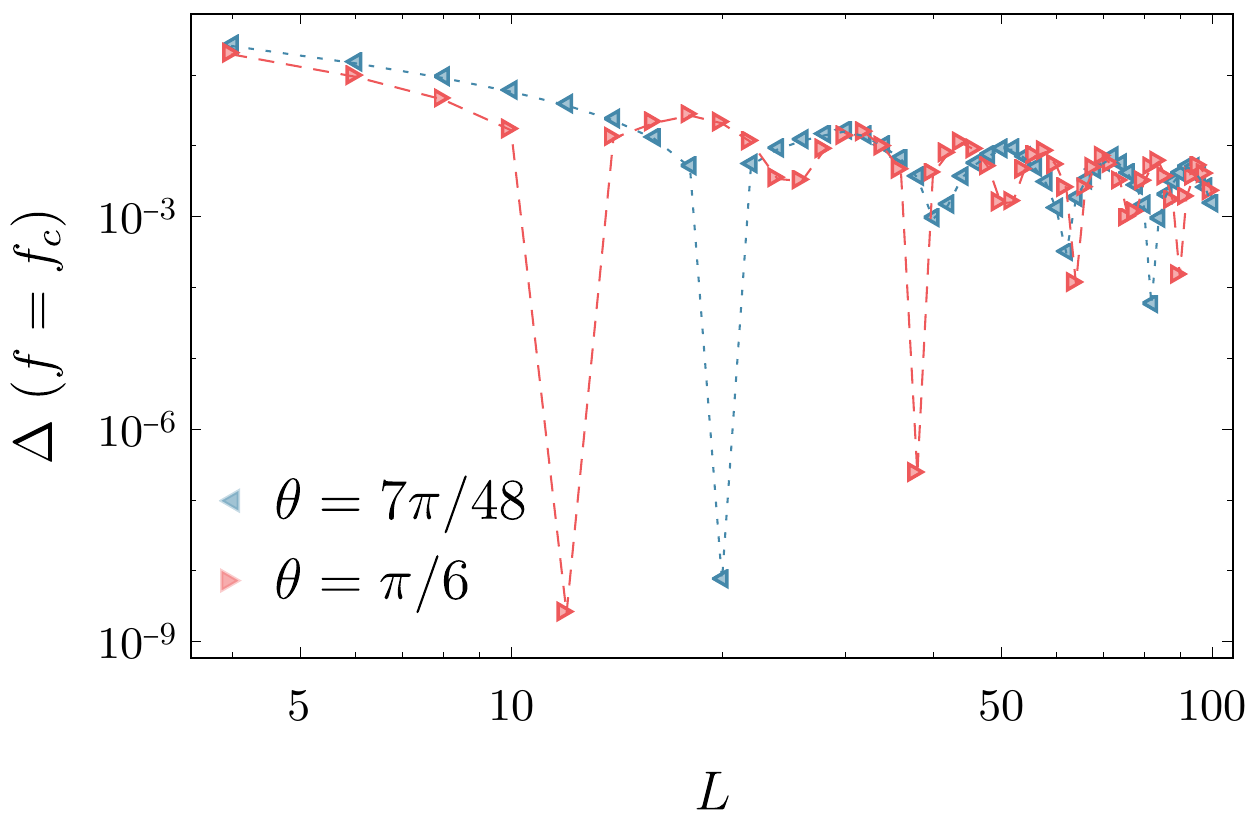}}
\end{picture}}
\caption{Scaling of the gap with the parameter $f$ along the self-dual line $f=J$, $\phi = \theta$ for $L = 100$. For small $\theta$, the gap seemingly closes as a power law while beyond $\pi/12$, such a relation no longer holds. (b) The energy gap as a function of system size along the self-dual line $\phi = \theta \le \pi/12$ at $f=f_c = 0.5$. The length scale of the Lifshitz oscillations for the chiral angles in this figure is greater than 100 sites. (c--d) The energy gap oscillates with the system size along the self-dual line $\phi = \theta > \pi/12$ at $f=f_c = 0.5$. [Inset]: The same, on a logarithmic scale. The minima of the Lifshitz oscillations occur at nonzero $\Delta$ i.e. the gap, although small, does not close.}
\end{figure}

Along the self-dual line, the FSS diagrams (Figs.~\ref{fig:Lifshitz2} and \ref{fig:Lifshitz3}) bring to light an interesting feature: the mass gap oscillates as the system size is varied, with a frequency that increases with $\theta$ up to $\phi = \theta = \pi/6$, beyond which the oscillation amplitudes die out as the system transitions to the incommensurate phase. It is perhaps worth mentioning that oscillatory energy gaps have been known to occur in other three-state \cite{howes1983quantum} and four-state \cite{henkel1984statistical} systems as well. Such features were carefully analyzed \cite{hoeger1985finite} for the one-dimensional XY model in a transverse field, where they can be attributed to analytically demonstrable level crossings. In the CCM, however, the same is owed to different origins. Similar oscillations were observed in the EE of this model by \citet{ZCTH} and studying the shapes of the EE curves, they proposed the empirical relation
\begin{equation}
\mathcal{L} = \phi^{-3.75}+1.16,
\end{equation}
for the length scale of the oscillations, $\mathcal{L}$. Above a certain point, $\phi = \theta > 0.29$ to be precise, $\mathcal{L}$ becomes comparable to (or smaller than) our system size $L = 100$. Hence, despite an immediate onset on tuning even slightly away from $\phi = \theta = 0$, it is only above $\theta = \pi/12$ (amongst the discrete values scanned) that the oscillations become manifestly observable. Since this length scale corresponds to that associated with the incommensurate order \cite{ZCTH}, it is reasonable to believe that the transition between the ordered and disordered phases proceeds through the incommensurate phase as previously suggested \cite{au1987commuting, albertini1989commensurate}. This constitutes evidence to support that there should indeed be a narrow sliver of an incommensurate phase extending all the way to $\phi = \theta = 0$ along the $f = J = 0.5$ line in the phase diagram. 

\section{Analysis as $\lambda \rightarrow 0$}
\label{app:luttinger}

This appendix will review the arguments of Ref.~\onlinecite{haldane1983phase}
for the presence of an intermediate incommensurate phase.

In our formalism, these arguments are most easily presented using the Bose gas action $\mathcal{S}_{\Psi}$ in Eq.~(\ref{SPsi}). We begin with the case $\lambda=0$. Then, $\mathcal{S}_\Psi$ describes a $z=2$ quantum phase transition at $T=0$ with decreasing $s$, associated with a nonanalyticity in the boson density \cite{ssbook}. Let the transition occur at $s=s_c$. Then, at length scales {\it larger\/} than the mean-particle spacing $\xi \sim (s_c -s)^{-1/2}$, we have a Luttinger liquid description of this dilute Bose gas \cite{SSS94}. Such a Luttinger liquid is described by the quantum fields $\theta$ and $\phi$ which obey the commutation relation (we use the notation of Ref.~\onlinecite{ssbook})
\begin{equation}
    [\phi(x), \theta(y)] = i \frac{\pi}{2} \mbox{sgn}(x-y)\,.
\end{equation}
Note that these variables bear no relation to those in the body of the paper.
The boson field is
\begin{equation}
    \Psi \sim e^{i \theta}
\end{equation}
and the action is
\begin{equation}
    \mathcal{S}_\theta = \frac{K}{2 \pi v} \int dx d\tau \left[(\partial_\tau \theta)^2 + v^2 (\partial_x \theta)^2 \right]\,,
\end{equation}
where $v$ is the sound velocity, and $K$ is the dimensionless Luttinger parameter. The main observation we shall need is that the Luttinger parameter $K \rightarrow 1$ in the $s \nearrow s_c$ limit of the $z=2$ quantum phase transition; the dilute Bose gas in this limit is a `Tonks gas', and is described as free fermions.

Now consider turning on a nonzero $\lambda$ in the Luttinger liquid regime. 
Then, the action $\mathcal{S}_\theta$ implies the scaling dimension
\begin{equation}
    \mbox{dim} \left[ \Psi^N \right] = \frac{N^2}{4 K}\,.
\end{equation}
So $\lambda$ is a relevant perturbation to the Luttinger liquid only if 
$N < \sqrt{8K}$. For $K=1$ and $N=3$, $\lambda$ is irrelevant, and so the Luttinger liquid phase ({\it i.e.,} incommensurate phase) is stable.

The weakness in the above argument is that it applies only to the Luttinger liquid phase present for $s<s_c$, and {\it not} to its $z=2$ critical endpoint at $s=s_c$.
To examine the stability of the critical endpoint, we have to study the regime of length scales {\it smaller} than $\xi \sim (s_c -s)^{-1/2}$, where the Luttinger liquid description is not valid \cite{SSS94}. In other words, there is an important issue in the order of limits: the arguments above are for $\lambda \rightarrow 0$ before $s \rightarrow s_c$, but these limits should be taken in the opposite order. 
\bibliographystyle{apsrev4-1_custom}
\bibliography{clock.bib}
\end{document}